\documentclass[reqno,11pt]{article}

\RequirePackage[OT1]{fontenc}
\RequirePackage{amsthm, amsmath, amssymb, amsfonts, natbib, xspace, graphicx, enumerate,multirow}

\usepackage{epsfig,amsmath,amssymb,euscript,amsfonts,mathrsfs}
\usepackage{colortbl,color,graphics,graphicx,epsfig,multirow,float,subfig}
\usepackage{framed}
\usepackage{dsfont}
\usepackage{bbm}
\usepackage{changepage}
\usepackage{nicefrac}
\usepackage{indentfirst} 
\usepackage{comment}
\usepackage{booktabs}
\usepackage{natbib}
\usepackage{arydshln}
\usepackage{xparse}

\NewDocumentCommand{\INTERVALINNARDS}{ m m }{
	#1 {,} #2
}
\NewDocumentCommand{\interval}{ s m >{\SplitArgument{1}{,}}m m o }{
	\IfBooleanTF{#1}{
		\left#2 \INTERVALINNARDS #3 \right#4
	}{
		\IfValueTF{#5}{
			#5{#2} \INTERVALINNARDS #3 #5{#4}
		}{
			#2 \INTERVALINNARDS #3 #4
		}
	}
}
\usepackage[top=1in, bottom=1in, left=1in, right=1in]{geometry}

\allowdisplaybreaks
\newtheorem{theorem}{Theorem}[section]

\newtheorem{assumption}{Assumption}[section]

\theoremstyle{plain}

\theoremstyle{remark}
\newtheorem{remark}{Remark}

\newtheoremstyle{assm} 
{\topsep}                    
{\topsep}                    
{}                               
{}                           
{\scshape}                   
{.}                          
{.5em}                       
{}  
\theoremstyle{assm}

\theoremstyle{definition}

\newcommand{\ignore}[1]{}

\usepackage{hyperref}
\usepackage{xr-hyper}

\DeclareMathOperator*{\argmax}{argmax}
\DeclareMathOperator*{\argmin}{argmin}

\begin{document}

\title{\vspace*{-2cm} Inference for Multiple Change-points in Piecewise Locally Stationary Time Series
}

\author{
Wai Leong Ng\thanks{Department of Mathematics, Statistics and Insurance, The Hang Seng University of Hong Kong, Shatin, N.T., Hong Kong. Email: {\ttfamily wlng@hsu.edu.hk, xinyitang@hsu.edu.hk}. 
Research has been supported in part by HKSAR-RGC-FDS Project Nos. UGC/FDS14/P01/20, UGC/FDS14/P04/21 and UGC/FDS14/P05/22.}, Xinyi Tang\\
The Hang Seng University of Hong Kong
\and
Mun Lau Cheung, Jiacheng Gao, Chun Yip Yau\thanks{Department of Statistics, The Chinese University of Hong Kong, Shatin, N.T., Hong Kong. 
Email: {\ttfamily 1155048823@link.cuhk.edu.hk, jcgao@link.cuhk.edu.hk, cyyau@sta.cuhk.edu.hk}. 
Supported in part by HKSAR-RGC-GRF Nos. 14302423, 14304221 and 14304425.} \\
The Chinese University of Hong Kong 
\and
Holger Dette\thanks{Fakultät für Mathematik, Ruhr-Universit{\"a}t Bochum, Germany. Email: {\ttfamily holger.dette@ruhr-uni-bochum.de}. Supported
by the Deutsche Forschungsgemeinschaft (DFG)  TRR
391 {\it Spatio-temporal Statistics for the Transition of Energy and Transport}, project number
520388526.}\\
Ruhr-Universit{\"a}t Bochum
}
\maketitle

\begin{abstract}

Change-point detection and locally stationary time series modeling are two major approaches for the analysis of 
non-stationary data. The former aims to identify stationary phases by detecting abrupt changes in the dynamics of a  time series model, 
while the latter employs (locally) time-varying models to describe smooth changes in  dependence structure of a time series. 
However, in some applications, abrupt and smooth changes can co-exist, and neither of the two approaches alone can model the data adequately. 
In this paper, we propose a novel likelihood-based procedure for the inference of multiple change-points in locally stationary time series. 
In contrast to traditional change-point analysis where an abrupt change occurs in a real-valued parameter,  
a change in locally stationary time series occurs in a parameter curve, and can be classified as a jump or a kink depending on whether the curve is discontinuous or not. 
We show that the proposed method can consistently estimate the number, locations, and the types of change-points. 
Two different asymptotic distributions corresponding respectively to jump and kink estimators are also established. 
Extensive simulation studies and a real data application to financial time series are provided.  	
	
\end{abstract}


\section{Introduction}\label{sec:intro}
\def\theequation{1.\arabic{equation}}
\setcounter{equation}{0}
Change-point analysis for time series has received considerable attention in recent decades. 
With estimated change-points, data can be segmented into pieces of stationary time series, which provides a parsimonious model for 
non-stationary time series. See, for example, the recent monograph by  \cite{horvath2024change} and the review article by \cite{cho:kirch:2024}, for a comprehensive review of change-point analysis in time series. 
However, the assumption of piecewise stationarity may be violated in some situations. 
For example, \cite{Adak1998} studied an earthquake dataset and discovered that there are excessive change-points upon fitting a piecewise stationary time series model. Instead, a piecewise locally stationary model provided a  much better fit to the data. 
Intuitively, a locally stationary time series models can be locally approximated by a stationary process, and 
its dynamics is continuously varying over time. 
The fundamental modelling  approach for the analysis of  non-stationary data is due to \cite{Dahlhaus1996a,Dahlhaus1997,Dahlhaus2000} who developed  a rigorous asymptotic theory based on time-varying spectral representations \citep[see also][among others]{Dahlhaus2006,dahlhaus2009empirical}. An alternative   modelling paradigm in the time domain was introduced by \cite{wu2005} and defines local stationarity through smoothly time-varying data-generating mechanisms and physical dependence measures \citep[see also][among others]{ZhouWu2010,WuZhou2011}.


When both gradually evolving features and structural breaks occur in a time series, neither a piecewise stationary model nor a locally stationary model is appropriate for modelling the time series. 
In such situations, change-point detection in piecewise locally stationary time series is necessary for understanding the general structure of the data. 
There is relatively little literature on detecting change points in such scenarios. Exemplary we mention   \cite{Shumway2007} who  studied earthquake records in Nicaragua, and discovered that sharp changes occurred in the time-varying power spectrum and   
\cite{vogt2015detecting} who developed a nonparametric method to estimate the point where a time series changes smoothly from a stationary to a locally stationary phase. More recently \cite{dettewuzhou} considered the problem of detecting one (abrupt) change point in the correlation structure of a piecewise locally stationary time series, and  \cite{CASINI2024} investigated the minimax-optimal testing problem for one change point in  the time varying spectral density.
In contrast to the classical  setting, detecting abrupt changes in locally stationary time series is very challenging. 
Specifically, the smooth parameter curve may change abruptly in the level, resulting in a discontinuity, or may change abruptly in the slope while the curve remains  continuous. These two situations are known as {\it jump} and {\it kink}, respectively, and will be shown to have different asymptotic properties. 
Indeed, the literature so far only focused on jumps, and did not consider kinks.    
To the best of our knowledge, the problem of inference on both jumps and kinks in locally stationary time series models remains open. 

In this paper, we develop a multiple change-point inference procedure for piecewise locally stationary time series with possibly jumps and kinks in its parameter curves. 
We aim at estimating the number and locations of change-points occurred in the time-varying spectrum, and construct a confidence interval for each change-point.
In particular, the proposed method consists of three steps: i) a quick detection of a set of potential change-points using scan statistics which measure the differences between different segments; ii) using a minimum description length (MDL) based criterion to estimate the number and locations of change-points; iii) making a refinement for the location of each change-point. Similar approaches on multiple change-point detection for traditional stationary time series can be found in \cite{LRSM2015}, \cite{Dette2015}, \cite{eichinger2018mosum} and \cite{detteecklevetter}. 
However, the simultaneous occurrence of jumps and kinks requires fundamentally new methodologies for their detection and differentiation. Addressing this problem in the setting of locally stationary time series  involves substantial and non-trivial analytical challenges.

This paper is organized as follows. Section \ref{sec:background} reviews some background on locally stationary time series, time-varying autoregressive models and parameter estimation. In Section \ref{sec:cpinference}, we introduce the multiple change-point problem settings and present  the proposed methodology. The asymptotic properties of the proposed method are discussed in Section \ref{sec:asymp_prop}. 
Sections \ref{sec:simulation} and  \ref{sec:realdata} present extensive simulation studies and a real data application to financial time series, respectively. Section \ref{sec:conclusion} concludes. Detailed simulation results and technical proofs of theorems and lemmas are provided in the Supplementary Materials.

\section{Backgrounds on Locally Stationary Time Series} \label{sec:background}
\def\theequation{2.\arabic{equation}}
\setcounter{equation}{0}

Following \cite{Dahlhaus1996a}, we introduce some backgrounds on locally stationary processes, including the time-varying autoregressive (tvAR) model and its estimation theory. 

\subsection{Locally Stationary Time Series}
A time series $\{X_{t,T}: t=1,2,\dots,T\}$ is called a locally stationary time series with a transfer function $A_{t,T}^0:\mathbb{R} \rightarrow \mathbb{C}$ and a trend $\mu:[0,1] \rightarrow \mathbb{R}$ if there exists a representation of the form
\begin{equation}\label{loc.stat.proc}
	X_{t,T}=\mu \Big(\frac{t}{T} \Big)+\int_{-\pi}^{\pi} e^{i \lambda t} A_{t,T}^0(\lambda)d \zeta(\lambda) \, ,
\end{equation}
where the following conditions hold:

\begin{enumerate}
	\item The quantity $\zeta(\cdot)$ is a complex-valued stochastic process  on $[-\pi , \pi]$ with $\overline{\zeta(\lambda)}=\zeta(-\lambda)$ and
	\begin{equation*}
		\mbox{cum} \{d \zeta(\lambda_1), d \zeta(\lambda_2), \dots, d \zeta(\lambda_k)\}=\nu \Big (\sum_{j=1}^{k} \lambda_j \big ) h_k(\lambda_1, \lambda_2, \dots, \lambda_{k-1}) d \lambda_1 d \lambda_2 \dots d \lambda_k \, ,
	\end{equation*}	
	where $\mbox{cum}\{\cdots\}$ denotes the cumulant of $k$-th order, $h_1=0$, $h_2(\lambda)=1$, $|h_k(\lambda_1,\lambda_2,\dots,\lambda_{k-1})|\leq c_k$ for some constant $c_k$ for all $k\in\mathbb{N}$, and $\nu(\lambda) = \sum_{j=-\infty}^{\infty} \delta(\lambda + 2\pi j)$ is the $2\pi$-periodic extension of the Dirac delta function.
	
	\item There exists a constant $K$ and a $2 \pi$-periodic function $A:[0, 1] \times \mathbb{R} \rightarrow \mathbb{C}$ with $A(u,-\lambda) = \overline{A(u, \lambda)}$ and
	\begin{eqnarray}\label{def.A}
		\sup_{t\in \{1,2,\ldots,T\},\lambda\in\mathbb{R}}|A_{t,T}^0(\lambda)-A(t/T,\lambda)| \leq KT^{-1}\,,
	\end{eqnarray}
	for all $T$. Also, the functions $A(u,\lambda)$ and $\mu(u)$ are continuous in $u$.
\end{enumerate}
The first condition ensures that $\zeta(\lambda)$ is an orthogonal zero-mean increment process for the Fourier-Stieltjes stochastic integral in \eqref{loc.stat.proc}.
The second condition on the smoothness of $A$ and $\mu$ guarantees that $X_{t,T}$ in 
\eqref{loc.stat.proc} is nearly stationary in a small neighborhood of $t$, which explains the term ``locally stationary".  
Note that $f(u,\lambda)=|A(u,\lambda)|^2$ is called the time-varying spectral density.
\cite{Dahlhaus1996a} showed that under some smoothness conditions on $A$, we have that 
\begin{equation*}
\int_{-\pi}^{\pi} |f_T(u,\lambda)-f(u,\lambda)|^2 d\lambda = o(1)\,,
\end{equation*}
where $f_T(u,\lambda) = \frac{1}{2\pi} \sum^{\infty}_{s=-\infty} \text{Cov}(X_{[uT-s/2],T},X_{[uT+s/2],T}) \exp(-i \lambda s)$ is the Wigner-Ville spectrum. 
This implies that if the spectral representation \eqref{loc.stat.proc} exists with $A_{t,T}^0(\lambda)$ satisfying \eqref{def.A} for a smooth $A(u,\lambda)$, then $|A(u,\lambda)|^2$ is uniquely determined by $\{X_{t,T}\}_{t=1,\dots,T}$.
For the sake of simplicity, we assume that the time series is centred (i.e. $\mu \equiv 0 $) throughout the paper.

As a locally stationary time series has a continuous time-varying spectral density which varies in both the frequency domain and time domain, we use a localized version of the periodogram to estimate the spectrum.   
Specifically, we choose a positive integer $h=h(T)$ as the window length. For each time point $t=h,h+1,\ldots,T$, the local periodogram is defined as
\begin{equation} \label{localperi}
	I_h(t,\lambda)=\frac{1}{2 \pi h} \Big |\sum_{k=t-h+1}^t X_{k,T} \exp^{-ik \lambda} \Big |^2\,,
\end{equation}
which is the usual periodogram computed from the local data $\{X_{t-h+1,T},X_{t-h+2,T},\ldots, X_{t,T}\}$.

\subsection{Time-varying Autoregressive Models}

Analogous to the autoregressive (AR) model for stationary time series, 
the time-varying autoregressive (tvAR) model provides a parsimonious way for modeling locally stationary time series. In particular, 
 tvAR($p$) is an AR($p$) model with AR coefficients changing continuously over time; see, for example, \cite{Eric2005}, \cite{Kim2001}  and \cite{Dahlhaus2012}. The tvAR($p$) process has a  representation of the form
\begin{equation} \label{eq_tvAR}
	X_{t,T}=\phi_1\left(\frac{t}{T}\right)X_{t-1,T}+\dots+\phi_p \left(\frac{t}{T}\right)X_{t-p,T}+\sigma \left(\frac{t}{T}\right) \epsilon_{t}\, ,
\end{equation}
where $\{\epsilon_{t}\}_{t \in \mathbb{Z}}$ is a centered unit-variance Gaussian white noise process, $\phi_1 , \dots , \phi_p:[0,1] \rightarrow \mathbb{R}$ and $\sigma : [0,1] \rightarrow (0,\infty)$ are  continuous on $[0,1]$ and differentiable on $(0,1)$ with bounded derivatives. \cite{Dahlhaus1996a} proves that the tvAR($p$) process \eqref{eq_tvAR} is locally stationary with a time-varying spectral density
\begin{equation}\label{tvspecden}
	f(u,\lambda)=\frac{\sigma^2(u)}{2\pi} \Big |1-\sum_{j=1}^p \phi_j(u)\exp(-i \lambda j)\Big |^{-2}\,,
\end{equation}
where $u \in [0,1]$ and $\lambda \in [-\pi,\pi]$. 
%
According to the Stone-Weierstress theorem \citep{Stone1948}, every continuous function defined on $[0,1]$ can be uniformly approximated  by a polynomial. Therefore, without much loss of generality, we focus on a  tvAR process of the form \eqref{eq_tvAR} with 
the time-varying parameters $\phi_j(\cdot)$ and $\sigma(\cdot)$ specified by polynomial functions of order $q$ on $[0,1]$. 
Thus, the time varying AR-process 
has the 
representation of the form
\begin{equation} \label{segmentmodel}
	X_{t,T}=\sum_{i=1}^{p} \Big [\sum_{j=0}^{q} \phi_{ij} \Big(\frac{t}{T}\Big)^j\Big] X_{t-i,T}+\Big[\sum_{j=0}^{q} \sigma_{j} \Big(\frac{t}{T}\Big )  \epsilon_{t} = \sum_{i=1}^{p} \phi_{i,\boldsymbol{\theta}}\Big (\frac{t}{T}\Big) X_{t-i,T}+\sigma_{\boldsymbol{\theta}} \Big(\frac{t}{T}\Big) \epsilon_{t}\,,
\end{equation}
where  $\boldsymbol{\theta}=(\phi_{10},\ldots,\phi_{1q},\phi_{20},\ldots,\phi_{2q},\ldots,\phi_{p0},\ldots,\phi_{pq},\sigma_{0},\ldots,\sigma_{q})'$ are the parameters.


To avoid the high computational complexity of evaluating the full log-likelihood function for the estimation of the parameter $\boldsymbol{\theta}$  in model\eqref{segmentmodel}, 
\cite{Dahlhaus1997} proposed the maximum conditional Gaussian log-likelihood estimation, which  
obtains the parameter estimate $\hat{\boldsymbol{\theta}}$ of $\boldsymbol{\theta}$ by maximizing 
\begin{eqnarray}\label{L.GL}
L^{GL}(\boldsymbol{\theta},\mathbf{X}) = -\frac{1}{2} \sum_{t=p_{max}+1}^{T} \Big[ \log \left(2\pi \sigma^2_{\boldsymbol{\theta}} \Big(\frac{t}{T}\right) \Big) + \frac{\left(X_{t,T}-\sum_{i=1}^{p_{max}}\phi_{i,\boldsymbol{\theta}}\left(\frac{t}{T}\right) X_{t-i,T} \right)^2}{\sigma^2_{\boldsymbol{\theta}} \left(\frac{t}{T}\right)}\Big]\,.
\end{eqnarray}
The asymptotic properties of $\hat{\boldsymbol{\theta}}$, including the consistency and asymptotic normality, have been derived  in \cite{Dahlhaus1997}. 

\section{Multiple Change-point Estimation for Piecewise tvAR models}
\def\theequation{3.\arabic{equation}}
\setcounter{equation}{0}
\label{sec:cpinference}

In this section, we describe the problem settings in Section \ref{sec:CPL}, and introduce the proposed method for multiple change-point estimation 
in Sections \ref{scan} - \ref{detection}.

\subsection{Problem Settings}\label{sec:CPL}\label{sec.3.1}
Assume that the observed time series $\{X_{t,T}: t=1,2,\dots,T\}$ is partitioned into $m+1$ segments by $m$ change-points located at $C = \{\tau_1,\tau_2,\dots,\tau_m\}$.   
We set $\tau_0 \triangleq 0$ and $\tau_{m+1} \triangleq T$, and denote the $k$-th segment as   $\mathbf{X}_{k,T}=\{X_{\tau_{k-1}+1,T},X_{\tau_{k-1}+2,T},\dots,X_{\tau_k,T}\}$.
We assume that $\mathbf{X}_{k,T}$ follows the tvAR($p_k$) model defined in \eqref{segmentmodel} with parameters 
\begin{eqnarray}\label{para.vec}
	\boldsymbol{\theta}_k=(\phi_{k10},\ldots,\phi_{k1q_{k}},\phi_{k20},\ldots,\phi_{k2q_{k}},\ldots,\phi_{kp_{k}0},\ldots,\phi_{kp_{k}q_{k}},\sigma_{k0},\ldots,\sigma_{kq_{k}})'\,,
\end{eqnarray}
which correspond to smooth parameter curves of polynomial functions $\phi_{k,i,\boldsymbol{\theta}_{k}}(u)=\sum_{j=0}^{q_k}\phi_{kij}u^{j}$, $i=1,\ldots,p_k$, and 
   $\sigma_{k,\boldsymbol{\theta}_{k}}(u)=\sum_{j=0}^{q_k}\sigma_{kj}u^{j}$, where $u \in (\tau_{k-1}/T, \tau_k/T] \triangleq (r_{k-1},r_k] \subset [0,1]$.   
For simplicity, we assume that the order of the tvAR processes $p_k$ and the degree of polynomial functions $q_k$ are bounded above by integers $p_{max}$ and $q_{max}$ respectively. It can include tvAR models with smaller model order and smaller degree of the polynomial function by setting the corresponding entries in $\boldsymbol{\theta}$ to zero. 

In contrast to change-points in stationary time series models, since locally stationary time series models are specified by smooth parameter curves,  
change-points 
can be classified into the following two types. 
We say that a change in segment $k$ at relative location $r_k$ is a {\it jump} if any of the parameter curves breaks, i.e.,
\begin{eqnarray}\label{def.jump}
\phi_{k,i,\boldsymbol{\theta}_{k}}(r_k)\neq \phi_{k+1,i,\boldsymbol{\theta}_{k+1}}(r_k)\,, \mbox{ for some } i=1,\ldots,p_{max}\,, \mbox{ or } 
\sigma_{k,\boldsymbol{\theta}_{k}}(r_k)\neq \sigma_{k+1,\boldsymbol{\theta}_{k+1}}(r_k)\,.
\end{eqnarray} 
On the other hand, we say that a change in segment $k$ at relative location $r_k$ is a {\it kink} if all curves are continuous but some of them have an abrupt change in the slope, i.e.,   
\begin{eqnarray}\label{def.kink0}
\phi_{k,i,\boldsymbol{\theta}_{k}}(r_k) &=& \phi_{k+1,i,\boldsymbol{\theta}_{k+1}}(r_k)\,, \mbox{ for all } i=1,\ldots,p_{max}\,, \mbox{ and } 
\sigma_{k,\boldsymbol{\theta}_{k}}(r_k) = \sigma_{k+1,\boldsymbol{\theta}_{k+1}}(r_k)\,,
\end{eqnarray}
	{but}
    \begin{eqnarray}
	\label{def.kink}
	\phi^{(1)}_{k,i,\boldsymbol{\theta}_{k}}(r_k)&\neq& \phi^{(1)}_{k+1,i,\boldsymbol{\theta}_{k+1}}(r_k)\,, \mbox{ for some } i=1,\ldots,p_{max}\,, \mbox{ or } 
	\sigma^{(1)}_{k,\boldsymbol{\theta}_{k}}(r_k)\neq \sigma^{(1)}_{k+1,\boldsymbol{\theta}_{k+1}}(r_k)\,,
\end{eqnarray} 
where the superscript $(1)$ stands for taking the first derivative. 

To describe the whole piecewise tvAR process, it requires the set of jumps $J \subseteq C$, the set of kinks $K=C \setminus J$, and the model parameter curves in each segment. For any segment between two jumps, the model parameters can freely be taken in the form of \eqref{para.vec}. On the other hand, if 
 $g_v$ consecutive kinks $\tau_{v},\ldots,\tau_{v+g_v-1} \in K$ lie between two jumps or end points of the data, i.e. $\tau_{v-1}, \tau_{v+g_v}\in J \cup \{0,T\}$ where $g_v \in \mathbb{N}$, then,  
  in view of continuity of parameter curves across kinks, the model parameters in segments $v$ to $v+g_v$ have to satisfy $g_v$ constraints given by \eqref{def.kink0} with $k=v,v+1\ldots,v+g_v-1$. 
 Thus, parameter estimation requires a constraint optimizations. To tackle this problem, we propose the following re-parametrization for the $g_v+1$ segments that covers $g_v$ consecutive kinks. Specifically, for $t\in\{\tau_{v-1}+1,\ldots,\tau_{v+g_v}\}$, the re-parametrized model is defined as
 \begin{eqnarray}\label{repara_kink_model}
 	X_{t,T}&=&
    \sum_{i=1}^{p_{max}} \phi^{(i)}_{\boldsymbol{\gamma},\boldsymbol{\alpha},v,g_v} \left(\frac{t}{T}\right) X_{t-i,T} + 
 	\sigma_{\xi,\boldsymbol{\beta},v,g_v}\left(\frac{t}{T}\right) \epsilon_{t,T}\\
 	& = & {\boldsymbol{X}_{t-p_{max}}^{t-1}}' \phi_{\boldsymbol{\gamma},\boldsymbol{\alpha},v,g_v} \left(\frac{t}{T}\right) + \sigma_{\xi,\boldsymbol{\beta},v,g_v}\left(\frac{t}{T}\right) \epsilon_{t,T} \nonumber\,,
 \end{eqnarray}
where $\boldsymbol{X}_{a}^{b}=(X_{b,T},X_{b-1,T},\dots,X_{a,T})'$, 
$\phi_{\boldsymbol{\gamma},\boldsymbol{\alpha},v,g_v} \left(\frac{t}{T}\right)=\left(\phi^{(1)}_{\boldsymbol{\gamma},\boldsymbol{\alpha},v,g_v} \left(\frac{t}{T}\right),\ldots,\phi^{(p_{max})}_{\boldsymbol{\gamma},\boldsymbol{\alpha},v,g_v} \left(\frac{t}{T}\right) \right)'$, and for $i=1,\ldots,p_{max}$, 
\begin{eqnarray*}
	\phi^{(i)}_{\boldsymbol{\gamma},\boldsymbol{\alpha},v,g_v} \Big (\frac{t}{T}\Big) & = & \gamma_{vi} + \sum_{j=1}^{q_{v}} \alpha_{vij}\Big(\frac{t}{T}-r_v\Big)_{-}^j + \sum_{k=v}^{v+g_v-2} \sum_{j=0}^{q_{k+1}} \Bigg(\prod_{\ell\neq j} \frac{t-\tau_{k,k+1}^{(\ell)}}{\tau_{k,k+1}^{(j)}-\tau_{k,k+1}^{(\ell)}}\Bigg) \alpha_{(k+1)ij} \mathbbm{1}_{\{\tau_{k} < {t} \leq \tau_{k+1}\}} \\
	& & \quad\quad\quad\quad + (\gamma_{(v+g_v-1)i}-\gamma_{vi})\mathbbm{1}_{\{\tau_{v+g_v-1} < {t} \}} +\sum_{j=1}^{q_{v+g_v}} \alpha_{(v+g_v)ij}\Big(\frac{t}{T}-r_{v+g_v-1}\Big)_{+}^j\,,\\
	\sigma_{\xi,\boldsymbol{\beta},v,g_v}\Big(\frac{t}{T}\Big) & = & \xi_{v}+ \sum_{j=1}^{q_{v}} \beta_{vj} \Big(\frac{t}{T}-r_v\Big)_{-}^j + \sum_{k=v}^{v+g_v-2} \sum_{j=0}^{q_{k+1}} \Bigg(\prod_{\ell\neq j} \frac{t-\tau_{k,k+1}^{(\ell)}}{\tau_{k,k+1}^{(j)}-\tau_{k,k+1}^{(\ell)}}\Bigg) \beta_{(k+1)j} \mathbbm{1}_{\{\tau_{k} < {t} \leq \tau_{k+1}\}}\\
& & \quad\quad\quad\quad + (\xi_{(v+g_v-1)i}-\xi_{vi})\mathbbm{1}_{\{\tau_{v+g_v-1} < {t} \}} +\sum_{j=1}^{q_{v+g_v}} \beta_{(v+g_v)j} \Big(\frac{t}{T}-r_{v+g_v-1}\Big)_{+}^j\,,
\end{eqnarray*}
with $(a)_{-}=\min\{a,0\}$, $(a)_{+}=\max\{a,0\}$, $r_k=\frac{\tau_k}{T}$,
$\alpha_{(k+1)i0} \triangleq \gamma_{ki}-\gamma_{vi}$, $\alpha_{(k+1)iq_{k+1}}\triangleq\gamma_{(k+1)i}-\gamma_{vi}=\alpha_{(k+2)i0}$,  
$\beta_{(k+1)0}\triangleq\xi_{k}-\xi_{v}$, $\beta_{(k+1)q_{k+1}}\triangleq\xi_{(k+1)}-\xi_{v}=\beta_{(k+2)0}$,  
$\tau_{k,k+1}^{(j)}=\big\lfloor\tau_k+(\frac{\tau_{k+1}-\tau_k}{q_{k+1}})j\big\rfloor$ for $k=v,\ldots,v+g_v-2$, and $\sum_{k=a}^{b} c_k \equiv 0$ for $a>b$.


The idea of the re-parametrization is to use Lagrange interpolating polynomials to represent the parameter curve of each segment, so we can control 
the end points of each curve without using constraints. 
The re-parametrized model \eqref{repara_kink_model} for $g_v$ consecutive segments has a parameter set  $\boldsymbol{\eta}_{v,g_v}=(\boldsymbol{\gamma}_{v},\ldots,\boldsymbol{\gamma}_{v+g_v-1},\xi_{v},\ldots,\xi_{v+g_v-1},\boldsymbol{\alpha}_v,\boldsymbol{\beta}_v,\ldots,\boldsymbol{\alpha}_{v+g_v},\boldsymbol{\beta}_{v+g_v})'$, 
where $\boldsymbol{\gamma}_{u}=(\gamma_{u1},\ldots,\gamma_{up_{max}})'$ for $u=v,\ldots,v+g_v-1$, $\boldsymbol{\alpha}_k=(\alpha_{k11},\ldots,\alpha_{kp_{max}q_{k}})'$ and
$\boldsymbol{\beta}_k=(\beta_{k1},\ldots,\beta_{kq_{k}})'$ for $k=v$ and $v+g_v$,
and $\boldsymbol{\alpha}_k=(\alpha_{k11},\ldots,\alpha_{kp_{max}(q_{k}-1)})'$, $\boldsymbol{\beta}_k=(\beta_{k1},\ldots,\beta_{k(q_{k}-1)})'$ for $k=v+1,\ldots,v+g_v-1$. 
Note that the re-parametrized model \eqref{repara_kink_model} can allow segments with tvAR model order less than $p_{max}$ by setting the corresponding parameters, $\gamma_{ki}$ and $\alpha_{kij}$, to zero in $\boldsymbol{\eta}_{v,g_v}$. 
It can be shown that the parameter $\boldsymbol{\eta}_{v,g_v}$ involved in the re-parametrized kink model are in one-to-one correspondence to the parameter $(\boldsymbol{\theta}_v,\ldots,\boldsymbol{\theta}_{v+g_v})$ in the original kink model under the continuity constraints. To facilitate presentation, we will
 only using the re-parametrized kink model when discussing parameter curves estimation. 


We remark that kink models in regression settings have received considerable attention in recent studies, see \cite{hansen2017kink1}, \cite{chen2020kink2} and \cite{zhong2022estimation}. However, for regression kink models, the kinks occur in the regression equation $\mathbb{E}(Y|X)=f(X)$ with respect to the covariate $X$, and hence $f(X)$ is continuous with an abrupt change in the slope at $X=\tau$. On the other hand, the kinks under consideration in this paper are in the parameter curves $\phi_{i,\boldsymbol{\theta}}(u),\sigma_{\boldsymbol{\theta}}(u)$ of the tvAR($p$) model with respect to time $u$, not the covariates $\{X_{[uT-1],T},X_{[uT-2],T},\ldots,X_{[uT-p],T}\}$.

Recall that the tvAR($p$) model in \eqref{eq_tvAR} has a time-varying spectral density in \eqref{tvspecden}. Hence, the time-varying spectral density for the $k$-th segment is given by  
\begin{equation*}
	f_k(u,\lambda)=\frac{\big(\sum_{l=0}^{q_k} \sigma_{kl} \left(u\right)^l\big)^2}{2\pi} \Big  | 1-\sum_{j=1}^{p_k}\Big  (\sum_{l=0}^{q_k} \phi_{kjl}\left(u\right)^l\Big  )\exp(-i \lambda j) \Big  |^{-2}\,.
\end{equation*}	
Denote the first derivative of the spectral density as $f^{(1)}_k(u,\lambda)=\frac{\partial{}}{\partial{u}}f_k(u,\lambda)$. Thus, the time-varying spectral density for the whole process is $f(u,\lambda)=\sum_{k=1}^{m+1} f_k(u,\lambda) \mathbbm{1}_{\{r_{k-1} < u \leq r_k\}}(u)$ and its first derivative is $f^{(1)}(u,\lambda)=\sum_{k=1}^{m+1} f^{(1)}_k(u,\lambda) \mathbbm{1}_{\{r_{k-1} < u \leq r_k\}}(u)$.
Note that jumps and kinks in parameter curves for tvAR models correspond to jumps and kinks in the spectral density function $f(u,\lambda)$ for some frequency $\lambda \in [-\pi,\pi]$ respectively, see Section \ref{jumpkinkillust} in the Supplementary Materials for some illustrations.
It will be shown in Section \ref{sec:asymp_prop} that jumps and kinks lead to different asymptotic behaviors of the change-point estimator.  

In the following subsections, we introduce the three-step procedure for multiple change-point estimation in piecewise locally stationary time series.

\subsection{First Step: Quick Scanning for Potential Change-points} \label{scan}
In the first step, we employ scan statistics to search for potential change-points in the spectral density with a low computational cost.  Specifically, we  
 consider the cumulative differences
\begin{equation*}
D_h(t,w)=\frac{1}{h}\sum_{k=-w}^w \Big [I_h \Big (t+h,\frac{2\pi k}{h} \Big )-I_h \Big (t,\frac{2\pi k}{h} \Big  ) \Big ]\,
\end{equation*}
to scan for potential jumps in the spectral density,
where $t=h,h+1,\ldots,T-h$, $w=0,1,2,\dots,\frac{h}{2}$, and $I_h$ is the local periodogram defined in \eqref{localperi}. 
The constant $h=h(T)$ is a pre-specified even integer called {\it window radius} which represents half of the sample size involved in $D_h$.    
Note that the local periodograms $I_h \left(t+h,\lambda \right)$ and $I_h \left(t,\lambda \right)$ can be viewed as estimates of 
the averaged time varying spectral densities $\frac{T}{h} \int_{{t}/{T}}^{{(t+h)}/{T}} f(u,\lambda)\, du$ and $\frac{T}{h} \int_{{(t-h)}{T}}^{{t}/{T}} f(u,\lambda)\, du$ on the intervals $[\frac{t}{T},\frac{t+h}{T}]$ and $[\frac{t-h}{T},\frac{t}{T}]$, respectively.   
Thus, $D_h(t,w)$ is an estimator of the difference 
\begin{eqnarray*}
D_{h,T}(t,w)=\frac{T}{h} \Big ( \int_{\frac{-2w\pi}{h}}^{\frac{2w\pi}{h}} \int_{\frac{t}{T}}^{\frac{t+h}{T}} f(u,\lambda) du d\lambda - \int_{\frac{-2w\pi}{h}}^{\frac{2w\pi}{h}} \int_{\frac{t-h}{T}}^{\frac{t}{T}} f(u,\lambda) du d\lambda\Big )\,.
\end{eqnarray*} 
between the averaged time varying spectral densities before and after time $t$.
Observe that no jump occurs at time $t$ if and only if $D_{h,T}(t,w)=0$ for all $w$. 
Therefore, a large value of $\sup_{w\in \{0,1,2,\dots,
{h}/{2}\}}|D_h(t,w)|$ indicates a jump in the spectral density at time $t$.  
Similar statistics for detecting abrupt change-points in frequency domain time series have been studied in \cite{Shumway2007} and \cite{Dette2015}.

On the other hand, to our best knowledge, estimation of kinks in the spectral density have not been investigated. 
To tackle this, we propose the cumulative difference
	\begin{eqnarray*}
	D^{(1)}_{\widetilde{h}}(t,w) & = & \frac{T}{\widetilde{h}} \Big [\frac{1}{{\widetilde{h}}}\sum_{k=-w}^w \Big \{ I_{\widetilde{h}} \Big(t+2{\widetilde{h}},\frac{2\pi k}{{\widetilde{h}}} \Big)-I_{\widetilde{h}} \Big(t+{\widetilde{h}},\frac{2\pi k}{{\widetilde{h}}} \Big) -I_{\widetilde{h}} \Big(t,\frac{2\pi k}{{\widetilde{h}}} \Big) + I_{\widetilde{h}} \Big(t-{\widetilde{h}},\frac{2\pi k}{{\widetilde{h}}} \Big)\Big \} \Big]\\
	&=& 
	\frac{T}{\widetilde{h}} \left[D_{\widetilde{h}}(t+{\widetilde{h}},w) - D_{\widetilde{h}}(t-{\widetilde{h}},w)\right]\,,
	\end{eqnarray*}
	where $t=2\widetilde{h}, 2\widetilde{h}+1,\ldots,T-2\widetilde{h}$, $w=0,1,2,\dots,\frac{{\widetilde{h}}}{2}$. 
	The window radius for $D^{(1)}_{\widetilde{h}}(t,w)$ is $2\widetilde{h}$, where $\widetilde{h}$ is also a pre-specified even integer. 	
	The statistic $D^{(1)}_{\widetilde{h}}(t,w)$ can be viewed as a numerical derivative of $D_h(t,w)$ and provides  an estimate of the difference 
    $$
    D^{(1)}_{\widetilde{h},T}(t,w)=\frac{T}{\widetilde{h}} \Big( \int_{\frac{-2w\pi}{\widetilde{h}}}^{\frac{2w\pi}{\widetilde{h}}} \int_{\frac{t}{T}}^{\frac{t+\widetilde{h}}{T}} f^{(1)}(u,\lambda) du d\lambda - \int_{\frac{-2w\pi}{\widetilde{h}}}^{\frac{2w\pi}{\widetilde{h}}} \int_{\frac{t-\widetilde{h}}{T}}^{\frac{t}{T}} f^{(1)}(u,\lambda) du d\lambda\Big).
    $$
 Therefore, analogous to using $D_h(t,w)$ to detect jumps, a large value of  $\sup_{w\in\{0,1,2,\dots,{\widetilde{h}}/{2}\}}|D^{(1)}_{\widetilde{h},T}(t,w)|$ 
 suggests that a kink in the spectral density has occurred at time $t$.
Putting together the information from all frequencies, we define the maximal difference statistics as 
\begin{equation*}
\mathcal{D}_h(t)=\sup_{w \in \{0,1,2,\dots,{h}/{2}\}} |D_h(t,w)|\,, \ ~{\rm and}~ \ 
\mathcal{D}^{(1)}_{\widetilde{h}}(t)=\sup_{w \in \{0,1,2,\dots,{{\widetilde{h}}}/{2}\}} |D^{(1)}_{\widetilde{h}}(t,w)| \,.
\end{equation*}
If $h(T)$ and ${\widetilde{h}}(T)$ are chosen such that $h(T)\rightarrow \infty$, ${\widetilde{h}}(T) \rightarrow \infty$ and $\frac{h(T)}{T} \rightarrow 0$, $\frac{{\widetilde{h}}(T)}{T} \rightarrow 0$ when $T \rightarrow \infty$, 
then the window is large enough to detect a structural change, and small enough to ensure that each window contains at most one change-point.  
Therefore, we classify a time point $t$ as a potential change-point if $\mathcal{D}_h(t)$ or $\mathcal{D}^{(1)}_{\widetilde{h}}(t)$ is the local maximum within an $h$ or $2\widetilde{h}$ neighborhood, respectively. 
Specifically, for jumps in the spectral density, the set of the potential change-points is defined as
\begin{equation} \label{eqt_C}
\hat{J}=\Big \{j \in \{h,h+1,\dots,T-h\}:\mathcal{D}_h(j)=\max_{t \in [j-h+1,j+h]} \mathcal{D}_h(t) \Big\} \,,
\end{equation}
where $\mathcal{D}_h(t)=0$ for $t<h$ and $t>T-h$.
Similarly, for kinks in the spectral density, the set of the corresponding potential change-points is defined as
\begin{equation*}
\hat{K}=\Big\{j \in \{2{\widetilde{h}},\dots,T-2{\widetilde{h}}\}:\mathcal{D}^{(1)}_{\widetilde{h}}(j)=\max_{t \in [j-{2\widetilde{h}}+1,j+{2\widetilde{h}}]} \mathcal{D}^{(1)}_{\widetilde{h}}(t) \Big\} {\Big\backslash} \Big\{\bigcup_{j \in \hat{J}} [j-h+1,j+h] \Big\} \,,
\end{equation*}
where $\mathcal{D}^{(1)}_{\widetilde{h}}(t)=0$ for $t<2{\widetilde{h}}$ and $t>T-2{\widetilde{h}}$. 
 Note that a jump in the spectral density may also induce a change in the first derivatives, and thus can also be detected by $\mathcal{D}^{(1)}_{\widetilde{h}}(\cdot)$.  
Therefore, in defining $\hat{K}$, we exclude the jumps detected in $\hat{J}$ to avoid missclassification.
To facilitate discussion, we denote the sets of potential jumps and kinks as $\hat{J} = \big \{\hat{\tau}_1^{(J)}, \hat{\tau}_2^{(J)} , \dots , \hat{\tau}_{\hat{m}^{(J)}}^{(J)}\big \}$ 
and $\hat{K} = \big \{\hat{\tau}_1^{(K)}, \hat{\tau}_2^{(K)} , \dots , \hat{\tau}_{\hat{m}^{(K)}}^{(K)}\big \}$, respectively.

\subsection{Second Step: Consistent Estimation by Model Selection Approach} \label{selection}

In the first step, 
the set of  potential change-points may be over-estimated since local maxima could occur by chance even if there is no change in the data.  Therefore, we introduce a second step to select the best subset of change-points, say  $\hat{C}^{(2)}$, from the set  $\hat{J}\cup\hat{K}$ of change-points by using 
a model selection criterion.   
The minimum description length (MDL) criterion for multiple change-point detection in piecewise stationary autoregressive processes was first introduced in \cite{Davis2005} and demonstrated excellent empirical performance.
Here, we develop an MDL-based framework for change-point detection in piecewise locally stationary time-varying autoregressive processes, which entails substantially richer temporal dynamics.

The MDL criterion is based on the minimum description length (MDL) principle, which aims to select the best-fitting model that requires the minimum amount of code length for encoding the observations \citep{Rissanen2012}. When a model, say $M$, is given, the MDL can be partitioned into two parts: 
\begin{equation}\label{MDLa}
\mbox{MDL}(M)=\mathcal{C}(M)+\mathcal{C}(\mathbf{e}|M)\,,
\end{equation}
where $\mathcal{C}(M)$  is the code length for encoding the model, and $\mathcal{C}(\mathbf{e}|M)$ is the code length for encoding the residual $\mathbf{e}$ given the model. Under the current setting, the model for the piecewise locally stationary time series can be represented by 
the number and locations of change-points, and the tvAR model in each segment. It is well known that 
encoding an integer $M$ and an estimator based on $N$ observations requires code lengths of $\log_2 M$ and $(\log_2 N)/2$, respectively \citep{Lee2001}.
Therefore, for a candidate subset of change-points $C=(\hat{\tau}^{(*)}_{1},\ldots,\hat{\tau}^{(*)}_{m})\subseteq \hat{J}\cup\hat{K}$, where $1=\hat{\tau}^{(*)}_{0}\leq \hat{\tau}^{(*)}_{1} <\cdots <\hat{\tau}^{(*)}_{m}\leq \hat{\tau}^{(*)}_{m+1}=T$, $|C|=m$, 
we have to encode the number of change-points $m$, the length $\hat{T}_k=\hat{\tau}^{(*)}_{k}-\hat{\tau}^{(*)}_{k-1}$ of each segment ($k=1,\ldots,m+1$), the model order $p_k$ of each segment, the degree of the polynomial function $q_k$ of each segment, and $(p_k+1)(q_k+\mathbbm{1}_{\{k \in \hat{J}_{cs}\cup\hat{K}_{cs}\}})$ parameters for the polynomial functions for the tvAR($p_k$) model \eqref{segmentmodel} in each segment, yielding
\begin{equation}\label{MDLb}
\mathcal{C}(M) = \log m+\sum_{k=1}^{m+1} \log p_k +\sum_{k=1}^{m+1} \log q_k + \sum_{k=1}^{m+1}\log \hat{T}_{k}+\sum_{k=1}^{m+1}\Big [\frac{(p_k+1) \big (q_k+\mathbbm{1}_{\{k \in \hat{J}_{cs}\cup\hat{K}_{cs}\}}\big )}{2} \log \hat{T}_k \Big ]\,, 
\end{equation}
where $\mathbbm{1}_A$ is defined as the indicator function of set $A$, $\hat{J}_{cs}=\{k:\hat{\tau}^{(*)}_{k-1},\hat{\tau}^{(*)}_k \in \hat{J}\}$ includes all indices of segments between two consecutive jumps in $C$, and $\hat{K}_{cs}=\big\{v:\hat{\tau}^{(*)}_{v},\ldots,\hat{\tau}^{(*)}_{v+g_v-1} \in \hat{K}~\text{with}~
\hat{\tau}^{(*)}_{v-1}, \hat{\tau}^{(*)}_{v+g_v} \in \hat{J} \cup \{0,T\}~\text{and}~g_v \in \mathbb{N} \big\}$ 
includes all indices of the first kink in a set of consecutive kinks and all indices of an isolated kink in $C$.
On the other hand, the code length for the residuals given a model $M$ can be approximated by the conditional Gaussian log-likelihood function 
\begin{equation}\label{MDLc}
\mathcal{C}(\mathbf{e}|M)= -\sum_{k \in \hat{J}_{cs}} \hat{L}_{T}^{(k)}(\hat{\boldsymbol{\theta}}_k,\hat{\mathbf{X}}_{k,T})-
\sum_{v \in \hat{K}_{cs}} \widetilde{L}_{T}^{(v,g_v)}(\hat{\boldsymbol{\eta}}_{v,g_v},\hat{\mathbf{X}}_{v:v+g_v,T})\, ,
\end{equation}
where 
$\hat{\mathbf{X}}_{k,T}=\{X_{\hat{\tau}^{(*)}_{k-1}+1,T},X_{\hat{\tau}^{(*)}_{k-1}+2,T},\ldots,X_{\hat{\tau}^{(*)}_k,T}\}$ and $\hat{\mathbf{X}}_{v:v+g_v,T} = \bigcup_{k=v}^{v+g_v} \hat{\mathbf{X}}_{k,T}$. 
As discussed in Section \ref{sec.3.1}, the likelihood for segments with consecutive kinks between two jumps have to be computed using the re-parametrized model. Thus,  \eqref{MDLc} involves two summations to distinguish between the cases of ordinary and re-parametrized model. 

The first summation in \eqref{MDLc} is taken over all segments that lie between two consecutive jumps in $C$. In particular, if  $\hat{\tau}^{(*)}_{k-1},\hat{\tau}^{(*)}_k \in \hat{J}$, then from \eqref{L.GL} the conditional Gaussian log-likelihood function for the $k$-th segment is defined as 
\begin{eqnarray*}\label{GLikelihood_step2}	
	\hat{L}_{T}^{(k)}(\boldsymbol{\theta},\hat{\mathbf{X}}_{k,T})  =  -\frac{1}{2} \sum_{t=\hat{\tau}^{(*)}_{k-1}+p_{k}+1}^{\hat{\tau}^{(*)}_k} \Big[ \log \Big(2\pi \sigma^2_{k,\boldsymbol{\theta}} \Big(\frac{t}{T}\Big)\Big) + \frac{\left(X_{t,T}-\sum_{i=1}^{p_{k}}\phi_{k,i,\boldsymbol{\theta}}\left(\frac{t}{T} \right) X_{t-i,T}  \right)^2 }{\sigma^2_{k,\boldsymbol{\theta}} \left(\frac{t}{T}\right)}\Big] \,,
\end{eqnarray*}
and the parameter estimate can be computed by $\hat{\boldsymbol{\theta}}_k=\argmax_{\boldsymbol{\theta}} \hat{L}_{T}^{(k)}(\boldsymbol{\theta},\hat{\mathbf{X}}_{k,T})$. 

The second summation in \eqref{MDLc} is taken over all segments with consecutive kinks or an isolated kink in $C$. In particular, 
if 
 $\hat{\tau}^{(*)}_{v},\ldots,\hat{\tau}^{(*)}_{v+g_v-1} \in \hat{K}$ 
 and $\hat{\tau}^{(*)}_{v-1}, \hat{\tau}^{(*)}_{v+g_v}\in \hat{J} \cup \{0,T\}$, then in terms of the re-parametrized model \eqref{repara_kink_model}, the conditional Gaussian log-likelihood function for all $g_v+1$ segments with $g_v$ consecutive kinks is 
\begin{eqnarray*} 
	\widetilde{L}_{T}^{(v,g_v)}(\boldsymbol{\eta}_{v,g_v},\hat{\mathbf{X}}_{v:v+g_v,T})=-\frac{1}{2}\sum_{t=\hat{\tau}^{(*)}_{v-1}+1}^{\hat{\tau}^{(*)}_{v+g_v}} \Big[ 
	\log \Big(2\pi \sigma_{\xi,\boldsymbol{\beta},v,g_v}^2\Big (\frac{t}{T}\Big )\Big) + \frac{\big(X_{t,T} - {\boldsymbol{X}_{t-p_{max}}^{t-1}}' \phi_{\boldsymbol{\gamma},\boldsymbol{\alpha},v,g_v} \left(\frac{t}{T}\right) \big )^2}{\sigma_{\xi,\boldsymbol{\beta},v,g_v}^2\left(\frac{t}{T}\right)} \Big] \,,
\end{eqnarray*}
and the parameter estimate can be computed by $\hat{\boldsymbol{\eta}}_{v,g} = \argmax_{\boldsymbol{\eta}_{v,g}}  \widetilde{L}_{T}^{(v,g)}(\boldsymbol{\eta}_{v,g},\hat{\mathbf{X}}_{v:v+g_v,T})$.

Combining \eqref{MDLa}, \eqref{MDLb} and \eqref{MDLc}, we define the MDL criterion as
\begin{eqnarray*}\label{MDL}
\mbox{MDL}(m,C,\boldsymbol{p},\boldsymbol{q}) & = & \log m+\sum_{k=1}^{m+1} \log p_k q_k + \sum_{k=1}^{m+1} \log \hat{T}_k\Big [\frac{(p_k+1) \big (q_k+\mathbbm{1}_{\{k \in \hat{J}_{cs}\cup\hat{K}_{cs}\}}\big)}{2}+1 \Big ] \nonumber\\
& & -\sum_{k \in \hat{J}_{cs}} \hat{L}_{T}^{(k)}(\hat{\boldsymbol{\theta}}_k,\hat{\mathbf{X}}_{k,T})-
\sum_{v \in \hat{K}_{cs}} \widetilde{L}_{T}^{(v,g_v)}(\hat{\boldsymbol{\eta}}_{v,g_v},\hat{\mathbf{X}}_{v:v+g_v,T})\,,
\end{eqnarray*}
where $\boldsymbol{p}=(p_1,\ldots,p_{m+1})$ and $\boldsymbol{q}=(q_1,\ldots,q_{m+1})$. We can then obtain the best subset of change-points $\hat{C}^{(2)}$ by the optimization
\begin{equation*}
(\hat{m}^{(2)},\hat{C}^{(2)},\hat{\boldsymbol{p}}^{(2)},\hat{\boldsymbol{q}}^{(2)})= \argmin_{\substack{C \subseteq \hat{J}\cup\hat{K},~ m = |C|, \\ \boldsymbol{p}\in \{1,\ldots,p_{max}\}^m,~ \boldsymbol{q}\in \{1,\ldots,q_{max}\}^m }}
 \mbox{MDL}(m,C,\boldsymbol{p},\boldsymbol{q})\, ,
\end{equation*}
where $\hat{m}^{(2)}$ is the number of elements in the set $\hat{C}^{(2)}$, and $\hat{\boldsymbol{p}}^{(2)}=(\hat{p}_1^{(2)},\ldots,\hat{p}_{\hat{m}^{(2)}+1}^{(2)})$ and $\hat{\boldsymbol{q}}^{(2)}=(\hat{q}_1^{(2)},\ldots,\hat{q}_{\hat{m}^{(2)}+1}^{(2)})$ are the estimated tvAR orders and the estimated degrees of the parameter polynomial functions respectively.

Note that theoretically we can omit the first step and apply the MDL criterion to estimate simultaneously the number, locations and types of change-points. However, the computational cost will increase exponentially with the length of the time series and thus it is computationally infeasible. In contrast, after the first step, we have selected a candidate set of potential change-points and the size of $\hat{J}\cup\hat{K}$ is much smaller than $T$, so the optimization can be conducted more efficiently.

\subsection{Third Step: Final Adjustment and Confidence Intervals Construction} \label{detection}

Recall that the set of estimated change-points in the second step is a subset  of the set of potential change-points obtained from the first step, 
which are estimated using small windows of sizes $h$ or $2{\widetilde{h}}$, and thus may not be very accurate. 
Therefore, in this third step, we make a final adjustment by using larger windows to estimate and construct a confidence interval for each change-point.

\subsubsection{Final Adjustment and Confidence Interval Construction for a Jump} \label{subsec_jump}
If the $k$-th estimated change-point is classified as a jump, i.e., $\hat{\tau}_k^{(2)} \in \hat{J}$, then the final estimate of the $k$-th change-point is defined as 
\begin{equation} \label{eq_final_esti}
	\hat{\tau}_k^{(3)}=\argmax_{\tau \in [\hat{\tau}_{k}^{(2)} - {h},\hat{\tau}_{k}^{(2)} + {h}]}\Bigg[\sum_{t=\hat{\tau}_{k-1}^{(2)} + {h_{k-1}}}^{\tau} \ell_k(\hat{\boldsymbol{\theta}}_k,X_{t,T}) + \sum_{t=\tau+1}^{\hat{\tau}_{k+1}^{(2)} - {h_{k+1}}} \ell_{k+1}(\hat{\boldsymbol{\theta}}_{k+1},X_{t,T}) \Bigg] \, ,
\end{equation}
where 
$h_{k}=h \mathbbm{1}_{\{\hat{\tau}_k^{(2)} \in \hat{J}\}}+ 2\widetilde{h} \mathbbm{1}_{\{\hat{\tau}_k^{(2)} \in \hat{K}\}}$,  
\begin{equation}\label{eq_condGauss}
	\ell_k(\boldsymbol{\theta}_k,X_{t,T})=-\frac{1}{2}\log \Big (2\pi \sigma_{k,\boldsymbol{\theta}_k}^2\Big(\frac{t}{T}\Big)\Big) - \frac{\big(X_{t,T}-\sum_{i=1}^{\hat{p}_k^{(2)}}\phi_{k,i,\boldsymbol{\theta}_k}(\frac{t}{T} ) X_{t-i,T} \big )^2  }{2 \sigma_{k,\boldsymbol{\theta}_k}^2(\frac{t}{T})}\,,
\end{equation}
is the conditional Gaussian log-likelihood function for $X_{t,T}$ following a tvAR($\hat{p}_k^{(2)}$) model, 
and $\ell_{k+1}(\boldsymbol{\theta}_{k+1},X_{t,T})$ is defined analogously for $X_{t,T}$ following a tvAR($\hat{p}_{k+1}^{(2)}$) model. The parameter estimates in \eqref{eq_final_esti} are defined as 
\begin{eqnarray*}
\hat{\boldsymbol{\theta}}_k=\hat{\boldsymbol{\theta}}_{k} (\tau) = \argmax_{\boldsymbol{\theta}_{k}}\sum_{t=\hat{\tau}_{k-1}^{(2)} + {h_{k-1}}}^{\tau} \ell_k(\boldsymbol{\theta}_k,X_{t,T})\,, \mbox{ and } 
\hat{\boldsymbol{\theta}}_{k+1}=\hat{\boldsymbol{\theta}}_{k+1} (\tau) = \argmax_{\boldsymbol{\theta}_{k+1}}\sum_{t=\tau+1}^{\hat{\tau}_{k+1}^{(2)} - {h_{k+1}}} \ell_{k+1}(\boldsymbol{\theta}_{k+1},X_{t,T})\,.
\end{eqnarray*}

We will shown in Section \ref{sec:asymp_prop} that  
{the $k$-th change-point can be asymptotically detected within its $h_k$-neighborhood.} 
Therefore, the {\it extended local window} $[\hat{\tau}_{k-1}^{(2)} + {h_{k-1}},\ldots,\hat{\tau}_{k+1}^{(2)} - {h_{k+1}}]$ of observations used in \eqref{eq_final_esti} is the largest one ensuring that 
only one change-point is contained inside. 
Also, it suffices to search for a change-point in the range $\tau\in [\hat{\tau}_{k}^{(2)} - {h},\hat{\tau}_{k}^{(2)} + {h}]$ in \eqref{eq_final_esti}. 

Denote the final estimates from \eqref{eq_final_esti} as $(\hat{\boldsymbol{\theta}}_k^{(3)},\hat{\boldsymbol{\theta}}_{k+1}^{(3)},\hat{\tau}_k^{(3)})$, where  
\begin{eqnarray*}
\hat{\boldsymbol{\theta}}_k^{(3)}=\hat{\boldsymbol{\theta}}_k (\hat{\tau}_k^{(3)})
=(\hat{\phi}_{k10},\ldots,\hat{\phi}_{k1\hat{q}_{k}^{(2)}},\hat{\phi}_{k20},\ldots,\hat{\phi}_{k2\hat{q}_{k}^{(2)}},\ldots,\hat{\phi}_{k\hat{p}_{k}^{(2)}0},\ldots,\hat{\phi}_{k\hat{p}_{k}^{(2)}\hat{q}_{k}^{(2)}},\hat{\sigma}_{k0},\ldots,\hat{\sigma}_{k\hat{q}_{k}^{(2)}})'\,,
\end{eqnarray*}
 and 
 define $\hat{\boldsymbol{\theta}}_{k+1}^{(3)}=\hat{\boldsymbol{\theta}}_{k+1} (\hat{\tau}_{k}^{(3)})$ 
 analogously. 
 
 In principle, we can use the asymptotic distribution  of  the final estimate $\hat{\tau}_k^{(3)}$ to  construct a confidence interval for each change-point (see Theorem  \ref{thm4.3}  in the following section). However, this distribution is highly model dependent. As an alternative    
 we propose a  bootstrap algorithm to construct a confidence interval for the $k$-th change-point  \citep[see also][for a similar approach in change-point inference for piecewise stationary time series]  {Yau2019}:

\noindent
{\it Step 1}: Specify a sufficiently large integer $B$. For each $i=1,2,\dots,B$, simulate a time series sample $\{\widetilde{X}_{\hat{\tau}_{k-1}^{(2)} + {h_{k-1}}-\hat{\tau}_k^{(3)}}^{(i)},\dots,\widetilde{X}_{0}^{(i)},\dots,\widetilde{X}_{\hat{\tau}_{k+1}^{(2)} - {h_{k+1}}-\hat{\tau}_k^{(3)}}^{(i)}\}$, where $\{\widetilde{X}_{\hat{\tau}_{k-1}^{(2)} + {h_{k-1}}-\hat{\tau}_k^{(3)}}^{(i)},\dots,\widetilde{X}_0^{(i)}\}$ and $\{\widetilde{X}_1^{(i)},\dots,\widetilde{X}_{\hat{\tau}_{k+1}^{(2)} - {h_{k+1}}-\hat{\tau}_k^{(3)}}^{(i)}\}$ follow a  tvAR($\hat{p}_{k}^{(2)}$) and tvAR($\hat{p}_{k+1}^{(2)}$) models with parameters  $\hat{\boldsymbol{\theta}}_{k}^{(3)}$ and $\hat{\boldsymbol{\theta}}_{k+1}^{(3)}$ respectively, under the relation \eqref{segmentmodel}
with a centred unit-variance Gaussian white noise process $\{\widetilde{\epsilon}^{(i)}_t\}$.
To be specific, for $t \leq 0$,
\begin{equation*}
	\widetilde{X}_{t}^{(i)}=\sum_{i=1}^{\hat{p}_{k}^{(2)}} \Big [\sum_{j=0}^{\hat{q}_{k}^{(2)}} \hat{\phi}_{kij}\Big (\frac{\hat{\tau}_k^{(3)}+t}{T}\Big)^j\Big ] \widetilde{X}_{t-i}^{(i)} + \Big [\sum_{j=0}^{\hat{q}_{k}^{(2)}} \hat{\sigma}_{kj} \Big(\frac{\hat{\tau}_k^{(3)}+t}{T}\Big)^j\Big ] \widetilde{\epsilon}^{(i)}_t\, , \hspace{0.5cm} \widetilde{\epsilon}^{(i)}_t \overset{\text{i.i.d.}}{\sim} N(0,1)\,,
\end{equation*}
and for $t > 0$,
\begin{equation*}
	\widetilde{X}_{t}^{(i)}=\sum_{i=1}^{\hat{p}_{k+1}^{(2)}} \Big[\sum_{j=0}^{\hat{q}_{k+1}^{(2)}} \hat{\phi}_{(k+1)ij}\Big(\frac{\hat{\tau}_k^{(3)}+t}{T}\Big)^j\Big ] \widetilde{X}_{t-i}^{(i)} + \Big [\sum_{j=0}^{\hat{q}_{k+1}^{(2)}} \hat{\sigma}_{(k+1)j} \Big(\frac{\hat{\tau}_k^{(3)}+t}{T}\Big)^j\Big ] \widetilde{\epsilon}^{(i)}_t\, , \hspace{0.5cm} \widetilde{\epsilon}^{(i)}_t \overset{\text{i.i.d.}}{\sim} N(0,1)\,, 
\end{equation*}	
where we can either set $\widetilde{X}_{t}^{(i)}=0$ for $t<\hat{\tau}_{k-1}^{(2)} + {h_{k-1}}-\hat{\tau}_k^{(3)}$ or simulate initial observations $\{\widetilde{X}_{t}^{(i)}\}_{t<\hat{\tau}_{k-1}^{(2)} + {h_{k-1}}-\hat{\tau}_k^{(3)}}$ as a burn-in.

\noindent 
{\it  Step 2}: For each $i=1,2,\dots,B$, calculate the conditional Gaussian log-likelihood functions $\ell_k(\hat{\boldsymbol{\theta}}^{(3)}_k, \widetilde{X}_t^{(i)})$ and $\ell_{k+1}(\hat{\boldsymbol{\theta}}^{(3)}_{k+1}, \widetilde{X}_t^{(i)})$ for $t=\hat{\tau}_{k-1}^{(2)} + {h_{k-1}}-\hat{\tau}_k^{(3)},\ldots,\hat{\tau}_{k+1}^{(2)} - {h_{k+1}}-\hat{\tau}_k^{(3)}$ based on the simulated time series sample $\{\widetilde{X}_{\hat{\tau}_{k-1}^{(2)} + {h_{k-1}}-\hat{\tau}_k^{(3)}}^{(i)},\dots,\widetilde{X}_{0}^{(i)},\dots,\widetilde{X}_{\hat{\tau}_{k+1}^{(2)} - {h_{k+1}}-\hat{\tau}_k^{(3)}}^{(i)}\}$.

\noindent
{\it Step 3}: For each $i=1,2,\dots,B$, set 
\begin{equation*}
	\widetilde{d}_k^{(i)}=\argmax_{d \in \{\hat{\tau}_{k}^{(2)} - {h}-\hat{\tau}_k^{(3)},\ldots,\hat{\tau}_{k}^{(2)} + {h}-\hat{\tau}_k^{(3)}\}}\Big [\sum_{t=\hat{\tau}_{k-1}^{(2)} + {h_{k-1}}-\hat{\tau}_k^{(3)}}^{d} \ell_k(\hat{\boldsymbol{\theta}}^{(3)}_k,\widetilde{X}_t^{(i)}) + \sum_{t=d+1}^{\hat{\tau}_{k+1}^{(2)} - {h_{k+1}}-\hat{\tau}_k^{(3)}} \ell_{k+1}(\hat{\boldsymbol{\theta}}^{(3)}_{k+1},\widetilde{X}_t^{(i)}) \Big ] \, .
\end{equation*}

\noindent
{\it Step 4}: Denote $\widetilde{l}_k$ and $\widetilde{u}_k$ as the $\alpha/2$ and the $1-\alpha/2$ percentiles of the samples $\{\widetilde{d}_k^{(1)},\widetilde{d}_k^{(2)},\dots,\widetilde{d}_k^{(B)} \}$, respectively. The bootstrap $100(1-\alpha)\%$ confidence interval for the $k$-th true change-point is $[\hat{\tau}_k^{(3)}-\widetilde{u}_k, \hat{\tau}_k^{(3)}-\widetilde{l}_k]$. Note that the confidence interval generated by the parametric bootstrap algorithm is asymmetric in general.

\subsubsection{Final Adjustment and Confidence Interval Construction for a Kink}\label{subsec_kink}
If the $k$-th estimated change-point is classified as a kink 
$\hat{\tau}_k^{(2)} \in \hat{K}$, then analogous to \eqref{eq_final_esti} we consider estimates from observations in the extended local window 
$[\hat{\tau}_{k-1}^{(2)}+h_{k-1}, \hat{\tau}_{k+1}^{(2)}-h_{k+1}]$ with  $h_{k}=h \mathbbm{1}_{\{\hat{\tau}_k^{(2)} \in \hat{J}\}}+ 2\widetilde{h} \mathbbm{1}_{\{\hat{\tau}_k^{(2)} \in \hat{K}\}}$ and denote the length of the extended local window as $\hat{T}^{(2)}_{k,k+1}= \hat{\tau}_{k+1}^{(2)}-h_{k+1}-\hat{\tau}_{k-1}^{(2)}-h_{k-1}$.
Without loss of generality, we consider the case of $\hat{p}_{k}^{(2)} < \hat{p}_{k+1}^{(2)}$, and 
let $\tau_k\in\{\hat{\tau}_{k}^{(2)} - {2\widetilde{h}},\ldots,\hat{\tau}_{k}^{(2)} + {2\widetilde{h}}\}$ be a generic kink point. 
For $t\in\{\hat{\tau}_{k-1}^{(2)} + h_{k-1},\ldots, \hat{\tau}_{k+1}^{(2)} - h_{k+1}\}$, 
we employ the re-parameterized model \eqref{repara_kink_model} with only one kink change-point, hence $g_k=1$, as  
\begin{eqnarray}
    && X_{t,T}=\sum_{i=1}^{\hat{p}_{k}^{(2)}} \Big [\gamma_{ki} + \sum_{j=1}^{\hat{q}_{k}^{(2)}} \alpha_{kij}\Big  (\frac{t}{T}-r_k\Big  )_{-}^j + \sum_{j=1}^{\hat{q}_{k+1}^{(2)}} \alpha_{(k+1)ij}\Big  (\frac{t}{T}-r_k\Big  )_{+}^j \Big  ] X_{t-i,T} \nonumber\\
	&&   +  \sum_{i=\hat{p}_{k}^{(2)}+1}^{\hat{p}_{k+1}^{(2)}} \Big [\sum_{j=1}^{\hat{q}_{k+1}^{(2)}} \alpha_{(k+1)ij}\Big  (\frac{t}{T}-r_k\Big  )_{+}^j \Big  ] X_{t-i,T} \nonumber  + \Big [\xi_{k}+ \sum_{j=1}^{\hat{q}_{k}^{(2)}} \beta_{kj} \Big  (\frac{t}{T}-r_k\Big  )_{-}^j + \sum_{j=1}^{\hat{q}_{k+1}^{(2)}} \beta_{(k+1)j} \Big  (\frac{t}{T}-r_k\Big  )_{+}^j\Big  ] \epsilon_{t,T}\\
	&  & = {\boldsymbol{X}_{t-\hat{p}_{k+1}^{(2)}}^{t-1}}' \phi_{\boldsymbol{\gamma},\boldsymbol{\alpha},k} \left(\frac{t}{T}\right) + \sigma_{\xi,\boldsymbol{\beta},k}\left(\frac{t}{T}\right) \epsilon_{t,T} \label{repara_kink_model_one} \,,
\end{eqnarray}
where $r_k=\frac{\tau_k}{T}$, $\phi_{\boldsymbol{\gamma},\boldsymbol{\alpha},k} \left(\frac{t}{T}\right)=\big (\phi^{(1)}_{\boldsymbol{\gamma},\boldsymbol{\alpha},k} \left(\frac{t}{T}\right),\ldots,\phi^{(\hat{p}_{k+1}^{(2)})}_{\boldsymbol{\gamma},\boldsymbol{\alpha},k} \left(\frac{t}{T}\right) \big)'$ with
\begin{eqnarray*}
\phi^{(i)}_{\boldsymbol{\gamma},\boldsymbol{\alpha},k} \Big (\frac{t}{T}\Big )= 
\left\{ \begin{array}{ll}
\gamma_{ki} + \sum_{j=1}^{\hat{q}_{k}^{(2)}} \alpha_{kij}\left(\frac{t}{T}-r_k\right)_{-}^j + \sum_{j=1}^{\hat{q}_{k+1}^{(2)}} \alpha_{(k+1)ij}\left(\frac{t}{T}-r_k\right)_{+}^j\,, & \mbox{for } \ i=1,\ldots,\hat{p}_{k}^{(2)}\,, \\ 
\sum_{j=1}^{\hat{q}_{k+1}^{(2)}} \alpha_{(k+1)ij}\left(\frac{t}{T}-r_k\right)_{+}^j\,, & \mbox{for } \ 
i=\hat{p}_{k}^{(2)}+1,\ldots,\hat{p}_{k+1}^{(2)}\,, 
	\end{array}
\right. 
\end{eqnarray*}
 and $\sigma_{\xi,\boldsymbol{\beta},k}\left(\frac{t}{T}\right)=\xi_{k}+ \sum_{j=1}^{\hat{q}_{k}^{(2)}} \beta_{kj} \left(\frac{t}{T}-r_k\right)_{-}^j + \sum_{j=1}^{\hat{q}_{k+1}^{(2)}} \beta_{(k+1)j} \left(\frac{t}{T}-r_k\right)_{+}^j$. The case of $\hat{p}_{k}^{(2)} \geq \hat{p}_{k+1}^{(2)}$ can be handled  analogously. 
 

The re-parameterized model parameters for the two segments around $\tau_k$ are collected in the vector  $\boldsymbol{\eta}_k=(\boldsymbol{\gamma}_{k},\xi_{k},\boldsymbol{\alpha}_k,\boldsymbol{\beta}_k,\boldsymbol{\alpha}_{k+1},\boldsymbol{\beta}_{k+1},r_k)'$,
 where $\boldsymbol{\gamma}_{k}=(\gamma_{k1},\ldots,\gamma_{k\hat{p}_{k}^{(2)}})'$,  $\boldsymbol{\alpha}_k=(\alpha_{k11},\ldots,\alpha_{k\hat{p}_{k}^{(2)}\hat{q}_{k}^{(2)}})'$, 
 $\boldsymbol{\beta}_k=(\beta_{k1},\ldots,\beta_{k\hat{q}_{k}^{(2)}})'$,  $\boldsymbol{\alpha}_{k+1}$ and  $\boldsymbol{\beta}_{k+1}$ are defined analogously. The corresponding true values are denoted as  $\boldsymbol{\eta}_k^0=(\boldsymbol{\gamma}^0_{k},\xi^0_{k},\boldsymbol{\alpha}^0_k,\boldsymbol{\beta}^0_k,\boldsymbol{\alpha}^0_{k+1},\boldsymbol{\beta}^0_{k+1},r^0_k)'$, where $r_k^0=\frac{\tau_k^0}{T}$. The final estimate $\hat{\boldsymbol{\eta}}_k^{(3)}$ of the re-parameterized model is defined as
\begin{equation} \label{eq_final_esti_kink}
	\hat{\boldsymbol{\eta}}_k^{(3)}=(\hat{\boldsymbol{\gamma}}^{(3)}_{k},\hat{\xi}^{(3)}_{k},\hat{\boldsymbol{\alpha}}^{(3)}_k,\hat{\boldsymbol{\beta}}^{(3)}_k,\hat{\boldsymbol{\alpha}}^{(3)}_{k+1},\hat{\boldsymbol{\beta}}^{(3)}_{k+1},\hat{r}^{(3)}_k)'=\argmax_{\boldsymbol{\eta}_k}\frac{1}{\hat{T}^{(2)}_{k,k+1}} \sum_{t=\hat{\tau}_{k-1}^{(2)} + {h_{k-1}}}^{\hat{\tau}_{k+1}^{(2)} - {h_{k+1}}} \ell(\boldsymbol{\eta}_k,X_{t,T})\,,
\end{equation}
where $\hat{r}_k^{(3)}=\frac{\hat{\tau}_k^{(3)}}{T}$ and
\begin{eqnarray}\label{obj_kink}
\ell(\boldsymbol{\eta}_k,X_{t,T}) & = & -\frac{1}{2}\log \Big (2\pi \sigma_{{\xi,\boldsymbol{\beta},k}}^2\Big (\frac{t}{T}\Big )\Big ) - \frac{\big (X_{t,T}- {\boldsymbol{X}_{t-\hat{p}_{k+1}^{(2)}}^{t-1}}' \phi_{\boldsymbol{\gamma},\boldsymbol{\alpha},k} \left(\frac{t}{T}\right) \big )^2}{2 \sigma_{{\xi,\boldsymbol{\beta},k}}^2(\frac{t}{T})}\,.
\end{eqnarray}
In contrast to the asymptotic distribution in the jump case, from Theorem \ref{thm4.3_kink} below, the estimated parameter $\hat{\boldsymbol{\eta}}_k^{(3)}$ 
is asymptotically normal with a $\sqrt{n}$ convergence rate. 
Hence, confidence intervals for the kinks can be constructed readily without using bootstrap. 
Specifically, define
\begin{equation}\label{sum_crit}
S_{k,T} (\boldsymbol{\eta}_k)= \frac{1}{\hat{T}^{(2)}_{k,k+1}} \sum_{t=\hat{\tau}_{k-1}^{(2)} + {h_{k-1}}}^{\hat{\tau}_{k+1}^{(2)} - {h_{k+1}}} \ell(\boldsymbol{\eta}_k,X_{t,T})\,,
\end{equation}
and hence $\hat{\boldsymbol{\eta}}_k^{(3)}=\argmax_{\boldsymbol{\eta}_k}S_{k,T} (\boldsymbol{\eta}_k)$. 
{
Denote the covariance of the first derivative of \( S_{k,T}(\boldsymbol{\eta}_k) \) with respect to \( \boldsymbol{\eta}_k \), and the second derivative of \( S_{k,T}(\boldsymbol{\eta}_k) \) with respect to \( \boldsymbol{\eta}_k \), as 
\( \nabla_{\boldsymbol{\eta}} S_{k,T}(\boldsymbol{\eta}_k) \nabla_{\boldsymbol{\eta}} S_{k,T}(\boldsymbol{\eta}_k)' \) and 
\( \nabla_{\boldsymbol{\eta}}^2 S_{k,T}(\boldsymbol{\eta}_k) \), respectively; see   
Section \ref{derivatives_L} in Supplementary Materials for their expressions.   
Then, from the asymptotic property in~\eqref{asym_kink}, the asymptotic variance $\mathbf{\Sigma}_k$ of $\hat{\boldsymbol{\eta}}_k^{(3)}$ can be estimated by the plug-in estimator $\hat{\mathbf{\Sigma}}_{k,T}=\hat{\mathbf{D}}_{k,T}^{-1}\hat{\mathbf{G}}_{k,T}\hat{\mathbf{D}}_{k,T}^{-1}$, where $\hat{\mathbf{G}}_{k,T}= \nabla_{\boldsymbol{\eta}} S_{k,T} (\hat{\boldsymbol{\eta}}_k^{(3)}) \nabla_{\boldsymbol{\eta}} S_{k,T} (\hat{\boldsymbol{\eta}}_k^{(3)})'$
and $\hat{\mathbf{D}}_{k,T}=\nabla_{\boldsymbol{\eta}}^2 S_{k,T} (\hat{\boldsymbol{\eta}}_k^{(3)})$
}. Hence, a marginal $100\times(1-\alpha)\%$ confidence interval for $\tau_k^0$ can be constructed by $\hat{\tau}_k^{(3)} \pm z_{\alpha/2} \frac{T\sqrt{[\hat{\mathbf{\Sigma}}_{k,T}]_{11}}}{\sqrt{\hat{\tau}_{k+1}^{(3)}-\hat{\tau}_{k-1}^{(3)}}}$, where $z_{\alpha/2}$ is the $(1-\alpha/2)$ quantile of standard normal distribution and $[\hat{\mathbf{\Sigma}}_{k,T}]_{11}$ is the element at the last row and last column of the matrix $\hat{\mathbf{\Sigma}}_{k,T}$.

	\begin{remark}\label{rem:choice_of_h}
	
 		Theoretically, as shown in Theorems \ref{thm4.1} and \ref{thm4.2}, the jumps and kinks can be detected almost surely asymptotically if the order of the window radii satisfies $h=O(T^{\delta})$ and $\widetilde{h}=O(T^{2/3+\widetilde{\delta}})$, respectively. Thus, the window radii $h$ and $\widetilde{h}$ can be chosen to be $h=CT^{\delta}$ and $\widetilde{h}=\widetilde{C}T^{2/3+\widetilde{\delta}}$ for some constants $C>0$, $\delta>0$, $\widetilde{C}>0$ and $\widetilde{\delta}>0$. The simulation studies in Section \ref{sec:simulation} suggest some the rule-of-thumb choices of  $C$, $\delta$, $\widetilde{C}$ and $\widetilde{\delta}$ which work satisfactorily.		
	\end{remark}

\section{Asymptotic Properties}
\def\theequation{4.\arabic{equation}}
\setcounter{equation}{0}
\label{sec:asymp_prop}
In this section, we investigate the asymptotic properties of the proposed inference procedure,  
including the estimation consistency of the number and locations of change-points, classification of jumps and kinks, and the validity of confidence intervals. 
For a precise statement of the results we  introduce the following assumptions.

\begin{assumption} \label{ass4.1}
~~
{\rm 
	\begin{enumerate}[(a)]
		\item Assume that the time series $\{X_{t,T}: t=1,2,\dots,T\}$ can be partitioned into $m_0+1$ locally stationary tvAR processes, where $C_0 = (\tau_1^0, \tau_2^0 , \dots , \tau_{m_0}^0)$ is the set of true change-points where $\tau_0^0=0$ and $\tau_{m_0+1}^0=T$. Denote the relative location of the $k$-th change-point by $r_k^0=\frac{\tau_k^0}{T}$ for $k=0,1,\dots,m_0+1$, and assume that $\min_{k=0,1,\dots,m}(r^0_{k+1}-r^0_k)>\epsilon_r$ for some $\epsilon_r > 0$.
		For $k=1,2,\dots,m_0+1$, the $k$-th segment $\{X_{\tau^0_{k-1}+1,T},X_{\tau^0_{k-1}+2,T},\dots,X_{\tau^0_{k},T}\}$ follows the tvAR$(p_k^0)$ model defined in \eqref{segmentmodel} with polynomial functions of degree $q_k^0$.
		Assume that the parameter of the $k$-th segment $\boldsymbol{\theta}_k^0=(\phi_{k10},\dots,\phi_{k{p^0_k}{q_k^0}},\sigma_{k0},\dots,\sigma_{k{q_k^0}})$ is an interior point of a compact space $\Theta_k$. Denote the set of true tvAR orders by $\boldsymbol{p}_0=(p_1^0,\ldots,p_{m_0+1}^0)$ and the set of true degrees of polynomial functions by $\boldsymbol{q}_0=(q_1^0,\ldots,q_{m_0+1}^0)$.
		
		\item Denote the set of true jumps in the spectral density as $J_0 \subseteq C_0$. For each true jump $\tau_k^0 \in J_0 \subseteq C_0$, there exists a $\lambda^* \in [-\pi,\pi]$ such that $f_k (\frac{\tau_k^0}{T},\lambda^* ) \neq \lim_{\delta\rightarrow 0^+} f_{k+1} (\frac{\tau_k^0}{T}+\delta,\lambda^* )$.
		
		\item Denote the set of true kinks in the spectral density as $K_0 = C_0 \setminus J_0$. For each true kink $\tau_k^0 \in K_0 = C_0 \setminus J_0$, there exists a $\lambda^* \in [-\pi,\pi]$ such that   $f^{(1)}_k\left(\frac{\tau_k^0}{T},\lambda^*\right) \neq \lim_{\delta\rightarrow 0^+} f^{(1)}_{k+1}\left(\frac{\tau_k^0}{T}+\delta,\lambda^*\right)$ and $f_k\left(\frac{\tau_k^0}{T},\lambda^*\right) =  f_{k+1}\left(\frac{\tau_k^0}{T},\lambda^*\right)$.
	\end{enumerate} 
}
\end{assumption}

\begin{assumption} \label{ass4.2} 
	{\rm For each $k=1,2,\dots,m_0+1$, assume that $1-\sum_{i=1}^{p^0_k} \sum_{j=0}^{q_k^0} \phi_{kij} \left(\frac{t}{T}\right)^j z^i \neq 0$ for all $t =\tau_{k-1}+1, \ldots, \tau_{k}$ and all $|z|\leq 1+c$ with some $c>0$.}
\end{assumption}
	


\begin{assumption} \label{ass4.4} 
	{\rm Assume that $\max_{k\in \{1,\ldots,m_0+1\}}p^{0}_k \leq p_{max}$, i.e, the order of the tvAR process in each segment is bounded by an integer $p_{max}$. Also, assume that
	$\max_{k\in \{1,\ldots,m_0+1\}}q^{0}_k \leq q_{max}$, i.e., the order of the parameter polynomial curves in each segment is bounded by an integer $q_{max}$. 
	 The minimum distance between consecutive change-points $T\epsilon_r>p_{max}(q_{max}+1)$ is assumed to ensure model identifiability.
}
\end{assumption}

\begin{assumption} \label{ass4.5} 
	{\rm The window radii $h$ and $\widetilde{h}$ satisfy the following conditions:
	\begin{itemize}
		\item[(i)] $h \rightarrow \infty$, $T^\epsilon/h\rightarrow 0$, $\frac{h}{T}\rightarrow 0$, and $3h<T\epsilon_r$ where $\epsilon_r>0$,
		\item[(ii)] $\widetilde{h} \rightarrow \infty$, $T^{2/3+\epsilon}/\widetilde{h}\rightarrow 0$, $\frac{\widetilde{h}}{T}\rightarrow 0$, and $6\widetilde{h}<T\epsilon_r$ where $\epsilon_r>0$,
	\end{itemize}
	as $T\rightarrow \infty$ for some $\epsilon>0$. 
}
\end{assumption}

\ignore{\color{blue}
	To develop asymptotic theory for multiple change-point estimation, we . To be precise, for $k=0,1,\dots,m+1$, the relative location of the $k$-th change-point $r_k$, which satisfies $\tau_k=[r_k T]$, is a constant in $(0,1)$.  
	We assume that any two change-points are well separated in the sense that $\min_{k=0,1,\dots,m}(r_{k+1}-r_k)>\epsilon_r$ for some $\epsilon_r > 0$, and hence the number of change-points is finite and bounded by $[1/\epsilon_r]+1$.
}

Assumptions \ref{ass4.1} and \ref{ass4.4} state the basic setting of the change-points configuration.
Assumption \ref{ass4.1} (a) assumes that the length of each segment is directly proportional to the length of the time series, and is common in change-point detection, see \cite{LRSM2015}, \cite{Shumway2007} and \cite{Dette2015}. 
 Assumptions \ref{ass4.1}(b) and \ref{ass4.1}(c) state that at every change-point, a discontinuity occurs either in the parameter curve or in the first derivative of the curve, corresponding to a jump and a kink, respectively. 
Assumption \ref{ass4.2} ensures that the tvAR processes can be represented by tvMA($\infty$) processes with coefficient functions satisfying the summability condition (3.1) stated in Theorems 3.1 and 3.2 of \cite{Dette2015}, which is used to establish asymptotic bounds on the maximum of $D_h(t,w)$ and $D^{(1)}_{\widetilde{h}}(t,w)$ in Theorem \ref{thm4.1}. 
Assumption \ref{ass4.5} involves technical criteria for the choices of window radii $h$ and $\widetilde{h}$ for the first step of the proposed method.
 
Our first result asserts that after the quick scanning in the first step, all the true jumps and kinks can be identified asymptotically within an $h$-neighbourhood and a $2\widetilde{h}$-neighbourhood of the potential change-points in $\hat{J}$ and $\hat{K}$, respectively.

\begin{theorem} \label{thm4.1} 
Let $\hat{J} =  \{\hat{\tau}_1^{(J)}, \hat{\tau}_2^{(J)} , \dots , \hat{\tau}_{\hat{m}^{(J)}}^{(J)}\}$ and $\hat{K} = \{\hat{\tau}_1^{(K)}, \hat{\tau}_2^{(K)} , \dots , \hat{\tau}_{\hat{m}^{(K)}}^{(K)}\}$ be the sets of potential change-points selected in the first step, where $\hat{m}^{(J)}=|\hat{J}|$ and $\hat{m}^{(K)}=|\hat{K}|$. If Assumptions \ref{ass4.1} -  \ref{ass4.5} hold, then, {as $T \to \infty$}, 
\begin{equation*}
	\mathbb{P} \Big  (\max_{\tau \in J_0} \min_{\hat{\tau} \in \hat{J}} |\tau - \hat{\tau}| < h \Big ) \rightarrow 1 ~~~\text{and}~~~ 	\mathbb{P} \Big  (\max_{\tau \in K_0} \min_{\hat{\tau} \in \hat{K}} |\tau - \hat{\tau}| < 2\widetilde{h} \Big  ) \rightarrow 1\,.
	\end{equation*}
\end{theorem}

Note that Theorem \ref{thm4.1} only guarantees that in the first step, an estimated change-point can be found in a neighborhood of each true change-point. 
In general, the number  $\hat{m}^{(J)}+\hat{m}^{(K)}$ of estimated change-points may larger than the true number $m_0$. 
Nevertheless, the following result shows that the MDL model selection approach in the second step yields the consistency of the number and locations of the change-points, and correct classification of jumps and kinks.

\begin{theorem} \label{thm4.2} 
If Assumptions \ref{ass4.1} -  \ref{ass4.5} hold, then 
the number of estimated change-points in the second step satisfies $\hat{m}^{(2)} \overset{p}{\rightarrow} m_0$. 
Furthermore, given that $\hat{m}^{(2)}=m_0$, and $\hat{\tau}_{1}^{(2)} < \hat{\tau}_{2}^{(2)}< \dots < \hat{\tau}_{m_0}^{(2)}$ are the estimated change-points in the second step, then we have
	\begin{eqnarray*}
	\mathbb{P} \Big ( \max_{\tau_k^0 \in J_0}  |\hat{\tau}_k^{(2)} - \tau_k^0| < h \Big  ) \rightarrow 1 \,,~~~\mathbb{P} \Big  ( \max_{\tau_k^0 \in K_0} |\hat{\tau}_k^{(2)} - \tau_k^0| < 2\widetilde{h} \Big ) \rightarrow 1 \,,\\
	\max_{k=1,2,\dots,m_0+1} |\hat{p}_k^{(2)} - p_k^0| \overset{p}{\rightarrow} 0 \,,~~~\text{and}~~~ \max_{k=1,2,\dots,m_0+1} |\hat{q}_k^{(2)} - q_k^0| \overset{p}{\rightarrow} 0 \,. 
	\end{eqnarray*}
\end{theorem}

From Theorem \ref{thm4.2}, the proposed method can estimate the number of change-points consistently. Also, we have consistency on the estimated relative locations of change-points in the sense that $\frac{\hat{\tau}_k^{(2)}}{T}\stackrel{p}{\rightarrow} r_k^0$ for all $k=1,2,\dots,m_0$. 
Next, the following two results provide the asymptotic distributions of the refined change-point and model parameter estimators in the third step for the cases of jumps and kinks, respectively. 
In particular, the refined estimators improve the convergence rates  
from $O_p(h)$ to $O_p(1)$ for jumps, and from $O_p(T^{2/3+\epsilon})$ to $O_p(T^{1/2})$ for kinks.

\begin{theorem} \label{thm4.3}
	Suppose that Assumptions \ref{ass4.1} -\ref{ass4.5} and the conditions of Theorem \ref{thm4.2} hold and $\hat{m}^{(2)}=m_0$. Then, for any $\tau_k^0 \in J_0$, we have
		\begin{equation*}
			\hat{\tau}_k^{(3)}-\tau_k^0 \xrightarrow{d} \argmax_{d} W_{k,d}\,,
		\end{equation*}
		where $W_{k,d}$ is the double-sided random walk defined as 
		\begin{equation}\label{asym_double}
			W_{k,d}=\begin{cases}
				\sum_{t=\tau_k^0+1}^{\tau_k^0+d} \left[\ell(\boldsymbol{\theta}_k^0, Y_t)-\ell(\boldsymbol{\theta}_{k+1}^0 ,Y_t) \right]\,, &d>0\,,\\
				0\,, & d=0\,, \\
				\sum_{t=\tau_k^0+d+1}^{\tau_k^0} \left[\ell(\boldsymbol{\theta}_{k+1}^0 ,Y_t)-\ell(\boldsymbol{\theta}_k^0, Y_t) \right]\,, &d<0\,,
			\end{cases}
		\end{equation}
		 $\ell(\boldsymbol{\theta}, Y_t)$ is the conditional Gaussian log-likelihood function for $Y_t$ at parameter $\boldsymbol{\theta}$ defined in \eqref{eq_condGauss}, 
 $\{Y_t: t \leq \tau_k^0\}$ and $\{Y_t: t>\tau_k^0\}$ follow the tvAR$(p_{k}^0)$ model with parameter $\boldsymbol{\theta}_k^0$ and tvAR$(p_{k+1}^0)$ model with $\boldsymbol{\theta}_{k+1}^0$ defined in \eqref{segmentmodel}, respectively. 
		In particular, we have $\hat{\tau}_k^{(3)} - \tau_k^0 = O_p(1)$.	
\end{theorem}

	To establish the asymptotic distribution of $\hat{\boldsymbol{\eta}}_k^{(3)}$ defined in Section \ref{subsec_kink}, we define $\mathbf{G}_{k,T}  =  \mathbb{E}\left( \nabla_{\boldsymbol{\eta}} S_{k,T} (\boldsymbol{\eta}_k^0)\nabla_{\boldsymbol{\eta}} S_{k,T} (\boldsymbol{\eta}_k^0)' \right)$ and $\mathbf{D}_{k,T}=\mathbb{E} \left(\nabla_{\boldsymbol{\eta}}^2 S_{k,T} (\boldsymbol{\eta}_k^0)\right)$, where $S_{k,T} (\boldsymbol{\eta}_k)$ is defined in \eqref{sum_crit}.
\begin{theorem} \label{thm4.3_kink} 
	Suppose that Assumptions \ref{ass4.1} - \ref{ass4.5} hold. 
Then, on the event $\{\hat{m}^{(2)}=m_0\}\cap\{\hat{p}_k^{(2)}=p_k^0~\text{and}~\hat{q}_k^{(2)}= q_k^0~\text{for}~k=1,2,\dots,m_0+1\}$, which occurs with probability approaching 1, 
for any $\tau_k^0 \in K_0$, 
the parameter estimator  $\hat{\boldsymbol{\eta}}_k^{(3)}$ defined in \eqref{eq_final_esti_kink}
of the true parameter $\boldsymbol{\eta}_k^0=(\boldsymbol{\gamma}^0_{k},\xi^0_{k},\boldsymbol{\alpha}^0_k,\boldsymbol{\beta}^0_k,\boldsymbol{\alpha}^0_{k+1},\boldsymbol{\beta}^0_{k+1},r^0_k)$ in the re-parameterized model \eqref{repara_kink_model_one} satisfies
			\begin{equation}\label{asym_kink}
				\sqrt{(r_{k+1}^0-r_{k-1}^0)T} \left(\hat{\boldsymbol{\eta}}_k^{(3)}-\boldsymbol{\eta}_k^0\right) \xrightarrow{d} N(0,\mathbf{\Sigma}_k)\,,
			\end{equation}	
			where $\mathbf{\Sigma}_k=\mathbf{D}_k^{-1}\mathbf{G}_k\mathbf{D}_k^{-1}$ with $\mathbf{D}_k=\lim_{T\rightarrow \infty}\mathbf{D}_{k,T}$ being a full rank matrix and 
 $\mathbf{G}_k=\lim_{T\rightarrow \infty}\mathbf{G}_{k,T}$ being a nonnegative definite matrix. In particular, we have $\hat{\tau}_k^{(3)} - \tau_k^0 = O_p(T^{1/2})$.		
\end{theorem}

Finally, our last result shows the asymptotic exactness of the bootstrap confidence interval for a jump in Section \ref{subsec_jump}, and the normal confidence interval for a kink in Section \ref{subsec_kink}.

\begin{theorem} \label{thm4.4} 
	Suppose that Assumptions \ref{ass4.1} - \ref{ass4.5} and the conditions of Theorem \ref{thm4.3} hold. Then, for any $\tau_k^0 \in J_0$, we have
	\begin{equation*}
	\mathbb{P} \left(\tau_k^0 \in [\hat{\tau}_k^{(3)}-\widetilde{u}_k,\hat{\tau}_k^{(3)}-\widetilde{l}_k]\right) \rightarrow 1-\alpha \,,
	\end{equation*}
	and for any $\tau_k^0 \in K_0$, we have
	\begin{equation*}
	\mathbb{P} \Big (\tau_k^0 \in \Big  [\hat{\tau}_k^{(3)}-z_{\alpha/2} {T\sqrt{[\hat{\mathbf{\Sigma}}_{k,T}]_{11}}} \Big /{\sqrt{\hat{\tau}_{k+1}^{(3)}-\hat{\tau}_{k-1}^{(3)}}},\hat{\tau}_k^{(3)}+z_{\alpha/2} {T\sqrt{[\hat{\mathbf{\Sigma}}_{k,T}]_{11}}}\Big /{\sqrt{\hat{\tau}_{k+1}^{(3)}-\hat{\tau}_{k-1}^{(3)}}} \Big ]\Big ) \rightarrow 1-\alpha \,.
	\end{equation*}	
\end{theorem}

In summary,  Theorems \ref{thm4.1} - \ref{thm4.4} show that the proposed methodology yields consistent estimation of the number, locations and types of the change-points and also provides asymptotically valid confidence intervals for the locations of the  change-points in piecewise tvAR models.

\section{Simulation Studies}
\def\theequation{5.\arabic{equation}}
\setcounter{equation}{0}
\label{sec:simulation}
In this section, we investigate the finite sample performance of the proposed method. 
Section \ref{sim:sensitivity} examines the choices of window radii $h$ and $\widetilde{h}$. Section \ref{sim:piece_localstat} studies the performance of multiple change-point detection in piecewise locally stationary processes, and compare it with methods in the literature. All reported  results are based on $1000$ simulation runs. We set $p_{max}=4$ and $q_{max}=2$, so each segment follows a tvAR(4) with a quadratic parameter curve.

\subsection{Choice of Window Radius}\label{sim:sensitivity}
In Step~1 (Section~\ref{scan}), we set scanning window radii $h$ and $\widetilde{h}$ to identify candidate change-points; these radii are also used in Step~3 (Section~\ref{detection}) for final refinement and confidence-interval construction. If $h$ and $\widetilde{h}$ are too small, discrepancy measures between adjacent segments become unstable; if they are too large, a single window may contain multiple change-points. Both cases reduce detection accuracy. We assess sensitivity to window sizes using the following models. Unless stated otherwise, the innovations 
$\{\epsilon_t\}$ are i.i.d. $\mathcal{N}\big(0, 1\big)$ variables.

\noindent 
{\it Model 1 (piecewise tvAR(1) model with a jump change-point):} 
\begin{multline*}
X_t = 
\left\{ \left[0.9 - 0.4\left(\frac{t}{T}\right)\right] \cdot \mathbbm{1}_{\{1 \leq t \leq \tau_0\}} 
+ \left[-0.7 + 0.2\left(\frac{t}{T}\right)\right] \cdot \mathbbm{1}_{\{\tau_0 + 1 \leq t \leq T\}} \right\} X_{t-1} \\
+ \left\{ \left[2 - \left(\frac{t}{T}\right)\right] \cdot \mathbbm{1}_{\{1 \leq t \leq \tau_0\}} 
+ \left[1 + \left(\frac{t}{T}\right)\right] \cdot \mathbbm{1}_{\{\tau_0 + 1 \leq t \leq T\}} \right\} \epsilon_t.
\end{multline*}
\noindent 
{\it 
Model 2 (piecewise tvAR(1) model with a kink change-point):}
\begin{equation*}
X_t = 
\left\{
\left[0.75 + 3.0\left(\frac{t - T/2}{T}\right)\right] \cdot \mathbbm{1}_{\{1 \leq t \leq \tau_0\}} 
+ \left[0.75 - 3.0\left(\frac{t - T/2}{T}\right)\right] \cdot \mathbbm{1}_{\{\tau_0 + 1 \leq t \leq T\}} 
\right\} X_{t-1} + \epsilon_t.
\end{equation*}

\begin{figure}[!ht]
    \centering
\includegraphics[width=.72\linewidth]{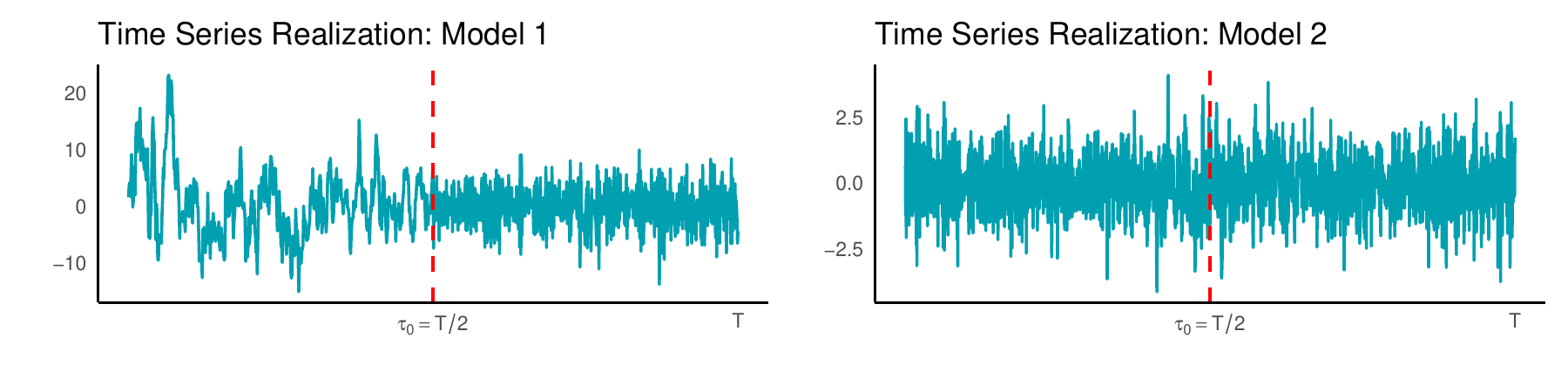}
    \caption{Realizations of time series and AR(1) coefficients for Models 1 and 2.}
    \label{fig:1_2_timeseries}
\end{figure}

The time series realizations for Model~1 (jump change-point) and Model~2 (kink change-point) are displayed in Figure~\ref{fig:1_2_timeseries}.
For each realization in each model, we conduct the proposed method and construct confidence intervals for the change-points under various $h$ and $\tilde{h}$. To ensure sufficient samples for detecting the change-points in~\eqref{eq_final_esti} and~\eqref{eq_final_esti_kink} of Step~3, the window radii should satisfy $h < T/4$ and $\widetilde{h} < T/8$, where $T$ is the sample size.  
To evaluate the performance of the proposed method under different window sizes, we consider the average coverage error (ACE), defined as
\begin{equation}\label{eq:ace}	
	\text{ACE}=\frac{1}{3} \sum_{\alpha \in \{0.80,\, 0.90,\, 0.95\}} \left| C_{\text{emp}}(\alpha) - \alpha \right|\,,
\end{equation} 
where \( C_{\text{emp}}(\alpha) \) denotes the empirical coverage rate of confidence intervals at the nominal level \( \alpha \). A smaller ACE indicates better overall coverage performance across confidence levels.




\begin{figure}[!ht]
    \centering
   \includegraphics[width=.73\linewidth]{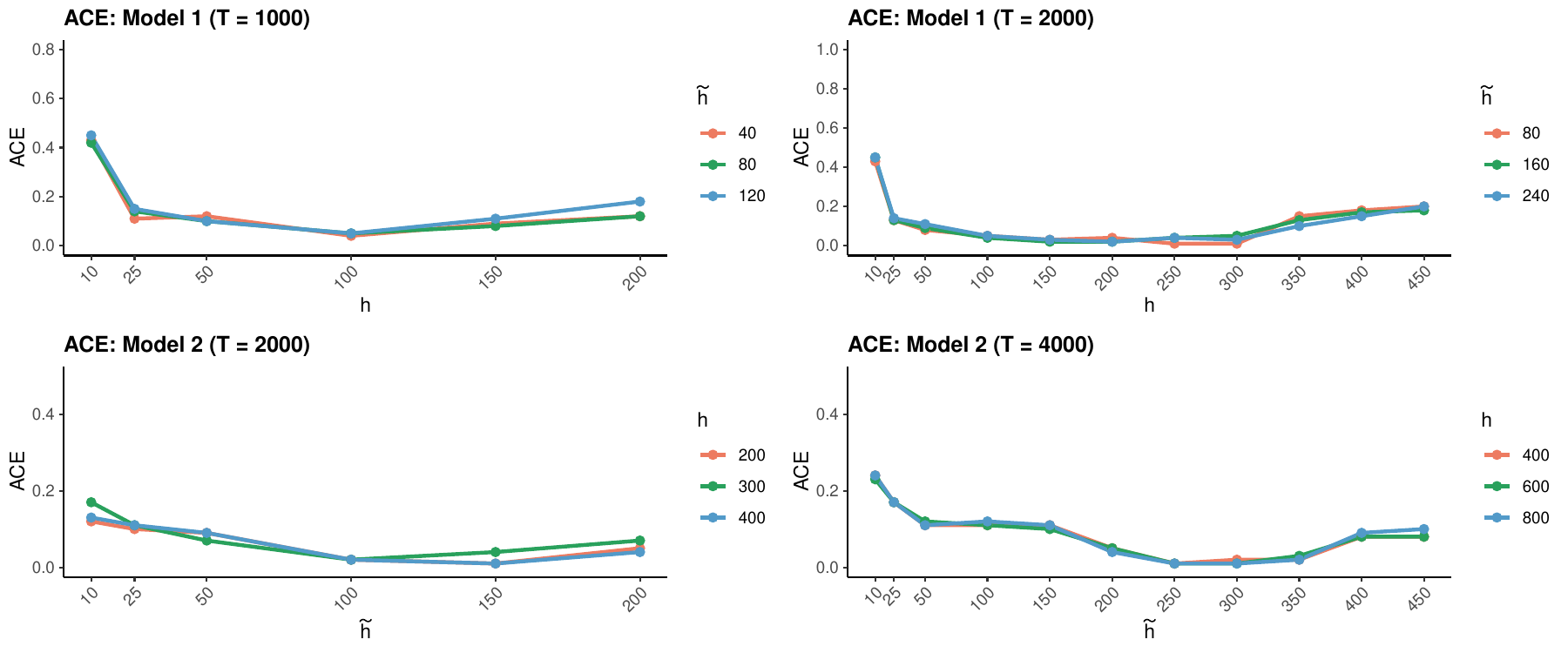}
    \caption{ACE for Model 1 and Model 2 under different radii $h$ and $\widetilde{h}$ and sample size $T$.} 
    \label{fig:line_ce_12}
\end{figure}

Figure~\ref{fig:line_ce_12} presents the average coverage error (ACE) under different choices of window radii and sample sizes for both Model~1 and Model~2. For the jump change-point model (Model~1), ACE increases when the jump window radius \( h \) is either too small or too large, which is consistent with the discussion at the beginning of this section. In contrast, the kink window radius \( \widetilde{h} \) has negligible influence on ACE in this setting, as the model contains only a jump-type change-point. For the kink change-point model (Model~2), a similar pattern is observed. ACE tends to increase when the kink window radius \( \widetilde{h} \) is too small or too large, while the jump window radius \( h \) has limited effect on the performance, since the model only features a kink-type change-point. Detailed numerical values of ACE under different settings are provided in the Supplementary Materials.

Based on these simulation results and in accordance with Assumption~\ref{ass4.5}, we adopt functional forms for the window radii given by \( h = C T^\delta \) and \( \widetilde{h} = \widetilde{C} T^{2/3 + \widetilde{\delta}} \) and  suggest the rule-of-thumb choices \( (C, \delta) = (1.76, 0.58) \) and \( (\widetilde{C}, \widetilde{\delta}) = (0.55, 0.07) \). Specifically, for Model~1, this choice corresponds to \( h = 100 \) when \( T = 1000 \), and \( h = 150 \) when \( T = 2000 \). For Model~2, this choice yields \( \widetilde{h} = 150 \) for \( T = 2000 \), and \( \widetilde{h} = 250 \) for \( T = 4000 \). These values coincide with the low ACE values observed in Figure~\ref{fig:line_ce_12}. 
In the subsequent simulation studies and real data applications, we adopt these optimal parameters to define the jump and kink window radii.

\subsection{Performance of Estimation and Inference}\label{sim:piece_localstat}

In this section, we examine the performance of the proposed method, and compare it with  \cite{Shumway2007} (LS),  which - to our best knowledge - is the only existing method for multiple  change-point detection  in  piecewise locally stationary time series. 
Beside, we benchmark our method to \cite{Dette2015} (MuBreD), which detects jumps in the spectrum of a piecewise stationary process, in order to investigate how this method  performs  when the process is actually piecewise locally stationary.  For this purpose we consider the following
 models. 

\noindent 
{\it  Model 3 (tvAR(1) model without change-points):}  for innovations $\epsilon_t \sim \mathcal{N}\big (0, 1\big )$,
\begin{equation*}
	X_t = \Big [0.99 - 1.98\Big (\frac{t}{2048}\Big  )\Big  ] X_{t-1} + \epsilon_t, \quad 1 \leq t \leq 2048\,.
\end{equation*}
\noindent 
{\it  Model 4 (tvAR(1) model with time-varying noise variance and no change-points): }  for innovations $\epsilon_t \sim \mathcal{N}\big (0, 100 (\frac{t}{2048} - 0.5)^2\big )$,
\begin{equation*}
	X_t = 0.5 X_{t-1} + \epsilon_t, \quad 1 \leq t \leq 2048 .
\end{equation*}
\noindent 
{\it Model 5 (piecewise tvAR(1) model with a single jump):}  for innovations $\epsilon_t \sim \mathcal{N}\big (0, 1\big )$,
\begin{multline*}
X_t = 
\left\{
\left[25.6\left(\frac{t}{2048}\right)^2 - 12.8\left(\frac{t}{2048}\right) + 0.8\right] \cdot \mathbbm{1}_{\{t \leq 1024\}}\right. \\
\left.+ \left[-1.6 \cos\left(\frac{\pi t}{2048}\right) - 0.8\right] \cdot \mathbbm{1}_{\{1025 \leq t \leq 2048\}}
\right\} X_{t-1} + \epsilon_t.
\end{multline*}
Note that the time-varying AR parameter curve in the second segment is a cosine function rather than a polynomial. This model evaluates the method's performance under model misspecfication.

\noindent 
{\it 
Model 6 (piecewise tvAR(1) model with two jumps):}   for innovations $\epsilon_t \sim \mathcal{N}\big (0, 1\big )$,
\begin{multline*}
X_t = 
\Bigl\{
\left[-0.75 + 3\left(\frac{t}{2048}\right)\right] \cdot \mathbbm{1}_{\{1 \leq t \leq 1024\}} 
+ \left[-3.75 + 6\left(\frac{t}{2048}\right)\right] \cdot \mathbbm{1}_{\{1025 \leq t \leq 1536\}} \\
+ \left[-5.25 + 6\left(\frac{t}{2048}\right)\right] \cdot \mathbbm{1}_{\{1537 \leq t \leq 2048\}} 
\Bigr\} X_{t-1} + \epsilon_t.
\end{multline*}
\noindent 
{\it  Model 7 (piecewise tvAR(1) model with two kinks):}  for innovations $\epsilon_t \sim \mathcal{N}\big (0, 1\big )$,
\begin{multline*}
X_t = 
\left\{
\left[0.75 + 3\left(\frac{t - 1024}{2048}\right)\right] \cdot \mathbbm{1}_{\{1 \leq t \leq 1024\}} 
+ \left[0.75 - 1.5\left(\frac{t - 1024}{512}\right)\right] \cdot \mathbbm{1}_{\{1025 \leq t \leq 2048\}}\right. \\
\left.+ \left[-0.75 + 6\left(\frac{t - 1536}{2048}\right)\right] \cdot \mathbbm{1}_{\{2049 \leq t \leq 3072\}} 
\right\} X_{t-1} + \epsilon_t.
\end{multline*}

Besides piecewise locally stationary processes, we also consider piecewise stationary
processes to investigate the efficiency of our general procedure
when stronger assumptions in the data generating process are satisfied.

\noindent 
{\it 
{Model 8} (two change-points in the AR(1) coefficient):}  for innovations $\epsilon_t \sim \mathcal{N}\big (0, 1\big )$,
\begin{equation*}
X_t=
0.75 X_{t-1} \cdot   \mathbbm{1}_{\{ 1 \leq t \leq 840\} } 
-0.75 X_{t-1}  \cdot \mathbbm{1}_{\{  841 \leq t \leq 1644\}} + 
0.75 X_{t-1} \cdot   \mathbbm{1}_{\{ 1645 \leq t \leq 2048\}} + \epsilon_t .
\end{equation*}
This model introduces sign reversals in the AR coefficient, with clearly identifiable structural breaks.

\begin{figure}[!ht]
    \centering
    \includegraphics[width=.9\linewidth]{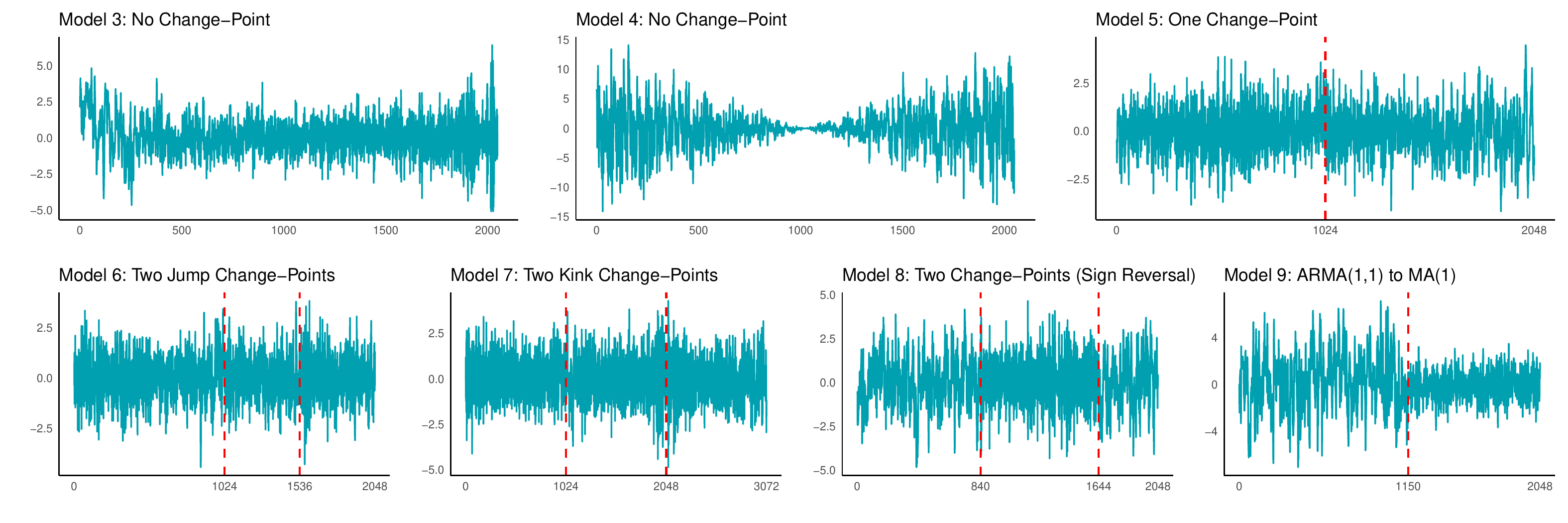}
    \caption{Realizations of time series for Model 3 -  9.}
    \label{fig:3_9_time_series}
\end{figure}

\noindent 
{\it  {Model 9} (Single change-point from ARMA(1,1) to MA(1)):}  for innovations $\epsilon_t \sim \mathcal{N}\big (0, 1\big )$,
\begin{equation*}
X_t = 
0.75 X_{t-1} \cdot \mathbbm{1}_{\{1 \leq t \leq 1150\}} 
+ \epsilon_t 
+ 0.75 \epsilon_{t-1} \cdot \mathbbm{1}_{\{1 \leq t \leq 2048\}}.
\end{equation*}
This model contains a structural change in the autoregressive component, transitioning from an ARMA(1,1) to an MA(1) process. It serves as a misspecified case for our tvAR-based procedure. Realizations of Models 3 t-  7 are shown in Figure~\ref{fig:3_9_time_series}. 

We investigate the performance of the proposed method, the LS  and the MuBreD method in  Models 3 - 9 considering two evaluation metrics. In addition to the average coverage error (ACE) defined in~\eqref{eq:ace}, we introduce the absolute error (AE), defined as
\begin{equation*}	
	\text{AE} = \left| p_{\text{correct}} - 1 \right|\,,
\end{equation*} 
where \( p_{\text{correct}} \) denotes the proportion of replications in which the number of change-points are correctly identified. A smaller AE indicates more accurate detection of the structural change configuration. 
The results of the estimated number of change-points are presented in Table~\ref{table3}.

In general, the proposed method has a higher accuracy than the LS and MuBreD methods in identifying the number of change-points. Also, the standard deviations of the estimated numbers of change-points are smaller, indicating 
the stability of the proposed method.
Note that LS fails in Model~4 since the method does not accommodate time-varying variance. Thus, it wrongly identifies the smoothly changing variance as abrupt changes in the autoregressive structure. Similarly, the MuBreD method, which assumes piecewise stationarity, is not well-suited for smoothly nonstationary models such as Models 3 and 4. In both cases, MuBreD severely overestimates the number of change-points, yielding maximum AE values of 1.000, as it falsely interprets smooth temporal evolution as structural breaks. Additionally,  for Model~5, although our method misspecifies the cosine parameter curve in the second segment 
as a polynomial, 
the change-point estimation remains stable and accurate, whereas MuBreD again overestimates due to its lack of flexibility in modeling time-varying dynamics.
Furthermore, for Model~7, the proposed method significantly outperforms the LS and MuBreD methods. The reason is that the 
LS method only detect jumps in the time varying spectrum, and therefore has no power in detecting kinks. MuBreD, on the other hand, misinterprets the smooth but nonlinear evolution as multiple stationary segments, leading to severe overestimation.  In Model 8 (piecewise stationary AR(1) process) our results indicate that all procedures consistently recover the correct number of change-points, with small and comparable AE and standard deviation (SD) values. The proposed method suffers only from a slight efficiency loss compared to MuBreD, which focuses on piecewise stationarity time series. 
 In contrast, Model 9 introduces structural misspecification by transitioning from an ARMA(1,1) to an MA(1) process, which violates the tvAR assumption underpinning the proposed method. Nevertheless, the proposed method remains robust, producing the most accurate results with the smallest AE (0.035) and the lowest SD (0.0184) among the three methods. 


\begin{table}[hptb]
\centering \caption{Estimated number of change-points for Models 3 to 9.}
\resizebox{0.8\textwidth}{!}{
\begin{tabular}{cc:cccc:cccc:cccc} 
\hline
\multicolumn{2}{c}{NO.CP}		  & \multicolumn{4}{c}{Proposed Method} & \multicolumn{4}{c}{LS}& \multicolumn{4}{c}{MuBreD} \\ 
\hline
Models & True 	& Mean & Median  & SD & AE & Mean & Median  & SD  & AE& Mean & Median  & SD  & AE   \\ 
\hline
3 &  0 &   0.030 & 0 & 0.171  & 0.030 & 0.112 & 0 & 0.328 & 0.108& 2.181    &      2     & 0.465      &   1.000   \\ 
4 &  0 &  0.385 & 0 & 0.607 &  0.320  & 1.870 & 2 & 0.632 & 1.000& 2.764   &       3  &   0.425       &  1.000   \\ 
5 &  1 &  1.155 & 1 & 0.415 & 0.265 & 1.318 & 1 & 0.558 & 0.298 & 2.907      &    3    & 0.304   &  0.999   \\ 
6 &  2 &   2.005 & 2 & 0.307 & 0.105  & 2.091 & 2 & 0.367 & 0.125 &  1.662  &         3   &   1.439 &     0.922  \\ 
7 &  2    &   1.967     &     2   &  0.333 &    0.111 &   0.016     &      0   &   0.178    &  0.992   & 3.377         &  4   &   1.993    &      1.000 \\ 
8  & 2 & 2.007     &    2 &  0.086  &  0.007 &2.002      &    2  &   0.045   &  0.002 &  2.000       &    2     &     0.000      &   0.000 \\
9 &  1 &    1.035      &    1    & 0.184  &   0.035	    & 0.577   & 1  & 0.541 & 0.465 & 1.194    &      1   &  0.396   &  0.194  \\ 
\hline
\end{tabular}
}
\label{table3}
\end{table}


Next, we examine the finite sample performance of the estimators and confidence intervals of change-points. Table~\ref{table4} summarizes the estimation results for Models 5 to 9 based on the realizations that the correct number of change-points are identified.  We report the mean, median, and standard deviation (SD) of estimated change-point locations, the coverage errors at the 80\%, 90\%, and 95\% nominal levels (CE(80\%), CE(90\%), CE(95\%)), and the average coverage error (ACE). For comparison, the results from LS and MuBreD are also included, except for the coverage errors since LS and MuBreD do not produce confidence intervals. 

For Model~5, although the second segment contains model misspecification (a non-polynomial AR coefficient), the proposed method accurately recovers the change-point location. The estimated mean and median (1019 and 1023, respectively) are very close to the true change-point (1024), and the standard deviation (SD) is substantially smaller (19.6) than that of the LS method (376.7). This highlights the robustness of the proposed approach even under deviations from model assumptions. In addition, the average coverage error (ACE = 0.117) of the bootstrap-based confidence intervals remains reasonably low, demonstrating the reliability of the inference procedure. For Model~6, which contains two jump-type change-points at 1024 and 1536, the proposed method consistently yields accurate estimates. The estimated means (1021 and 1536) are nearly identical to the true values, with small SDs of 15.9 and 4.5, respectively. In contrast, LS exhibits larger variability (SDs of 30.8 and 32.8), and MuBreD produces estimates that are heavily biased away from the true locations. The corresponding ACE values (0.040 and 0.023) are also small, confirming the accuracy and stability of the proposed bootstrap-based procedure when model assumptions are satisfied. For Model~7, which involves two kink-type change-points, asymptotic inference is employed to construct confidence intervals. The estimated means (1053 and 2046) are reasonably close to the true locations (1024 and 2048), with variability (SDs of 59.4 and 46.7) higher than the results of jump-dections in the models with jump change-points.  This is reasonable since detecting kinks is in general more difficult than detecting jumps. 
No estimate is reported for MuBreD because none of the 1000 replications correctly identify the true number of change-points. The estimated means (663 and 1797) and SDs (663 and 1797) by the LS method are also not accurate since LS and MuBreD are designed only for detecting jumps but not kinks. 
The small ACE values (0.029 and 0.042) of our methods demonstrate the effectiveness of the asymptotic confidence intervals for kink-type structural changes. For the piecewise stationary Models 8 and 9, the proposed method exhibits superior localization accuracy compared to LS and MuBreD in terms of estimated means, medians and standard deviations. This is somewhat surprising since MuBreD focuses on piecewise stationary time series and the proposed method allows the more general setting of piecewise local-stationarity. 
One possible explanation for the accuracy is as follows.  Both the proposed procedure and MuBreD estimate changes points using scan statistics. The estimates may not be very accurate since each scanning window involves limited data. However, the proposed procedure has a final refinement step which greatly improves preliminary estimates obtained from the scanning step, and results in much better final estimates.



In summary, the simulation experiments in this section demonstrate that the proposed method outperforms the LS and MuBreD approaches in both detection accuracy, especially when model assumptions hold, and maintains robustness even under misspecification. 

\begin{table}[hptb]
\centering \caption{Estimation results and coverage error of confidence intevals in Models 5 to 9. 
}
\resizebox{0.82\textwidth}{!}{
\begin{tabular}{cc:ccccccc:ccc:ccc} \hline
    \multicolumn{2}{c}{True Locations}		  & \multicolumn{7}{c}{Proposed Method} & \multicolumn{3}{c}{LS}& \multicolumn{3}{c}{MuBreD} \\ \hline
Models & Location	&  Mean & Median  & SD  &CE(80\%) &CE(90\%) &CE(95\%) & ACE                & Mean & Median  & SD  & Mean & Median  & SD   \\ \hline
5 & 1024  & 1019  &  1023  & 19.6    &  0.200   &   0.100  &   0.050 & 0.117 & 1015  & 983   & 376.7 & 637     &   637    &    NA  \\
6 & 1024    & 1021 & 1023 & 15.9  & 0.077 & 0.033 & 0.011 & 0.040 & 1022  & 1022  & 30.8 &  688    &    750  &   150.5 \\
6 & 1536    & 1536 & 1535 & 4.5 & 0.018 & 0.009 & 0.041 & 0.023 & 1534  & 1535  & 32.8 &  1465  &     1450     & 164.6\\ 
7&  1024  &   1053  &  1058  & 59.4  &  0.056  &  0.019  &  0.013 & 0.029&  663  &    554 &  460.4 & NA    &     NA  &      NA \\
7& 2048  &   2046  &  2043 &  46.7  &  0.038  &  0.044    & 0.044 & 0.042  & 1797   &   1523   & 683.1 & NA    &     NA  &      NA \\ 
8  & 840 & 840 &    840   & 3.1 &  0.055   &   0.024    &  0.016   & 0.032  &  838    &    839    &  19.4  & 819    &     833  &     34.1 \\
8  & 1644 &  1644    & 1644   &  3.0  &   0.051   &   0.008    &  0.000  &  0.020 & 1645     &  1646   &   19.3 & 1532 &       1535  &     27.2 \\ 
9 & 1150  &   1151    & 1150  & 15.7 & 0.033 &    0.050  &  0.059 & 0.047	    & 1144   & 1149 & 117 & 1131     &  1142    &  28.8  \\\hline
\end{tabular}
}
\label{table4}
\end{table}

\section{Real Data Analysis}
\def\theequation{6.\arabic{equation}}
\setcounter{equation}{0}
\label{sec:realdata}

We apply the method propsed in this paper to the log-return time series of the Hong Kong Hang Seng Index (HSI), using daily closing price data from January 1996 to June 1998, which consists of 612 observations. The goal is to detect structural changes in financial time series and compare the performance of our method with two existing approaches: the LS method of \cite{Shumway2007}, which assumes piecewise local stationarity and is designed to detect only abrupt jump-type changes; and the MuBreD method of \cite{Dette2015}, which is based on a piecewise stationarity assumption.

\begin{figure}[ht]
\centering
\includegraphics[width=0.8\textwidth]{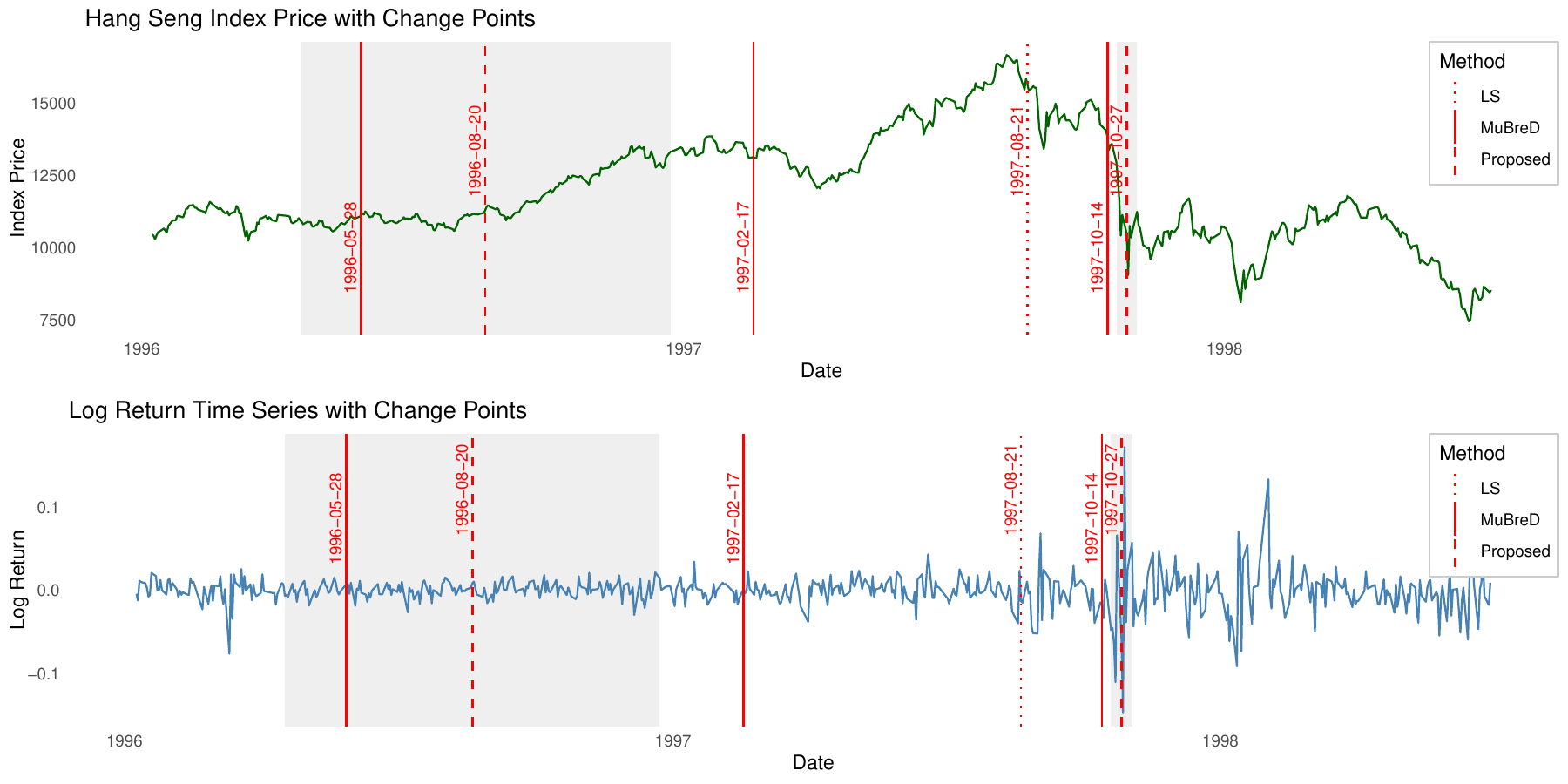}
\caption{Hang Seng Index price and corresponding log-return time series with detected change-points from both the proposed method, the LS and the MuBreD. The Kink (dashed line) and Jump (dashed line) change-points, along with their corresponding confidence intervals (gray shaded areas), are estimated using our proposed method. The LS and MuBreD change-points are represented by dotted lines and solid lines respectively.}
\label{fig:LOC_HS}
\end{figure}

The estimation results of LS, MuBreD and the proposed method are reported in Table~\ref{tab:LOC_HS}. 
The results are also depicted In Figure~\ref{fig:LOC_HS}, which presents the HSI price series and log-return series, with the detected change-points indicated by vertical lines and the corresponding 90\% confidence intervals shown as gray shaded areas. 

Our three-step procedure detects two change-points .The first is a kink-type change-point estimated at index 153, corresponding to August 20, 1996, with a 90\% confidence interval ranging from index 66 (April 17, 1996) to index 240 (December 23, 1996). The second is a jump-type change-point estimated at index 444, corresponding to October 27, 1997, with a 90\% confidence interval from index 439 (October 20, 1997) to index 449 (November 3, 1997). The kink-type change reflects a continuous but non-differentiable change in the slope of the time-varying parameter curves, while the jump reflects a discontinuity in the parameter level.

\begin{table}[ht]
\centering
\caption{Detected change-points and 90\% confidence intervals by the proposed method.}
\resizebox{0.7\textwidth}{!}{
\begin{tabular}{cccccc}
\toprule
Method & CP (Index) & 90\% CI (Index) & CP (Date) & 90\% CI (Date) & Type \\
\midrule
\multirow{2}{*}{Proposed} 
&  153 & [66, 240] & 1996-08-20 & [1996-04-17, 1996-12-23] & Kink \\
&  444 & [439, 449] & 1997-10-27 & [1997-10-20, 1997-11-03] & Jump \\
\hdashline
LS
&  401 &NA &1997-08-21&NA&Jump \\
\hdashline
\multirow{3}{*}{MuBreD} 
& 95 &NA &1996-05-28 &NA&Jump \\
& 275 &NA &1997-02-17 &NA&Jump \\
& 435 &NA &1997-10-14&NA&Jump \\
\bottomrule
\end{tabular}
}
\label{tab:LOC_HS}
\end{table}

The first change-point aligns closely with a politically sensitive period during the final year leading up to the 1997 Handover. Although no abrupt market crash occurred around August 1996, this period was marked by growing political uncertainty and structural transition. Earlier that year, the Preparatory Committee for the Hong Kong Special Administrative Region was established on January 26, 1996, and by mid-1996, preparations for the first Chief Executive election were already underway. These developments may have gradually influenced investor sentiment and market dynamics, resulting in a structural change in the underlying dependence structure. Our method identifies this as a kink-type change, which reflects a continuous but non-differentiable shift in the time-varying spectral density. This type of change is not detectable by methods such as LS, which are designed to capture only abrupt jump-type changes.

The second change-point corresponds directly to the 1997 Hong Kong financial crisis. On October 21 and 22, 1997, the HSI dropped by more than 9\% on each day. On October 23, the index crashed over 10\%, closing at 10,426.3 points, which was a sharp decline from its September level of around 15,000 points.  
This was one of the most severe market crises in Hong Kong’s history, driven by speculative attacks and international financial instability. Our method detects this change-point precisely as a jump, and the confidence interval tightly brackets the major crash days, demonstrating the accuracy and reliability of the refined estimation procedure.

For comparison, from Table \ref{tab:LOC_HS}, the LS method detects only one change-point at time 401, corresponding to August 21, 1997. This change-point occurs nearly two months before the actual market collapse and may reflect early signs of volatility, but it fails to capture the true timing of the structural break. More importantly, the LS method does not detect the kink-type change in 1996, as it is designed only for identifying jumps. In addition, the LS method does not provide confidence intervals, which limits its interpretability and temporal precision.
The MuBreD method detects three jump-type change-points, overestimating the true number of structural breaks. This over-detection results from its assumption of piecewise stationarity, which makes it prone to misinterpreting smoothly time-varying or continuous structural changes as multiple abrupt shifts. In addition, LS and MuBreD do not provide confidence intervals, making their results less interpretable.

\section{Conclusion}\label{sec:conclusion}
In this paper, we have developed a novel three-step procedure for multiple change-point detection and inference in the temporal dynamics of a  piecewise locally stationary time series, allowing for jumps and kinks in the parameter curve. Our method  efficiently estimates  the number, locations and types of change-points. We have theoretically established the consistency for these estimates and the validity of the associated confidence intervals. Simulations and a real data application  show a satisfactory  finite sample performance of the new procedure. Moreover, the methods remains effective in practice even when the piecewise tvAR assumption is mildly violated. The raw data and the source codes are available on GitHub (https://github.com/XinyiTANGaca/LocStaCP).



\section{Supplementary Material}
This section provides additional simulation results, technical proofs and the derivation of the first and second order derivatives of $S_{k,T} (\boldsymbol{\eta}_k)$ with respect to $\boldsymbol{\eta}_k$.
\subsection{Simulation results}\label{sec:detail_sim}
Tables \ref{tab:sen_1a} to \ref{tab:sen_2b}  report the simulation results for Model 1 and 2 under varying window radius \( h \) and \( \widetilde{h} \). The table summarizes the coverage error (CE) of confidence intervals under  nominal levels of 80\%, 90\%, and 95\%, and the average coverage error (ACE). 

\begin{table}[htbp]
\scriptsize
\hspace{3cm}
\begin{tabular}{c*{7}{c}}
\toprule
\hline
\multirow{8}{*}{\normalsize $\widetilde{h} = 40$} & $h$ & 10 & 25 & 50 & 100 & 150 & 200 \\ \cline{2-8}
~ & CE(80CI) & 0.4 & 0.08 & 0.11 & 0.06 & 0.07 & 0.12 \\ 
~ & CE(90CI) & 0.44 & 0.12 & 0.13 & 0 & 0.1 & 0.12 \\ 
~ & CE(95CI) & 0.46 & 0.13 & 0.12 & 0.05 & 0.1 & 0.13 \\ 
~ & AVE.CE & 0.43 & 0.11 & 0.12 & 0.04 & 0.09 & 0.12 \\ \hline
\multirow{8}{*}{\normalsize $\widetilde{h} = 80$} & $h$ & 10 & 25 & 50 & 100 & 150 & 200 \\ \cline{2-8}
~ & CE(80CI) & 0.4 & 0.11 & 0.1 & 0 & 0.07 & 0.11  \\ 
~ & CE(90CI) & 0.43 & 0.15 & 0.11 & 0.06 & 0.05 & 0.14 \\ 
~ & CE(95CI) & 0.44 & 0.17 & 0.1 & 0.1 & 0.12 & 0.1 \\ 
~ & AVE.CE & 0.42 & 0.14 & 0.1 & 0.05 & 0.08 & 0.12 \\ \hline
\multirow{8}{*}{\normalsize $\widetilde{h} = 120$} & $h$ & 10 & 25 & 50 & 100 & 150 & 200 \\ \cline{2-8}
~ & CE(80\%) & 0.42 & 0.11 & 0.1 & 0.01 & 0.08 & 0.14 \\ 
~ & CE(90\%) & 0.46 & 0.16 & 0.1 & 0.06 & 0.10 & 0.18 \\ 
~ & CE(95\%) & 0.47 & 0.19 & 0.11 & 0.09 & 0.15 & 0.21 \\ 
~ & ACE & 0.45 & 0.15 & 0.1 & 0.05 & 0.11 & 0.18 \\ \hline
\bottomrule
\end{tabular}
\caption{{\small  CE and ACE for Model 1 ($T$ = 1000).}}
\label{tab:sen_1a}
\end{table}



\begin{table}[]
\scriptsize
\hspace{0cm}
\begin{tabular}{c*{12}{c}}
\toprule
\hline
\multirow{8}{*}{\normalsize $\widetilde{h} = 80$} &$h$&10&25&50&100&150&200&250&300&350&400&450\\\cline{2-13}
~ & CE(80CI) & 0.39 & 0.09 & 0.02 & 0.03 & 0.05 & 0.08 & 0.02 & 0.03 & 0.15 & 0.11 & 0.21 \\ 
~ & CE(90CI) & 0.44 & 0.14 & 0.1 & 0.06 & 0.02 & 0.03 & 0.01 & 0.01 & 0.13 & 0.22 & 0.23 \\ 
~ & CE(95CI) & 0.47 & 0.15 & 0.11 & 0.07 & 0.01 & 0.01 & 0.01 & 0 & 0.16 & 0.22 & 0.17 \\ 
~ & AVE.CE & 0.43 & 0.13 & 0.08 & 0.05 & 0.03 & 0.04 & 0.01 & 0.01 & 0.15 & 0.18 & 0.2 \\ \hline
\multirow{8}{*}{\normalsize $\widetilde{h} = 160$} &$h$&10&25&50&100&150&200&250&300&350&400&450\\\cline{2-13}
~ & CE(80CI) & 0.41 & 0.1 & 0.05 & 0.02 & 0.02 & 0.04 & 0.01 & 0.03 & 0.11 & 0.11 & 0.18 \\ 
~ & CE(90CI) & 0.46 & 0.14 & 0.09 & 0.04 & 0 & 0.01 & 0.05 & 0.06 & 0.13 & 0.20 & 0.23 \\ 
~ & CE(95CI) & 0.48 & 0.15 & 0.12 & 0.07 & 0.05 & 0.02 & 0.06 & 0.06 & 0.15 & 0.20 & 0.12 \\ 
~ & AVE.CE & 0.45 & 0.13 & 0.09 & 0.04 & 0.02 & 0.02 & 0.04 & 0.05 & 0.13 & 0.17 & 0.18 \\ \hline
\multirow{8}{*}{\normalsize $\widetilde{h} = 240$} &$h$&10&25&50&100&150&200&250&300&350&400&450\\\cline{2-13}
~ & CE(80\%) & 0.41 & 0.11 & 0.06 & 0.02 & 0.01 & 0.03 & 0.01 & 0.02 & 0.11 & 0.09 & 0.15 \\ 
~ & CE(90\%) & 0.46 & 0.16 & 0.12 & 0.06 & 0.02 & 0.01 & 0.05 & 0.01 & 0.07 & 0.16 & 0.22 \\ 
~ & CE(95\%) & 0.49 & 0.16 & 0.14 & 0.08 & 0.05 & 0.03 & 0.06 & 0.05 & 0.12 & 0.19 & 0.23 \\ 
~ & AVE.CE & 0.45 & 0.14 & 0.11 & 0.05 & 0.03 & 0.02 & 0.04 & 0.03 & 0.1 & 0.15 & 0.2 \\ \hline
 \bottomrule
\end{tabular}
\caption{{\small CE and ACE for Model 1 ($T$ = 2000).}}
\label{tab:sen_1b}
\end{table}

\begin{table}[htbp]
\scriptsize
\hspace{3cm}
\begin{tabular}{c*{7}{c}}
\toprule
\hline
\multirow{8}{*}{\normalsize $h = 200$} & $\widetilde{h}$ & 10 & 25 & 50 & 100 & 150 & 200  \\ \cline{2-8}
~ & CE(80\%) & 0.12 & 0.12 & 0.12 & 0.02 & 0.04 & 0.11 \\ 
~ & CE(90\%) & 0.11 & 0.12 & 0.09 & 0.02 & 0.00 & 0.04 \\  
~ & CE(95\%) & 0.13 & 0.07 & 0.06 & 0.03 & 0.00 & 0.00 \\ 
~ &  AVE.CE  & 0.12 & 0.10 & 0.09 & 0.02 & 0.01 & 0.05 \\\hline
\multirow{8}{*}{\normalsize $h = 300$} & $\widetilde{h}$ & 10 & 25 & 50 & 100 & 150 & 200  \\ \cline{2-8}
~ & CE(80CI) & 0.20 & 0.09 & 0.07 & 0.01 & 0.02 & 0.11 \\ 
~ & CE(90CI) & 0.15 & 0.13 & 0.07 & 0.01 & 0.07 & 0.07 \\ 
~ & CE(95CI) & 0.16 & 0.11 & 0.06 & 0.03 & 0.03 & 0.03 \\ 
~ & AVE.CE & 0.17 & 0.11 & 0.07 & 0.02 & 0.04 & 0.07 \\ \hline
\multirow{8}{*}{\normalsize $h = 400$} & $\widetilde{h}$ & 10 & 25 & 50 & 100 & 150 & 200 \\ \cline{2-8}
~ & CE(80CI) & 0.13 & 0.12 & 0.12 & 0.02 & 0.03 & 0.10  \\ 
~ & CE(90CI) & 0.15 & 0.12 & 0.09 & 0.01 & 0.00 & 0.03  \\ 
~ & CE(95CI) & 0.12 & 0.08 & 0.07 & 0.02 & 0.00 & 0.00  \\ 
~ &  AVE.CE  & 0.13 & 0.11 & 0.09 & 0.02 & 0.01 & 0.04  \\\hline
 \bottomrule
\end{tabular}
\caption{{\small CE and ACE for Model 2 ($T$ = 2000).}}
\label{tab:sen_2a}
\end{table}

\begin{table}[]
\scriptsize
\hspace{0cm}
\begin{tabular}{c*{13}{c}}
\toprule
\hline
\multirow{8}{*}{\normalsize $h = 400$} &$\widetilde{h}$&10&25&50&100&150&200&250&300&350&400&450\\\cline{2-13}
&CE (80CI)       &0.25&0.22&0.13&0.13& 0.12 & 0.08 & 0.01 & 0.03 & 0.03 & 0.12 & 0.14 \\
&CE (90CI)    &0.24&0.18&0.12&0.12& 0.11 & 0.06 & 0.00 & 0.01 & 0.02 & 0.08 & 0.07 \\
&CE (95CI)   &0.24&0.1&0.08&0.07& 0.10 & 0.01 & 0.02 & 0.01 & 0.02 & 0.04 & 0.02 \\
&AVE.CE  & 0.24 & 0.17 & 0.11 & 0.11 & 0.11 & 0.05 & 0.01 & 0.02 & 0.02 & 0.08 & 0.08\\\hline
\multirow{8}{*}{\normalsize $h = 600$} &$\widetilde{h}$&10&25&50&100&150&200&250&300&350&400&450\\\cline{2-13}
~ & CE(80CI) &0.25 & 0.22 & 0.14 & 0.13 & 0.11 & 0.09 & 0.01 & 0.02 & 0.04 & 0.13 & 0.14 \\ 
~ & CE(90CI) &0.24 & 0.18 & 0.13 & 0.13 & 0.1 & 0.06 & 0.01 & 0.01 & 0.02 & 0.08 & 0.07 \\ 
~ & CE(95CI) &0.20 & 0.1 & 0.09 & 0.08 & 0.1 & 0.01 & 0.02 & 0.01 & 0.02 & 0.04 & 0.02 \\ 
~ & AVE.CE &0.23 & 0.17 & 0.12 & 0.11 & 0.1 & 0.05 & 0.01 & 0.01 & 0.03 & 0.08 & 0.08 \\ \hline
\multirow{8}{*}{\normalsize $h = 800$} &$\widetilde{h}$&10&25&50&100&150&200&250&300&350&400&450\\\cline{2-13}
&CE(80CI) &0.25&0.22&0.13&0.14&0.11&0.07&0.00&0.01&0.03&0.14&0.14 \\ 
&CE(90CI) &0.23&0.18&0.12&0.14&0.1&0.05&0.01&0.02&0.02&0.09&0.10\\ 
&CE(95CI)  &0.23&0.10&0.08&0.09&0.11&0.01&0.03&0.01&0.02&0.04&0.06\\ 
&AVE.CE   & 0.24 & 0.17 & 0.11 & 0.12 & 0.11 & 0.04 & 0.01 & 0.01 & 0.02 & 0.09 & 0.10 \\   \hline
 \bottomrule
\end{tabular}
\caption{{\small CE and ACE for Model 2 ($T$ = 4000).}}
\label{tab:sen_2b}
\end{table}

\newpage
\subsection{Proofs}
\begin{proof}[\bf{Proof of Theorem \ref{thm4.1}}]
	We first prove the asymptotic results for change-points with a jump in the spectral density. Let $A_t=$ \{some point in the $t$-th local-window $W_{h}(t)$ is a local change-point estimate in $\hat{J}$\} and let $A=\bigcap_{t\in J_0}A_t$. The proof is completed if we can prove that $\mathbb{P}(A)\to 1$ as $T\to\infty$. Let $\mathbb{Z}_T=\{1,2,\dots,T\}$ and define $\mathcal{E}=\mathbb{Z}_T \backslash\{\bigcup_{t\in J_0} W_{h}(t)\}$ as the set of all points outside the $h$-neighborhood of the true change-points with a jump. 
	
	By Assumptions \ref{ass4.1}, \ref{ass4.2} and the proof of Proposition 2.4 in \cite{dahlhaus2009empirical}, {
     if all the coefficient functions $\phi_j(\cdot)$ and $\sigma^2 \left(\cdot\right)$ in \eqref{eq_tvAR} are of bounded variation, and $1-\sum_{i=1}^p \phi_i(u)z^i \neq 0$ for all $u\in [0,1]$ and all $|z|\leq 1+c$ for some $c>0$, then the tvAR($p$) process $\{X_{t,T}\}$ in \eqref{eq_tvAR} can be represented by a one-sided tvMA($\infty$) process of the form
\begin{equation} \label{eq_tvMA}
	X_{t,T}=\sum_{j=0}^\infty \theta_{t,T}(j) \epsilon_{t-j}\,,
\end{equation}
where $\{\epsilon_{t}\}_{t \in \mathbb{Z}}$ denotes a centered unit-variance Gaussian white noise process, and the coefficients $\theta_{t,T}(j)$'s satisfy $\sup_{t,T} |\theta_{t,T}(j)| \leq K \rho^j$  with some constants $K>0$ and $0<\rho<1$:
Also, there exists a function $\theta:[0,1]\times \{0,1,\ldots\}\rightarrow \mathbb{R}$ satisfying 
\begin{equation*}
	\sup_{u} |\theta(u,j)| \leq K \rho^j\,,~~~\sum^T_{t=1} \Big| \theta_{t,T}(j)-\theta\Big (\frac{t}{T},j\Big) \Big| \leq Kj^2\rho^{j-1}\,,~~~\text{and}~~~ V(\theta\left(\cdot,j\right)) \leq Kj^2 \rho^{j-1} \,,
\end{equation*}
where $V(\theta\left(\cdot,j\right))$ is the total variation of the function $\theta\left(\cdot,j\right)$. This representation~\eqref{eq_tvMA} satisfies the summability condition (3.1) in Theorem 3.1 of \cite{Dette2015}, ensuring the validity of the theoretical framework.} Hence, by Assumption \ref{ass4.5}, and similar arguments in the proof of Theorems 3.2 and 3.5 of \cite{Dette2015}, we can show that for any $0<\gamma< 1/2$, we have
	\begin{equation}\label{detteeq1}
		h^\gamma \sup_{t \in \{h,h+1,\ldots,T-h\}} \sup_{w \in \{0,1,2,\ldots,h/2\}} \left| D_h(t,w) - D_{h,T}(t,w) \right| = o_p(1)\,,
	\end{equation}
	where $$D_{h,T}(t,w)=\frac{T}{h} \left( \int_{\frac{-2w\pi}{h}}^{\frac{2w\pi}{h}} \int_{\frac{t}{T}}^{\frac{t+h}{T}} f(u,\lambda) du d\lambda - \int_{\frac{-2w\pi}{h}}^{\frac{2w\pi}{h}} \int_{\frac{t-h}{T}}^{\frac{t}{T}} f(u,\lambda) du d\lambda\right)\,,$$
	Since by Assumption \ref{ass4.1}(b), we have $\sup_{w \in \{0,1,2,\ldots,h/2\}} |D_{h,T}(\tau^0_k,w)| > 0$ for all $\tau^0_k \in J_0$, and by the continuity of the spectral density for each segment, we have 
	\begin{eqnarray*}
		\sup_{w \in \{0,1,2,\ldots,h/2\}} |D_{h,T}(t,w)| =o(1)\,,
	\end{eqnarray*} 
	as $T\rightarrow \infty$ for all $t \in \mathcal{E}$. 
	Also, the followings hold:
	\begin{itemize}
		\item[(i)] There exists a constant $C \in \mathbb{R}^{+}$ such that for all true change-point with a jump $\tau^0_k \in J_0$,
		\begin{equation*}
		\lim_{T \rightarrow \infty} \mathbb{P} \left( \mathcal{D}_h(\tau^0_k) > C \right)  = 1\,.
		\end{equation*}
		
		\item[(ii)] For any $C^* \in \mathbb{R}^{+}$, we have
		\begin{equation*}
		\lim_{T \rightarrow \infty}\mathbb{P}\left(\max_{t \in \mathcal{E}} \mathcal{D}_h(t)\geq C^* \right) = 0 \,. 
		\end{equation*}
	\end{itemize}	
	Note that one sufficient condition for the event $A$ to occur is that
	\begin{equation} \label{thm1_1}
	\min_{t\in C_0} \mathcal{D}_h(t)>\max_{t\in\mathcal{E}} \mathcal{D}_h(t)\,.
	\end{equation}	
	Note that from \eqref{thm1_1}, we have	
	\begin{equation} \label{thm1_2}
	\mathbb{P}(A)\geq \mathbb{P}\bigg(\min_{t\in C_0} \mathcal{D}_h(t)> \frac{C}{2} >\max_{t\in\mathcal{E}} \mathcal{D}_h(t)\bigg)\,.
	\end{equation}
	where $C>0$ satisfying (i).	Thus, we have
	\begin{equation}\label{thm1_3}
	\begin{aligned}
	\mathbb{P}\left(\min_{t\in C_0}\mathcal{D}_h(t)>\frac{C}{2} \right)&=1-\mathbb{P}\left(\bigcup_{t\in C_0}\{\mathcal{D}_h(t)\leq C/2 \}\right) \\ & \geq 1-\sum_{t\in C_0}\mathbb{P} \left(\mathcal{D}_h(t)\leq \frac{C}{2} \right) \to 1\,.
	\end{aligned}
	\end{equation}
	Also, by similar arguments in the proof of Theorem 3.5 of \cite{Dette2015}, we have
	\begin{equation}\label{thm1_4}
	\begin{aligned}
	\mathbb{P}\left(\frac{C}{2} >\max_{t\in\mathcal{E}} \mathcal{D}_h(t)\right)&=1-\mathbb{P}\left(\bigcup_{t\in\mathcal{E}}\{\mathcal{D}_h(t)\geq C/2\}\right)\\  & \geq 1-\mathbb{P}\left(\max_{t \in \mathcal{E}} \mathcal{D}_h(t)\geq C/2 \right) \to 1\,.
	\end{aligned}
	\end{equation}
	Combining \eqref{thm1_2}, \eqref{thm1_3}, and \eqref{thm1_4} yields $\mathbb{P}(A)\to 1$, completing the proof of first half of Theorem \ref{thm4.1}.
	
	We now prove the asymptotic results for change-points with a kink in the spectral density. Similar to the proof above, define $A^{(1)}_t=$ \{some point in the $t$-th local-window $W_{2\widetilde{h}}(t)$ is a local change-point estimate in $\hat{K}$\} and let $A^{(1)}=\bigcap_{t\in K_0}A^{(1)}_t$. The proof is completed if we can prove $\mathbb{P}(A^{(1)})\to 1$ as $T\to\infty$. Let $\mathbb{Z}_T=\{1,2,\dots,T\}$ and define $\mathcal{E}^{(1)}=\mathbb{Z}_T \backslash\{\bigcup_{t\in K_0} W_{2\widetilde{h}}(t)\}$ as the set of all points outside the $2\widetilde{h}$-neighborhood of the true change-points with a kink. Similar to the above \eqref{detteeq1}, we have for any $0<\gamma< 1/2$, we have
	\begin{equation}\label{detteeq2}
	\widetilde{h}^\gamma \sup_{t \in \{\widetilde{h},\widetilde{h}+1,\ldots,T-\widetilde{h}\}} \sup_{w \in \{0,1,2,\ldots,\widetilde{h}/2\}} \left| D_{\widetilde{h}}(t,w) - D_{\widetilde{h},T}(t,w) \right| = o_p(1)\,,
	\end{equation}
	where $$D_{\widetilde{h},T}(t,w)=\frac{T}{\widetilde{h}} \left( \int_{\frac{-2w\pi}{\widetilde{h}}}^{\frac{2w\pi}{\widetilde{h}}} \int_{\frac{t}{T}}^{\frac{t+\widetilde{h}}{T}} f(u,\lambda) du d\lambda - \int_{\frac{-2w\pi}{\widetilde{h}}}^{\frac{2w\pi}{\widetilde{h}}} \int_{\frac{t-\widetilde{h}}{T}}^{\frac{t}{T}} f(u,\lambda) du d\lambda\right)\,.$$
	Since by definition $D^{(1)}_{\widetilde{h}}(t,w) = \frac{T}{\widetilde{h}}
	[D_{\widetilde{h}}(t+{\widetilde{h}},w) - D_{\widetilde{h}}(t-{\widetilde{h}},w)]$. 
	Define $$D^{(1)}_{\widetilde{h},T}(t,w)=\frac{T}{\widetilde{h}} \left( \int_{\frac{-2w\pi}{\widetilde{h}}}^{\frac{2w\pi}{\widetilde{h}}} \int_{\frac{t}{T}}^{\frac{t+\widetilde{h}}{T}} f^{(1)}(u,\lambda) du d\lambda - \int_{\frac{-2w\pi}{\widetilde{h}}}^{\frac{2w\pi}{\widetilde{h}}} \int_{\frac{t-\widetilde{h}}{T}}^{\frac{t}{T}} f^{(1)}(u,\lambda) du d\lambda\right)\,,$$
	Since we have for $t \in \{2\widetilde{h},2\widetilde{h}+1,\ldots,T-2\widetilde{h}\}$ and $w \in \{0,1,2,\ldots,\widetilde{h}/2\}$,
	\begin{eqnarray*}
		& & \frac{D_{\widetilde{h},T}(t+{\widetilde{h}},w) - D_{\widetilde{h},T}(t-{\widetilde{h}},w)}{\widetilde{h}/T} \\
		&=& \frac{T^2}{\widetilde{h}^2} \left( \int_{\frac{-2w\pi}{\widetilde{h}}}^{\frac{2w\pi}{\widetilde{h}}} \int_{\frac{t+\widetilde{h}}{T}}^{\frac{t+2\widetilde{h}}{T}} f(u,\lambda) du d\lambda - \int_{\frac{-2w\pi}{\widetilde{h}}}^{\frac{2w\pi}{\widetilde{h}}} \int_{\frac{t}{T}}^{\frac{t+\widetilde{h}}{T}} f(u,\lambda) du d\lambda\right.\\
		& & \quad\quad\quad\quad - \left.\int_{\frac{-2w\pi}{\widetilde{h}}}^{\frac{2w\pi}{\widetilde{h}}} \int_{\frac{t-\widetilde{h}}{T}}^{\frac{t}{T}} f(u,\lambda) du d\lambda + \int_{\frac{-2w\pi}{\widetilde{h}}}^{\frac{2w\pi}{\widetilde{h}}} \int_{\frac{t-2\widetilde{h}}{T}}^{\frac{t-\widetilde{h}}{T}} f(u,\lambda) du d\lambda\right)\\
		&=& \frac{T^2}{\widetilde{h}^2} \left( \int_{\frac{-2w\pi}{\widetilde{h}}}^{\frac{2w\pi}{\widetilde{h}}} \int_{\frac{t}{T}}^{\frac{t+\widetilde{h}}{T}} f\left(u+\frac{\widetilde{h}}{T},\lambda \right) du d\lambda - \int_{\frac{-2w\pi}{\widetilde{h}}}^{\frac{2w\pi}{\widetilde{h}}} \int_{\frac{t}{T}}^{\frac{t+\widetilde{h}}{T}} f(u,\lambda) du d\lambda\right.\\
		& & \quad\quad\quad\quad - \left.\int_{\frac{-2w\pi}{\widetilde{h}}}^{\frac{2w\pi}{\widetilde{h}}} \int_{\frac{t-\widetilde{h}}{T}}^{\frac{t}{T}} f(u,\lambda) du d\lambda + \int_{\frac{-2w\pi}{\widetilde{h}}}^{\frac{2w\pi}{\widetilde{h}}} \int_{\frac{t-\widetilde{h}}{T}}^{\frac{t}{T}} f\left(u-\frac{\widetilde{h}}{T},\lambda \right) du d\lambda\right)\\
		&=& \frac{T}{\widetilde{h}} \left( \int_{\frac{-2w\pi}{\widetilde{h}}}^{\frac{2w\pi}{\widetilde{h}}} \int_{\frac{t}{T}}^{\frac{t+\widetilde{h}}{T}} \frac{f(u+\widetilde{h}/T,\lambda) - f(u,\lambda)}{\widetilde{h}/T} du d\lambda\right. - \left.\int_{\frac{-2w\pi}{\widetilde{h}}}^{\frac{2w\pi}{\widetilde{h}}} \int_{\frac{t-\widetilde{h}}{T}}^{\frac{t}{T}} \frac{f(u,\lambda) - f(u-\widetilde{h}/T,\lambda)}{\widetilde{h}/T} du d\lambda\right)\\
		&=&\frac{T}{\widetilde{h}} \left( \int_{\frac{-2w\pi}{\widetilde{h}}}^{\frac{2w\pi}{\widetilde{h}}} \int_{\frac{t}{T}}^{\frac{t+\widetilde{h}}{T}} f^{(1)}(u,\lambda) du d\lambda - \int_{\frac{-2w\pi}{\widetilde{h}}}^{\frac{2w\pi}{\widetilde{h}}} \int_{\frac{t-\widetilde{h}}{T}}^{\frac{t}{T}} f^{(1)}(u,\lambda) du d\lambda\right)+o(1)\\
		&=& D^{(1)}_{\widetilde{h},T}(t,w)+o(1)\,.
	\end{eqnarray*}
	and the convergence is uniform for all $(t,w)$. As a result, by Assumption \ref{ass4.5}(ii) we have for any $0<\gamma< 1/2$,
	\begin{equation}\label{detteeq3}
	 \sup_{t \in \{2\widetilde{h},\widetilde{h}+1,\ldots,T-2\widetilde{h}\}} \sup_{w \in \{0,1,2,\ldots,\widetilde{h}/2\}} \left| D^{(1)}_{\widetilde{h}}(t,w) - D^{(1)}_{\widetilde{h},T}(t,w) \right| = o_p \left(\frac{T}{\widetilde{h}^{1+\gamma}}\right)=o_p(1)\,,
	\end{equation}

Since by Assumption \ref{ass4.1}(c), we have $\sup_{w \in \{0,1,2,\ldots,h/2\}} |D^{(1)}_{\widetilde{h},T}(\tau^0_k,w)| > 0 $ for all $\tau^0_k \in K_0$, and by the continuity of the first derivative of the spectral density for each segment, we have for all $t \in \mathcal{E}^{(1)}$, $\sup_{w \in \{0,1,2,\ldots,h/2\}} |D^{(1)}_{\widetilde{h},T}(t,w)| =o(1)$ as $T\rightarrow \infty$, and hence we have the followings hold:
\begin{itemize}
\item[(i)] There exists a constant $C \in \mathbb{R}^{+}$ such that for all true change-point $\tau^0_k \in K_0$,
\begin{equation*}
	\lim_{T \rightarrow \infty} \mathbb{P} \left( \mathcal{D}^{(1)}_{\widetilde{h}}(\tau^0_k) > C \right)  = 1\,.
\end{equation*}

\item[(ii)] For any $C^* \in \mathbb{R}^{+}$, we have
\begin{equation*}
	\lim_{T \rightarrow \infty}\mathbb{P}\left(\max_{t \in \mathcal{E}^{(1)}} \mathcal{D}^{(1)}_{\widetilde{h}}(t)\geq C^* \right) = 0 \,. 
\end{equation*}
\end{itemize}	
Note that one sufficient condition for the event $A^{(1)}$ to occur is that
\begin{equation} \label{thm1_11}
\min_{t\in K_0} \mathcal{D}^{(1)}_{\widetilde{h}}(t)>\max_{t\in\mathcal{E}^{(1)}} \mathcal{D}^{(1)}_{\widetilde{h}}(t)\,.
\end{equation}	
Note that from \eqref{thm1_11}, we have	
\begin{equation} \label{thm1_22}
\mathbb{P}(A^{(1)})\geq \mathbb{P}\bigg(\min_{t\in K_0} \mathcal{D}^{(1)}_{\widetilde{h}}(t)> \frac{C}{2} >\max_{t\in\mathcal{E}^{(1)}} \mathcal{D}^{(1)}_{\widetilde{h}}(t)\bigg)\,.
\end{equation}
where $C>0$ satisfying (i).	Thus, we have
\begin{equation}\label{thm1_33}
\begin{aligned}
	\mathbb{P}\left(\min_{t\in K_0} \mathcal{D}^{(1)}_{\widetilde{h}}(t)>\frac{C}{2} \right)&=1-\mathbb{P}\left(\bigcup_{t\in K_0}\{\mathcal{D}^{(1)}_{\widetilde{h}}(t)\leq C/2 \}\right) \\ & \geq 1-\sum_{t\in K_0}\mathbb{P} \left(\mathcal{D}^{(1)}_{\widetilde{h}}(t)\leq \frac{C}{2} \right) \to 1\,.
\end{aligned}
\end{equation}
and 
\begin{equation}\label{thm1_44}
\begin{aligned}
	\mathbb{P}\left(\frac{C}{2} >\max_{t\in\mathcal{E}^{(1)}} \mathcal{D}^{(1)}_{\widetilde{h}}(t)\right)&=1-\mathbb{P}\left(\bigcup_{t\in\mathcal{E}^{(1)}}\{\mathcal{D}^{(1)}_{\widetilde{h}}(t)\geq C/2\}\right)\\  & \geq 1-\mathbb{P}\left(\max_{t \in \mathcal{E}^{(1)}} \mathcal{D}^{(1)}_{\widetilde{h}}(t)\geq C/2 \right) \to 1\,.
\end{aligned}
\end{equation}
Combining \eqref{thm1_22}, \eqref{thm1_33}, and \eqref{thm1_44} yields $\mathbb{P}(A^{(1)})\to 1$, completing the proof of Theorem \ref{thm4.1}.

\end{proof}

\begin{proof}[\bf{Proof of Theorem \ref{thm4.2}}]
	Let $\boldsymbol{\theta}_k=(\phi_{k10},\dots,\phi_{k{p_k}{q_k}},\sigma_{k0},\dots,\sigma_{k{q_k}})$ be the set of parameters in the $k$-th segment and $\mathbf{X}_{k,T}=(X_{\tau_{k-1}+1,T},X_{\tau_{k-1}+2,T},\dots,X_{\tau_k,T})$ be the $k$-th segment of the time series, we can then express the conditional Gaussian log-likelihood function for observation $X_{t,T}$ in the $k$-th segment as	
	\begin{equation*}
		\ell_k(\boldsymbol{\theta}_k,X_{t,T})=-\frac{1}{2}\log \left(2\pi \sigma_{k,\boldsymbol{\theta}_k}^2\left(\frac{t}{T}\right)\right) - \frac{\left(X_{t,T}-\sum_{i=1}^{p_{k}}\phi_{k,i,\boldsymbol{\theta}_k}\left(\frac{t}{T}\right) X_{t-i,T}  \right)^2}{2 \sigma_{k,\boldsymbol{\theta}_k}^2\left(\frac{t}{T}\right)}\,,
	\end{equation*} 
	and the conditional Gaussian log-likelihood for the $k$-th segment as 	
	\begin{equation*}
	L_T^{(k)}(\boldsymbol{\theta}_k,\mathbf{X}_{k,T}) =  \sum_{t=\tau_{k-1}+p_k+1}^{\tau_k} \ell_k(\boldsymbol{\theta}_k,X_{t,T})\,.
	\end{equation*}
	The proof of Theorem \ref{thm4.2} is based on an extension of Theorem 2 of \cite{LRSM2015} from piecewise stationary AR models to piecewise locally stationary tvAR models, and the proof of Theorem 2 of \cite{LRSM2015} is based on Theorem 5 of \cite{Yau2013}. Hence, Theorem \ref{thm4.2} can be proved if Assumptions 1(1), 1*, 2(2), 3 and 5 in \cite{Yau2013} are fulfilled. From Theorem 4.2 of \cite{Dahlhaus1997}, we can see that the maximum conditional Gaussian log-likelihood estimation can yield consistent estimation at a convergence rate of $O_p(T^{-1/2})$ for the parameters in the sense that all the estimated parameters $\hat{\phi}_{kij}$ and $\hat{\sigma}_{kj}$ will converge to $\phi_{kij}^0$ and $\sigma_{kj}^0$ with convergence rate $O_p(T^{-1/2})$ when the length of each segment of the time series tends to infinity. Theorem 2.7 of \cite{Dahlhaus2017}  established a generalization of the ergodic theorem to locally stationary process. Together with the compactness of parameter spaces, it follows that Assumption 3 in \cite{Yau2013} is fulfilled. 
	From the conditional Gaussian log-likelihood function of the model which only depends on $(X_{t,T},\dots,X_{t-p_k,T})$ with finite length $p_k+1$,  and Assumptions \ref{ass4.1} to \ref{ass4.5} which state the requirement of moment existence and model identifiability, we can see that our model setting also satisfy Assumptions 1(1), 1*, 2(2) and 5 in \cite{Yau2013}. As a result, in \cite{Yau2013}, it is shown that optimizing $\mbox{MDL}(m,C,\boldsymbol{p},\boldsymbol{q})$ over all possible change-point locations gives consistent estimates. Essentially, the proof is based on the following two arguments. First, for any estimate $(\hat{m},\hat{C},\hat{\boldsymbol{p}},\hat{\boldsymbol{q}})$, if no element in $\hat{C}$ is close to some $\tau^0_k$, then the lack-of-fit term in $\mbox{MDL}(\hat{m},\hat{C},\hat{\boldsymbol{p}},\hat{\boldsymbol{q}})$ dominates and is greater than $\mbox{MDL}(m_0,C_0,\boldsymbol{p}_0,\boldsymbol{q}_0)$, contradicting that the estimates minimize the $\mbox{MDL}$. Second, if the locations are consistently estimated but there are more estimated change-points, i.e., $\hat{m}>m_0$, then the penality term dominates and $\mbox{MDL}(\hat{m},\hat{C},\hat{\boldsymbol{p}},\hat{\boldsymbol{q}})$ is greater than $\mbox{MDL}(m_0,C_0,\boldsymbol{p}_0,\boldsymbol{q}_0)$, arriving at the same contraditciton. Thus, the same argument remains valid if the optimization domain is replaced by $\hat{J}\cup\hat{K}$, since $\tau^0_k$ can be consistently estimated by some elements in $\hat{J}\cup\hat{K}$. This completes the proof of Theorem \ref{thm4.2}.
\end{proof}

\begin{proof}[\bf{Proof of Theorem \ref{thm4.3}}]
	In estimating the final location $\hat{\tau}_k^{(3)}$ of each change-point with a jump in \eqref{eq_final_esti} from $\hat{\tau}_k^{(2)} \in \hat{C}^{(2)} \cap \hat{J}$, we consider the observations within the extended local window $\{ \hat{\tau}_{k-1}^{(2)} + {h_{k-1}}, \dots, \hat{\tau}_{k}^{(2)}-1, \hat{\tau}_{k}^{(2)}, \hat{\tau}_{k}^{(2)}+1, \dots, \hat{\tau}_{k+1}^{(2)} - {h_{k+1}}\}$ of each potential change-point. From Theorem 4.2 of \cite{Dahlhaus1997} and that $h/T \rightarrow 0$, we have consistent parameter estimates in \eqref{eq_final_esti}, $\hat{\boldsymbol{\theta}}_k=\hat{\boldsymbol{\theta}}_{k} (\tau) \rightarrow \boldsymbol{\theta}_k^0$ and $\hat{\boldsymbol{\theta}}_{k+1}=\hat{\boldsymbol{\theta}}_{k+1} (\tau) \rightarrow \boldsymbol{\theta}_{k+1}^0$ in probability for all $\tau \in [\hat{\tau}_{k}^{(2)} - {h},\hat{\tau}_{k}^{(2)} + {h}]$. Since the ratio $h/T$ tends to 0 as $T \rightarrow \infty$, we can approximate the process in the neighborhood of the true change-point $\tau^0_k$ locally as a piecewise stationary AR process with a jump in the AR parameters. Then based on stationary AR($\hat{p}_{k}^{(2)}$) models and AR($\hat{p}_{k+1}^{(2)}$) models, there exists a function $G(\mathbf{X}_t)$ that dominates $|\ell_k(\boldsymbol{\theta}_k,X_t)|$ and $|\ell_{k+1}(\boldsymbol{\theta}_{k+1},X_t)|$ for all $\boldsymbol{\theta}_k\in\boldsymbol{\Theta}_k$ and $\boldsymbol{\theta}_{k+1}\in\boldsymbol{\Theta}_{k+1}$. We can take $G(X_t,\dots,X_{t-\hat{p}_{k}^{(2)}})=|\frac{1}{2}\log(2\pi \sigma^2)|+(\sum_{j=0}^{\hat{p}_{k}^{(2)}}X_j^2+2\sum_{i\neq j}|X_i X_j|)/2\sigma^2$ and $G(X_t,\dots,X_{t-\hat{p}_{k+1}^{(2)}})=|\frac{1}{2}\log(2\pi \sigma^2)|+(\sum_{j=0}^{\hat{p}_{k+1}^{(2)}}X_j^2+2\sum_{i\neq j}|X_i X_j|)/2\sigma^2$ respectively. By using the uniform law of large number (ULLN) in \cite{jennrich1969asymptotic}, we have as $h\to\infty$, $\frac{1}{h}\sum_{t=1}^{h}\ell_k(\boldsymbol{\theta}_k,X_t)$ converges uniformly to $\mathbb{E}_{\boldsymbol{\theta}_k^0}(\ell_k(\boldsymbol{\theta}_k,X_{t,T})), \forall \boldsymbol{\theta}_k\in\boldsymbol{\Theta}_k$ and also similarly for the $(k+1)$-th segments. The remaining part of the proof follows the same as that of Theorem 3 in \cite{LRSM2015}.
\end{proof}

\begin{proof}[\bf{Proof of Theorem \ref{thm4.3_kink}}]
	To show the asymptotic normality of $\hat{\boldsymbol{\eta}}_k^{(3)}$, we first derive its consistency. By Theorems 3.2 and 4.2 in \cite{Dahlhaus1997}, we have consistent parameter estimation and its asymptotic normality for every subsegment without the true change-point $r_k^0$ within the extended window. Hence if there exists a subsequence $T_m \subset \mathbb{N}$ such that $\hat{r}^{(3)}_k \rightarrow r^*$ where $|r^*-r_k^0|>c$ for some $c>0$ when $T_m\rightarrow\infty$, then the argmax condition in \eqref{eq_final_esti_kink} is violated because the sum of $\ell(\boldsymbol{\eta}_k,X_{t,T})$ in the segment between $r^*$ and $r_k^0$ is not attaining the maximum. Hence, we have  $\hat{\boldsymbol{\eta}}_k^{(3)} \overset{p}{\to} \boldsymbol{\eta}_k^0$. Next, we show the asymptotic normality of $\hat{\boldsymbol{\eta}}_k^{(3)}$. Theorems 2.7 and 2.9 of \cite{Dahlhaus2017} established the large of large numbers and the central limit theorem for the locally stationary sequence $\{\ell(\boldsymbol{\eta}_k,X_{t,T})\}$, then by the compactness of the parameter space and similar arguments in the proof of Theorem 2.1 in \cite{zhong2022estimation}, it follows that $\sqrt{(r_{k+1}^0-r_{k-1}^0)T} \left(\hat{\boldsymbol{\eta}}_k^{(3)}-\boldsymbol{\eta}_k^0\right)$ is asymptotically normal with mean $0$ and variance matrix $\mathbf{\Sigma}_k=\mathbf{D}_k^{-1}\mathbf{G}_k\mathbf{D}_k^{-1}$. This completes the proof of Theorem \ref{thm4.3_kink}.
\end{proof}

\begin{proof}[\bf{Proof of Theorem \ref{thm4.4}}]
	Denote $\widetilde{\mathbb{P}}$ as the probability measure conditional on the original sample $\{X_{1,T},\dots,X_{T,T}\}$ under the bootstrap scheme. Below we drop the superscript $(i)$ indicating the $i$-th bootstrap sample for notational simplicity. Note that 
	\begin{eqnarray*}
	\widetilde{d}_k & = & \argmax_{d \in \{\hat{\tau}_{k}^{(2)} - {h}-\hat{\tau}_k^{(3)},\ldots,\hat{\tau}_{k}^{(2)} + {h}-\hat{\tau}_k^{(3)}\}}\left[\sum_{t=\hat{\tau}_{k-1}^{(2)} + {h_{k-1}}-\hat{\tau}_k^{(3)}}^{d} \ell_k(\hat{\boldsymbol{\theta}}^{(3)}_k,\widetilde{X}_t) + \sum_{t=d+1}^{\hat{\tau}_{k+1}^{(2)} - {h_{k+1}}-\hat{\tau}_k^{(3)}} \ell_{k+1}(\hat{\boldsymbol{\theta}}^{(3)}_{k+1},\widetilde{X}_t) \right]\\
	&=& 
	\argmax_{d \in \{\hat{\tau}_{k}^{(2)} - {h}-\hat{\tau}_k^{(3)},\ldots,\hat{\tau}_{k}^{(2)} + {h}-\hat{\tau}_k^{(3)}\}} \widetilde{W}_{k,d} 
	\end{eqnarray*}	
	where 
	\begin{equation}
	\widetilde{W}_{k,d} =\begin{cases}
	\sum_{t=1}^{d} \left(\ell_k (\hat{\boldsymbol{\theta}}^{(3)}_k, \widetilde{X}_t)-\ell_{k+1}(\hat{\boldsymbol{\theta}}^{(3)}_{k+1},\widetilde{X}_t) \right)\,, &d>0\,,\\
	0\,, & d=0\,, \\
	\sum_{t=d+1}^{0} \left(\ell_{k+1} (\hat{\boldsymbol{\theta}}^{(3)}_{k+1} ,\widetilde{X}_t)-\ell_k (\hat{\boldsymbol{\theta}}^{(3)}_k, \widetilde{X}_t) \right)\,, &d<0\,.
	\end{cases}
	\end{equation}
	Hence, it is sufficient to show that for any fixed ${s} \in \mathbb{N}$ when $T \rightarrow \infty$, we have
	\begin{equation*}
	\sup_{x \in \mathbb{R}} \left|  \widetilde{\mathbb{P}} \left(\mathop{\arg\max}\limits_{d \in \{-{s},\ldots,{s}\}}\widetilde{W}_{k,d} \leq x\right) -  \mathbb{P} \left(\mathop{\arg\max}\limits_{d \in \{-{s},\ldots,{s}\}}{W}_{k,d} \leq x \right) \right| \rightarrow 0~\text{in probability}\,,
	\end{equation*}
	where ${W}_{k,d}$ is defined in \eqref{asym_double}, 
	Moreover, for ${s} \rightarrow \infty$ when $T \rightarrow \infty$, we have
	\begin{equation*}
	\sup_{x \in \mathbb{R}} \left|  \widetilde{\mathbb{P}} \left(\widetilde{d}_k \leq x\right) -  \mathbb{P} \left(\mathop{\arg\max}\limits_{d \in \mathbb{Z}}{W}_{k,d} \leq x \right) \right| \rightarrow 0~\text{in probability}\,.
	\end{equation*}	
	
	Denote $\mathcal{L}(\mathbf{Y})$ and $\widetilde{\mathcal{L}}(\mathbf{X})$ as the distribution of random variables $\mathbf{Y}$ and $\mathbf{X}$ under probability measures $\mathbb{P}$ and $\widetilde{\mathbb{P}}$ respectively.
	We first prove the finite dimensional convergence for double-sided random walk obtained by parametric bootstrap, i.e., conditional on the sample $\{X_{1,T},\dots,X_{T,T}\}$, for any finite $m \in \mathbb{N}$, we have
	\begin{equation*}
	\widetilde{\mathcal{L}}(\boldsymbol{\widetilde{W}})\triangleq \widetilde{\mathcal{L}}(\widetilde{W}_{k,-m},\widetilde{W}_{k,-m+1},\dots, \widetilde{W}_{k,m})\overset{d}{\to} \mathcal{L}(W_{k,-m},W_{k,-m+1},\dots,W_{k,m})\triangleq \mathcal{L}(\boldsymbol{W})~\text{in probability}\,.
	\end{equation*}
	For $t>0$, the conditional Gaussian log-likelihood functions can be denoted as $\ell_{k+1}(\hat{\boldsymbol{\theta}}^{(3)}_{k+1},\widetilde{X}_t)=\widetilde{\ell}_{t}(\hat{\boldsymbol{\theta}}^{(3)}_{k+1})$, and $\ell_k(\hat{\boldsymbol{\theta}}_k^{(3)},\widetilde{X}_t)=\widetilde{\ell}_{t}(\hat{\boldsymbol{\theta}}^{(3)}_k)$. Similarly, we have $\ell_{k+1}(\boldsymbol{\theta}^0_{k+1},X_{t,T})=\ell_{t}(\boldsymbol{\theta}^0_{k+1})$, and $\ell_k(\boldsymbol{\theta}^0_k,X_{t,T})=\ell_{t}(\boldsymbol{\theta}^0_k)$. Similar expressions can be obtained for $t<0$.	It suffices to show that for all $\boldsymbol{\alpha}=(\alpha_{-m},\alpha_{-m+1},\dots,\alpha_m)^T\in\mathbb{R}^{2m+1}$, 
	$\widetilde{\mathcal{L}}(\boldsymbol{\alpha}^T\boldsymbol{\widetilde{W}})\overset{d}{\to}\mathcal{L}(\boldsymbol{\alpha}^T\boldsymbol{W})~\text{in probability}$. That is,  
	\begin{eqnarray*}
		& &\widetilde{\mathcal{L}}\left(\sum_{j=1}^{m}\left(\sum_{t=-m}^{-j} \alpha_t(\widetilde{\ell}_{-j}(\hat{\boldsymbol{\theta}}^{(3)}_{k+1})-\widetilde{\ell}_{-j}(\hat{\boldsymbol{\theta}}^{(3)}_{k}))\right)+0-\sum_{j=1}^{m}\left(\sum_{t=j}^{m}\alpha_{t}(\widetilde{\ell}_{j}(\hat{\boldsymbol{\theta}}^{(3)}_{k+1})-\widetilde{\ell}_{j}(\hat{\boldsymbol{\theta}}^{(3)}_{k}))\right)\right)\\
		&\overset{d}{\to}& \mathcal{L}\left(\sum_{j=1}^{m}\left(\sum_{t=-m}^{-j} \alpha_t(\ell_{-j}(\boldsymbol{\theta}^0_{k+1})-\ell_{-j}(\boldsymbol{\theta}^0_{k}))\right)+0-\sum_{j=1}^{m}\left(\sum_{t=j}^{m}\alpha_{t}(\ell_{j}(\boldsymbol{\theta}^0_{k+1})-\ell_{j}(\boldsymbol{\theta}^0_{k}))\right)\right)~\text{in probability}\,.
	\end{eqnarray*}
	Since $\boldsymbol{\alpha}\in\mathbb{R}^{2m+1}$ is arbitrary, it in turn suffices to show that
	\begin{eqnarray}\label{ll.convergence}
	& &\widetilde{\mathcal{L}}\big(\widetilde{\ell}_{-m}(\hat{\boldsymbol{\theta}}^{(3)}_{k+1})-\widetilde{\ell}_{-m}(\hat{\boldsymbol{\theta}}^{(3)}_{k}),\dots,\widetilde{\ell}_{m}(\hat{\boldsymbol{\theta}}^{(3)}_{k+1})-\widetilde{\ell}_{m}(\hat{\boldsymbol{\theta}}^{(3)}_{k})
	\big)\nonumber\\
	& \overset{d}{\to}& \mathcal{L}\big(\ell_{-m}(\boldsymbol{\theta}_{k+1}^0)-\ell_{-m}(\boldsymbol{\theta}_{k}^0),\dots,\ell_{m}(\boldsymbol{\theta}_{k+1}^0)-\ell_{m}(\boldsymbol{\theta}_{k}^0)\big)~\text{in probability}\,.
	\end{eqnarray}
	By Theorem 4.2 of \cite{Dahlhaus1997}, we have $\hat{\boldsymbol{\theta}}^{(3)}_{k}\overset{p}{\to}\boldsymbol{\theta}_{k}^0$ and $\hat{\boldsymbol{\theta}}^{(3)}_{k+1}\overset{p}{\to}\boldsymbol{\theta}_{k+1}^0$.	
	Moreover, $\ell_t(\boldsymbol{\theta})$ is continuous with respect to $\boldsymbol{\theta}$ and $X_{t,T}$, 
%
	Therefore, by a similar argument in the proof of Theorem 2.1 in \cite{Yau2019}, it follows that $\widetilde{\mathcal{L}}(\boldsymbol{\widetilde{W}})\overset{d}{\to}\mathcal{L}(\boldsymbol{W})~\text{in probability}$. Hence, the finite dimensional convergence for double-sided random walk is proved for any $m \in \mathbb{N}$. 
	By the argmax continuous mapping theorem, we have for any finite $s \in \mathbb{N}$ that
	\begin{equation*}
	\mathop{\arg\max}_{d\in\{-{s},\dots,{s}\}}\widetilde{W}_{k,d}\overset{d}{\to}\mathop{\arg\max}_{d\in\{-{s},\dots,{s}\}}W_{k,d}~\text{in probability}\,.
	\end{equation*}
	Next, \cite{billingsley99} gives a characterization of weak convergence in the space of sequence of real numbers by convergence of the finite-dimensional distributions, see Example 1.2 in P.9 and Example 2.4 in P.19 in \cite{billingsley99} which state that the finite dimensional sets is a convergence-determining class. It implies that, for ${s} \rightarrow \infty$ when $T \rightarrow \infty$, $\{\widetilde{W}_{k,d}: d \in \{-{s},\dots,{s}\}\} \overset{d}{\to} \{W_{k,d}: d \in \mathbb{Z}\}~\text{in probability}$.
	Then, by using the argmax continuous mapping theorem, we have $\widetilde{d}_k \overset{d}{\to} \mathop{\arg\max}_{d\in \mathbb{Z}}W_{k,d}~\text{in probability}$. Hence, the proof of Theorem \ref{thm4.4} is completed because the second part of the theorem follows directly from Theorem \ref{thm4.3_kink}.
\end{proof}


\subsection{Derivation of the first and second order derivatives of \texorpdfstring{$S_{k,T}(\boldsymbol{\eta}k)$}{S{k,T}(eta_k)} with respect to \texorpdfstring{$\boldsymbol{\eta}_k$}{eta_k}}\label{derivatives_L}

\begin{proof}[\unskip\nopunct]
Recall that the covariance of the first derivative of \( S_{k,T}(\boldsymbol{\eta}_k) \) with respect to \( \boldsymbol{\eta}_k \) is given by
\begin{align*}
\nabla_{\boldsymbol{\eta}} S_{k,T} (\boldsymbol{\eta}_k) \nabla_{\boldsymbol{\eta}} S_{k,T} (\boldsymbol{\eta}_k)' = \frac{1}{\hat{T}^{(2)}_{k,k+1}} \sum_{t=\hat{\tau}_{k-1}^{(2)} + {h_{k-1}}}^{\hat{\tau}_{k+1}^{(2)} - {h_{k+1}}}\left( \frac{\partial \ell(\boldsymbol{\eta}_k,X_{t,T})}{\partial \boldsymbol{\eta}_k}\right)\left(\frac{\partial \ell(\boldsymbol{\eta}_k,X_{t,T})}{\partial \boldsymbol{\eta}_k}\right)'.
\end{align*}
Similarly, the second derivative of \( S_{k,T}(\boldsymbol{\eta}_k) \), evaluated at \( \boldsymbol{\eta}_k \), is
$$\nabla_{\boldsymbol{\eta}}^2 S_{k,T} (\boldsymbol{\eta}_k)=\frac{1}{\hat{T}^{(2)}_{k,k+1}} \sum_{t=\hat{\tau}_{k-1}^{(2)} + {h_{k-1}}}^{\hat{\tau}_{k+1}^{(2)} - {h_{k+1}}} \frac{\partial^2 \ell(\boldsymbol{\eta}_k,X_{t,T})}{\partial \boldsymbol{\eta}_k \partial \boldsymbol{\eta}_k'}.$$
	Next, we derive the first and second order derivatives of $\ell(\boldsymbol{\eta}_k,X_{t,T})$ with respect to $\boldsymbol{\eta}_k$. For notation simplicity, we define
	$\phi(u)=\phi_{\boldsymbol{\gamma},\boldsymbol{\alpha},k}(u)$ and $\phi^{(i)}(u)=\phi^{(i)}_{\boldsymbol{\gamma},\boldsymbol{\alpha},k}(u)$ for $i=1,\ldots,p^*$. Similarly, define $\sigma(u)=\sigma_{\xi,\boldsymbol{\beta},k}(u)$. Without loss of generality, we only consider the case $\hat{p}_{k}^{(2)} = \hat{p}_{k+1}^{(2)}$ and $\hat{q}_{k}^{(2)} = \hat{q}_{k+1}^{(2)}$. For the cases  $\hat{p}_{k}^{(2)} \neq \hat{p}_{k+1}^{(2)}$ and $\hat{q}_{k}^{(2)} \neq \hat{q}_{k+1}^{(2)}$, the results can be derived analogously with more involved notations.
	
	Define $p^*=\hat{p}_{k}^{(2)} = \hat{p}_{k+1}^{(2)}$, $q^*=\hat{q}_{k}^{(2)} = \hat{q}_{k+1}^{(2)}$, vector $\boldsymbol{0}_m=(0,\ldots,0)'$ with size $m$, vector $\boldsymbol{1}_i=(0,\ldots,1,\ldots,0)'$ with size $p^*$ and the only non-zero element is the $i$-th element which is $1$, $m\times n$ matrix $\boldsymbol{0}_{m,n}$ with all elements equal $0$, $z=p^*+1+p^*q^*+q^*+p^*q^*+q^*+1$ which is the size of the parameter vector $\boldsymbol{\eta}_k$, vector $\Lambda_-=\left((\frac{t}{T}-r_k)_{-},\ldots,(\frac{t}{T}-r_k)_{-}^{q^*}\right)'$, vector $\Lambda_+=\left((\frac{t}{T}-r_k)_{+},\ldots,(\frac{t}{T}-r_k)_{+}^{q^*}\right)'$, vector $\Lambda^{(1)}_-=\frac{d \Lambda_-}{d r_k}=\left(- \mathbbm{1}_{\{\frac{t}{T}<r_k\}},-2 (\frac{t}{T}-r_k)_{-},\ldots,-q^* (\frac{t}{T}-r_k)_{-}^{q^*-1}\right)'$, and vector $\Lambda^{(1)}_+=\frac{d \Lambda_+}{d r_k}=\left(- \mathbbm{1}_{\{\frac{t}{T}>r_k\}},-2 (\frac{t}{T}-r_k)_{+},\ldots,-q^* (\frac{t}{T}-r_k)_{+}^{q^*-1}\right)'$. 
	
	We also define vectors $\Lambda^{\phi,i}_-=\left(\boldsymbol{0}_{(i-1)q^*},\Lambda_-,\boldsymbol{0}_{(p^*-i)q^*}\right)$ and  $\Lambda^{\phi,i}_+=\left(\boldsymbol{0}_{(i-1)q^*},\Lambda_+,\boldsymbol{0}_{(p^*-i)q^*}\right)$ both of size $p^*q^*$ for $i=1,\ldots,p^*$, and vectors $\Lambda^{\phi,i,(1)}_-=\frac{d \Lambda^{\phi,i}_-}{d r_k}=\left(\boldsymbol{0}_{(i-1)q^*},\Lambda^{(1)}_-,\boldsymbol{0}_{(p^*-i)q^*}\right)'$ and $\Lambda^{\phi,i,(1)}_+=\frac{d \Lambda^{\phi,i}_+}{d r_k}=\left(\boldsymbol{0}_{(i-1)q^*},\Lambda^{(1)}_+,\boldsymbol{0}_{(p^*-i)q^*}\right)'$. 
	
	By the chain rule, the first derivative of $\ell(\boldsymbol{\eta}_k,X_{t,T})$ with respect to $\boldsymbol{\eta}_k$ is

	\begin{eqnarray*}
	\nabla_{\boldsymbol{\eta}} \ell(\boldsymbol{\eta}_k,X_{t,T}) & = & \frac{\partial \ell(\boldsymbol{\eta}_k,X_{t,T})}{\partial \boldsymbol{\eta}_k} =   \nabla_{\sigma} \ell(\boldsymbol{\eta}_k,X_{t,T}) \nabla_{\boldsymbol{\eta}} \sigma + \sum_{i=1}^{p^*} \nabla_{\phi^{(i)}} \ell(\boldsymbol{\eta}_k,X_{t,T}) \nabla_{\boldsymbol{\eta}} \phi^{(i)} \\
		& = & \left[-\frac{1}{\sigma\left(\frac{t}{T}\right)}+\frac{\left(X_{t,T}- {\boldsymbol{X}_{t-p^*}^{t-1}}' \phi\left(\frac{t}{T}\right) \right)^2}{ \sigma^3\left(\frac{t}{T}\right)}\right] \left(\boldsymbol{0}'_{p^*},1,\boldsymbol{0}'_{p^*q^*},\Lambda'_-,\boldsymbol{0}'_{p^*q^*},\Lambda'_+,\frac{d \sigma}{d r_k}\right)'\\
		& & \quad\quad\quad\quad\quad + \sum_{i=1}^{p^*} \left[\frac{\left(X_{t,T}- {\boldsymbol{X}_{t-p^*}^{t-1}}' \phi\left(\frac{t}{T}\right) \right)}{ \sigma^2\left(\frac{t}{T}\right)} X_{t-i,T}\right] \left(\boldsymbol{1}_i,0,\Lambda^{\phi,i}_-,\boldsymbol{0}'_{q^*} ,\Lambda^{\phi,i}_+,\boldsymbol{0}'_{q^*},\frac{d \phi^{(i)}}{d r_k}\right)'
	\end{eqnarray*}
where the first derivatives of $\sigma$ and $\phi^{(i)}$ with respect to the relative change-point $r_k$ are respectively 
\begin{equation*}
\frac{d \sigma}{d r_k} = - \beta_{k1}\mathbbm{1}_{\{\frac{t}{T}<r_k\}} - \beta_{(k+1)1}\mathbbm{1}_{\{\frac{t}{T}>r_k\}} - \sum_{j=2}^{q^*} j \beta_{kj} \left(\frac{t}{T}-r_k\right)_{-}^{j-1} - \sum_{j=2}^{q^*} j \beta_{(k+1)j} \left(\frac{t}{T}-r_k\right)_{+}^{j-1} \,,
\end{equation*}
and
\begin{equation*}
\frac{d \phi^{(i)}}{d r_k} = - \alpha_{ki1} \mathbbm{1}_{\{\frac{t}{T}<r_k\}} - \alpha_{(k+1)i1}\mathbbm{1}_{\{\frac{t}{T}>r_k\}}  - \sum_{j=2}^{q^*} j \alpha_{kij}\left(\frac{t}{T}-r_k\right)_{-}^{j-1} - \sum_{j=2}^{q^*} j  \alpha_{(k+1)ij}\left(\frac{t}{T}-r_k\right)_{+}^{j-1} \,.
\end{equation*}

Similarly, by the chain rule and the product rule, the second derivative of $\ell(\boldsymbol{\eta}_k,X_{t,T})$ with respect to $\boldsymbol{\eta}_k$ is
\begin{eqnarray*}
	& & \nabla^2_{\boldsymbol{\eta}} \ell(\boldsymbol{\eta}_k,X_{t,T})  =  \frac{\partial^2 \ell(\boldsymbol{\eta}_k,X_{t,T})}{\partial \boldsymbol{\eta}_k \partial \boldsymbol{\eta}_k'}\\
	& = &   \nabla^2_{\sigma} \ell(\boldsymbol{\eta}_k,X_{t,T}) {\nabla_{\boldsymbol{\eta}} \sigma  \left(\nabla_{\boldsymbol{\eta}} \sigma\right)'}+ \sum_{i=1}^{p^*} \nabla^2_{\sigma,\phi^{(i)}} \ell(\boldsymbol{\eta}_k,X_{t,T}) \nabla_{\boldsymbol{\eta}} \sigma  \nabla_{\boldsymbol{\eta}'} \phi^{(i)} + \nabla_{\sigma} \ell(\boldsymbol{\eta}_k,X_{t,T}) \nabla^2_{\boldsymbol{\eta}} \sigma \\
	& & \quad + \sum_{i=1}^{p^*} \left\{\nabla^2_{\phi^{(i)},\sigma} \ell(\boldsymbol{\eta}_k,X_{t,T})\nabla_{\boldsymbol{\eta}} \phi^{(i)}  \nabla_{\boldsymbol{\eta}'} \sigma + \sum_{j=1}^{p^*} \left[ \nabla^2_{\phi^{(i)},\phi^{(j)}} \ell(\boldsymbol{\eta}_k,X_{t,T})\nabla_{\boldsymbol{\eta}} \phi^{(i)} \nabla_{\boldsymbol{\eta}'} \phi^{(j)} \right] \right.\\
	& & \quad\quad \left. + \nabla_{\phi^{(i)}} \ell(\boldsymbol{\eta}_k,X_{t,T}) \nabla^2_{\boldsymbol{\eta}'} \phi^{(i)}  \right\}\\
	& = &  \left[\frac{1}{\sigma^2\left(\frac{t}{T}\right)} - \frac{3\left(X_{t,T}- {\boldsymbol{X}_{t-p^*}^{t-1}}' \phi\left(\frac{t}{T}\right) \right)^2}{ \sigma^4\left(\frac{t}{T}\right)}\right]\times \\
&&\quad\quad\quad{\left(\boldsymbol{0}'_{p^*},1,\boldsymbol{0}'_{p^*q^*},\Lambda'_-,\boldsymbol{0}'_{p^*q^*},\Lambda'_+,\frac{d \sigma}{d r_k}\right)'\left(\boldsymbol{0}'_{p^*},1,\boldsymbol{0}'_{p^*q^*},\Lambda'_-,\boldsymbol{0}'_{p^*q^*},\Lambda'_+,\frac{d \sigma}{d r_k}\right)} \\
	& &  + \sum_{i=1}^{p^*} \left[- \frac{2\left(X_{t,T}- {\boldsymbol{X}_{t-p^*}^{t-1}}' \phi\left(\frac{t}{T}\right) \right)}{ \sigma^3\left(\frac{t}{T}\right)} X_{t-i,T}\right] \times \\
	& & \quad\quad\quad\quad\quad\quad\quad\quad\quad \left(\boldsymbol{0}'_{p^*},1,\boldsymbol{0}'_{p^*q^*},\Lambda'_-,\boldsymbol{0}'_{p^*q^*},\Lambda'_+,\frac{d \sigma}{d r_k}\right)' \left(\boldsymbol{1}_i,0,\Lambda^{\phi,i}_-,\boldsymbol{0}'_{q^*} ,\Lambda^{\phi,i}_+,\boldsymbol{0}'_{q^*},\frac{d \phi^{(i)}}{d r_k}\right) \\
	& & \quad + \left[-\frac{1}{\sigma\left(\frac{t}{T}\right)}+\frac{\left(X_{t,T}- {\boldsymbol{X}_{t-p^*}^{t-1}}' \phi\left(\frac{t}{T}\right) \right)^2}{ \sigma^3\left(\frac{t}{T}\right)}\right] \begin{pmatrix}
		\boldsymbol{0}_{z-1,z-1} & \frac{d \nabla_{\boldsymbol{\eta}} \sigma_{[z-1]}}{d r_k} \\
		\frac{d \nabla_{\boldsymbol{\eta}} \sigma_{[z-1]}}{d r_k} ' & \frac{d^2 \sigma}{d r_k^2}
	\end{pmatrix} \\
	& & \quad\quad + \sum_{i=1}^{p^*} \left\{\left[- \frac{2\left(X_{t,T}- {\boldsymbol{X}_{t-p^*}^{t-1}}' \phi\left(\frac{t}{T}\right) \right)}{ \sigma^3\left(\frac{t}{T}\right)} X_{t-i,T}\right] \right. \times\\
	& & \quad\quad\quad\quad\quad\quad\quad\quad\quad\quad\quad \left(\boldsymbol{1}_i,0,\Lambda^{\phi,i}_-,\boldsymbol{0}'_{q^*} ,\Lambda^{\phi,i}_+,\boldsymbol{0}'_{q^*},\frac{d \phi^{(i)}}{d r_k}\right)' \left(\boldsymbol{0}'_{p^*},1,\boldsymbol{0}'_{p^*q^*},\Lambda'_-,\boldsymbol{0}'_{p^*q^*},\Lambda'_+,\frac{d \sigma}{d r_k}\right)\\
	& & \quad\quad\quad + \sum_{j=1}^{p^*} \left[- \frac{X_{t-i,T} X_{t-j,T}}{ \sigma^2\left(\frac{t}{T}\right)} \right]\left(\boldsymbol{1}_i,0,\Lambda^{\phi,i}_-,\boldsymbol{0}'_{q^*} ,\Lambda^{\phi,i}_+,\boldsymbol{0}'_{q^*},\frac{d \phi^{(i)}}{d r_k}\right)'\left(\boldsymbol{1}_j,0,\Lambda^{\phi,j}_-,\boldsymbol{0}'_{q^*} ,\Lambda^{\phi,j}_+,\boldsymbol{0}'_{q^*},\frac{d \phi^{(j)}}{d r_k}\right) \\
	& & \quad\quad\quad\quad \left. + \left[\frac{\left(X_{t,T}- {\boldsymbol{X}_{t-p^*}^{t-1}}' \phi\left(\frac{t}{T}\right) \right)}{ \sigma^2\left(\frac{t}{T}\right)} X_{t-i,T}\right] \begin{pmatrix}
		\boldsymbol{0}_{z-1,z-1} & \frac{d \nabla_{\boldsymbol{\eta}} \phi^{(i)}_{[z-1]}}{d r_k} \\
		\frac{d \nabla_{\boldsymbol{\eta}} \phi^{(i)}_{[z-1]}}{d r_k} ' & \frac{d^2 \phi^{(i)}}{d r_k^2}
	\end{pmatrix}  \right\}
\end{eqnarray*}
where
$\nabla_{\boldsymbol{\eta}} \sigma_{[z-1]}$ and $\nabla_{\boldsymbol{\eta}} \phi^{(i)}_{[z-1]}$ are the first $z-1$ elements of vectors $\nabla_{\boldsymbol{\eta}} \sigma$ and $\nabla_{\boldsymbol{\eta}} \phi^{(i)}$ respectively and hence
\begin{eqnarray*}
	\frac{d \nabla_{\boldsymbol{\eta}} \sigma_{[z-1]}}{d r_k}  =\left(\boldsymbol{0}'_{p^*+1+p^*q^*},{\Lambda^{(1)}_-}',\boldsymbol{0}'_{p^*q^*},{\Lambda^{(1)}_+}'\right)' ,~\text{and}~
	\frac{d \nabla_{\boldsymbol{\eta}} \phi^{(i)}_{[z-1]}}{d r_k}  =\left(\boldsymbol{0}'_{p^*+1},{\Lambda^{\phi,i,(1)}_-}',\boldsymbol{0}'_{q^*},{\Lambda^{\phi,i,(1)}_+}',\boldsymbol{0}'_{q^*}\right)' \,,
\end{eqnarray*}
and the second derivatives of $\sigma$ and $\phi^{(i)}$ with respect to the relative change-point $r_k$ are respectively
\begin{equation*}
	\frac{d^2 \sigma}{d r_k^2} =  2\beta_{k2}\mathbbm{1}_{\{\frac{t}{T}<r_k\}} + 2\beta_{(k+1)2}\mathbbm{1}_{\{\frac{t}{T}>r_k\}} + \sum_{j=3}^{q^*} (j-1)j \beta_{kj} \left(\frac{t}{T}-r_k\right)_{-}^{j-2} + \sum_{j=3}^{q^*} (j-1)j \beta_{(k+1)j} \left(\frac{t}{T}-r_k\right)_{+}^{j-2} \,,
\end{equation*}
and
\begin{equation*}
\frac{d^2 \phi^{(i)}}{d r_k^2} = 2 \alpha_{ki2} \mathbbm{1}_{\{\frac{t}{T}<r_k\}} + 2 \alpha_{(k+1)i2}\mathbbm{1}_{\{\frac{t}{T}>r_k\}}  + \sum_{j=3}^{q^*} (j-1)j \alpha_{kij}\left(\frac{t}{T}-r_k\right)_{-}^{j-2} + \sum_{j=3}^{q^*} (j-1)j  \alpha_{(k+1)ij}\left(\frac{t}{T}-r_k\right)_{+}^{j-2} \,.
\end{equation*}
\end{proof}

\subsection{Correspondence of jumps and kinks in the spectral density function $f(u,\lambda)$ to jumps and kinks in the parameter curves}\label{jumpkinkillust}

\begin{proof}[\unskip\nopunct]
In this section, we show that jumps and kinks in parameter curves for tvAR models respectively correspond to jumps and kinks in the spectral density function $f(u,\lambda)$ in \eqref{tvspecden} for some frequency $\lambda \in [-\pi,\pi]$. To facilitate discussion, we illustrate the main idea using a tvAR($1$) model. For general tvAR($p$) models, the same rationale can be applied.

Recall a change in segment $k$ at relative location $\frac{\tau_k}{T}$ is a {\it jump} if any of the parameter curves breaks, i.e., $\phi_{k,i,\boldsymbol{\theta}_{k}}(\frac{\tau_k}{T})\neq \phi_{k+1,i,\boldsymbol{\theta}_{k+1}}(\frac{\tau_k}{T})\,,\mbox{ for some }i=1,\ldots,p_{max}\,,\mbox{ or }\sigma_{k,\boldsymbol{\theta}_{k}}(r_k)\neq\sigma_{k+1,\boldsymbol{\theta}_{k+1}}(r_k)$. 
Consider the tvAR($1$) model as illustration. Define $\phi_{k,1}=\phi_{k,1,\boldsymbol{\theta}_{k}}(\frac{\tau_k}{T})$ and $\sigma_{k}=\sigma_{k,\boldsymbol{\theta}_{k}}(r_k)$, and similarly for $\phi_{k+1,1}$ and $\sigma_{k+1}$. The spectral density function at relative location $r_k$ for the two segments are 
$$f_k\left(\frac{\tau_k}{T},\lambda\right)=\frac{\sigma_{k}^2}{2\pi} \frac{1}{1+\phi_{k,1}^2-2\phi_{k,1}\cos(\lambda)}\,,$$ and $$f_{k+1}\left(\frac{\tau_k}{T},\lambda\right)=\frac{\sigma_{k+1}^2}{2\pi} \frac{1}{1+\phi_{k+1,1}^2-2\phi_{k+1,1}\cos(\lambda)}\,,$$ respectively.
Assume $\phi_{k,1} \neq \phi_{k+1,1}$ or $\sigma_{k} \neq \sigma_{k+1}$. Suppose the spectral density is continuous at relative location $\frac{\tau_k}{T}$ for all frequency $\lambda$, i.e., $f_k(\frac{\tau_k}{T},\lambda)=f_{k+1}(\frac{\tau_k}{T},\lambda)$ for all $\lambda \in [-\pi,\pi]$, then after rearranging terms, we obtain a linear equation in terms of $\cos(\lambda)$, i.e., $A\cos(\lambda)+B=0$, where 
\begin{eqnarray*}
	A & = & 2\sigma_{k+1}^2\phi_{k,1}-2\sigma_{k}^2\phi_{k+1,1} \,,\\
	B & = & \sigma_{k}^2-\sigma_{k+1}^2+\sigma_{k}^2\phi_{k+1,1}^2-\sigma_{k+1}^2\phi_{k,1}^2 \,.
\end{eqnarray*}
Since there are infinitely many solutions for the above linear equation, we have $A \equiv 0 $ and $B \equiv 0$, and hence we have $\sigma_{k} = \sigma_{k+1}$ and $\phi_{k,1} = \phi_{k+1,1}$, which contradict the assumptions.

On the other hand, recall that a change in segment $k$ at relative location $r_k$ is a {\it kink} if all curves are continuous but some of them have an abrupt change in the slope, i.e.,  \eqref{def.jump} does not hold, but $\phi^{(1)}_{k,i,\boldsymbol{\theta}_{k}}(r_k)\neq \phi^{(1)}_{k+1,i,\boldsymbol{\theta}_{k+1}}(r_k)\,, \mbox{ for some } i=1,\ldots,p_{max}\,,\mbox{ or }\sigma^{(1)}_{k,\boldsymbol{\theta}_{k}}(r_k)\neq \sigma^{(1)}_{k+1,\boldsymbol{\theta}_{k+1}}(r_k)$, where the superscript stands for taking the first derivative. 
Consider the tvAR($1$) model as illustration. By continuity we have $\phi_{k,1}=\phi_{k+1,1}$ and $\sigma_{k}=\sigma_{k+1}$. Define $\phi_{k,1}^{(1)}=\phi_{k,1,\boldsymbol{\theta}_{k}}^{(1)}(\frac{\tau_k}{T})$ and $\sigma_{k}^{(1)}=\sigma_{k,\boldsymbol{\theta}_{k}}^{(1)}(r_k)$, and similarly for $\phi_{k+1,1}^{(1)}$ and $\sigma_{k+1}^{(1)}$. The first derivative of the spectral density function at relative location $r_k$ for the two segments are 
\begin{eqnarray*}
f_k^{(1)}\left(\frac{\tau_k}{T},\lambda\right) & = & \frac{\partial{f_k(u,\lambda)}}{\partial{\sigma_{k}}}\Big|_{(\phi_{k,1},\sigma_{k})}\frac{\partial{\sigma_{k}}}{\partial{u}}\Big|_{\frac{\tau_k}{T}}+\frac{\partial{f_k(u,\lambda)}}{\partial{\phi_{k,1}}}\Big|_{(\phi_{k,1},\sigma_{k})}\frac{\partial{\phi_{k,1}}}{\partial{u}}\Big|_{\frac{\tau_k}{T}}\\
& = & \frac{\sigma_{k}}{\pi} \frac{\sigma_{k}^{(1)}}{1+\phi_{k,1}^2-2\phi_{k,1}\cos(\lambda)}-\frac{\sigma_{k}^2}{\pi} \frac{(\phi_{k,1}-\cos(\lambda))\phi_{k,1}^{(1)}}{(1+\phi_{k,1}^2-2\phi_{k,1}\cos(\lambda))^2}\\
& = & \frac{\sigma_{k}}{\pi} \frac{(1+\phi_{k,1}^2-2\phi_{k,1}\cos(\lambda))\sigma_{k}^{(1)}}{(1+\phi_{k,1}^2-2\phi_{k,1}\cos(\lambda))^2}-\frac{\sigma_{k}^2}{\pi} \frac{(\phi_{k,1}-\cos(\lambda))\phi_{k,1}^{(1)}}{(1+\phi_{k,1}^2-2\phi_{k,1}\cos(\lambda))^2}\,,
\end{eqnarray*}
and
\begin{eqnarray*}
	f_{k+1}\left(\frac{\tau_k}{T},\lambda\right) & = & \frac{\partial{f_{k+1}(u,\lambda)}}{\partial{\sigma_{k+1}}}\Big|_{(\phi_{{k+1},1},\sigma_{k+1})}\frac{\partial{\sigma_{k+1}}}{\partial{u}}\Big|_{\frac{\tau_k}{T}}+\frac{\partial{f_{k+1}(u,\lambda)}}{\partial{\phi_{{k+1},1}}}\Big|_{(\phi_{{k+1},1},\sigma_{k+1})}\frac{\partial{\phi_{{k+1},1}}}{\partial{u}}\Big|_{\frac{\tau_k}{T}}\\
	& = & \frac{\sigma_{k+1}}{\pi} \frac{\sigma_{k+1}^{(1)}}{1+\phi_{{k+1},1}^2-2\phi_{{k+1},1}\cos(\lambda)}-\frac{\sigma_{k+1}^2}{\pi} \frac{(\phi_{{k+1},1}-\cos(\lambda))\phi_{{k+1},1}^{(1)}}{(1+\phi_{{k+1},1}^2-2\phi_{{k+1},1}\cos(\lambda))^2}\\
	& = & \frac{\sigma_{k}}{\pi} \frac{(1+\phi_{k,1}^2-2\phi_{k,1}\cos(\lambda))\sigma_{k+1}^{(1)}}{(1+\phi_{k,1}^2-2\phi_{k,1}\cos(\lambda))^2}-\frac{\sigma_{k}^2}{\pi} \frac{(\phi_{k,1}-\cos(\lambda))\phi_{{k+1},1}^{(1)}}{(1+\phi_{k,1}^2-2\phi_{k,1}\cos(\lambda))^2}\,,
\end{eqnarray*}
respectively. Assume $\phi_{k,1}^{(1)} \neq \phi_{k+1,1}^{(1)}$ or $\sigma_{k}^{(1)} \neq \sigma_{k+1}^{(1)}$. Suppose the first derivative of the spectral density is continuous at relative location $\frac{\tau_k}{T}$ for all frequency $\lambda$, i.e., $f_k^{(1)}(\frac{\tau_k}{T},\lambda)=f_{k+1}^{(1)}(\frac{\tau_k}{T},\lambda)$ for all $\lambda \in [-\pi,\pi]$, then after rearranging terms, we obtain a linear equation in terms of $\cos(\lambda)$, i.e., $A\cos(\lambda)+B=0$, where 
\begin{eqnarray*}
	A & = & \sigma_{k}^2(\phi_{k,1}^{(1)}-\phi_{k+1,1}^{(1)})-2\sigma_{k}\phi_{k,1} (\sigma_{k}^{(1)}-\sigma_{k+1}^{(1)})\,,\\
	B & = & (\sigma_{k}+\sigma_{k}\phi_{k,1}^2)(\sigma_{k}^{(1)}-\sigma_{k+1}^{(1)})-\sigma_{k}^2\phi_{k,1}(\phi_{k,1}^{(1)}-\phi_{k+1,1}^{(1)}) \,.
\end{eqnarray*}
Since there are infinitely many solutions for the above linear equation, we have $A \equiv 0 $ and $B \equiv 0$, and hence we have $\phi_{k,1}^{(1)} = \phi_{k+1,1}^{(1)}$ and $\sigma_{k}^{(1)} = \sigma_{k+1}^{(1)}$, which contradict the assumptions.

\end{proof}

\bibliographystyle{apalike}
\bibliography{literature}

\end{document}